\documentclass[twoside,fleqn]{ActaStyle}
\usepackage[colorlinks=true, pdfstartview=FitV, linkcolor=black,
            citecolor=black, urlcolor=black]{hyperref}
\usepackage{graphicx}
\usepackage{dcolumn}
\usepackage{rotating}
\usepackage{times,cite}
\usepackage{amsmath}
\usepackage{amssymb}
\usepackage{amsfonts}
\usepackage{amsthm}
\usepackage{multicol}
\usepackage{dsfont}
\usepackage{float}
\usepackage{color}
\usepackage{wrapfig}

\usepackage{fancyhdr}
\renewcommand{\subsectionmark}[1]{}
\renewcommand{\thesection}{\arabic{section}}
\renewcommand{\thesubsection}{\thesection.\arabic{subsection}}

\renewcommand{\thefigure}{\thesection.\arabic{figure}}
\renewcommand{\thetable}{\thesection.\arabic{table}}

\def\authorheading{J\'{a}n Ivan\v{c}o}
\def\shorttitle{Practical photoemission characterization of molecular films and related interfaces}

\begin{document}
\pagerange{207}{260}

\title{PRACTICAL PHOTOEMISSION CHARACTERIZATION OF MOLECULAR FILMS AND RELATED INTERFACES}

\author{J\'{a}n Ivan\v{c}o\email{Jan.Ivanco@savba.sk}}
{Institute of Physics, Slovak Academy of Sciences, D\'ubravsk\'a cesta 9, Bratislava, Slovakia}

\abstract{
Even though the term `organic electronics' evokes rather organic devices, 
a significant part of its scope deals with physical properties of `active 
elements' such as organic films and interfaces. Examination of the film 
growth and the evolution of the interface formation are particularly needful 
for the understanding a mechanism controlling their final properties. 
Performing such experiments in an ultra-high vacuum allows both to `stretch' 
the time scale for pseudo real-time observations and to control properties 
of the probed systems on the atomic level. Photoemission technique probes 
directly electronic and chemical structure and it has thereby established 
among major tools employed in the field.
This review primarily focuses to electronic properties of oligomeric 
molecular films and their interfaces examined by photoemission. Yet, it does 
not aspire after a complete overview on the topic; it rather aims to 
otherwise standard issues encountered at the photoemission characterization 
and analysis of the organic materials, though requiring to consider 
particularities of molecular films in terms of the growth, electronic 
properties, and their characterization and analysis. In particular, the 
fundamental electronic parameters of molecular films such as the work 
function, the ionization energy, and the interfacial energy level alignment, 
and their interplay, will be pursued with considering often neglected 
influence of the molecular orientation. Further, the implication on the band 
bending in molecular films based on photoemission characterization, and a 
model on the driving mechanism for the interfacial energy level alignment 
will be addressed.}

\vspace{0.3cm}

\begin{minipage}{2.5cm}
\quad{\small {\sf KEYWORDS:}}
\end{minipage}
\begin{minipage}{10cm}
Photoemission, Molecular films, Organic 
semiconductors, Molecular orientation, Ionization energy, Work function, 
Energy level alignment, Band bending, Interfacial dipole, Organic-metal 
interface, Organic hetero\-interface, Organic-inorganic interface.
\end{minipage}

\newpage
\tableofcontents

\newpage
\section*{List of symbols}
\fancyhead[LO]{List of symbols}
\addcontentsline{toc}{section}{List of symbols}

\begin{tabular}{ll}
$e$ & elementary charge, 1.602$\times $10$^{ - 19}$ C \\

$E_{B}$ & electron binding energy \\

$E_{EA}$ & electron affinity \\

$E_{F}$ & Fermi level \\

$E_{G}$ & energy gap \\

$E_{I}$ & ionization energy \\

$E_{I,el}$ & (first) ionization energy of elements \\

$E_{I,film}$ & (first) ionization energy of a film \\

$E_{I,org}$ & (first) ionization energy of organics \\

$E_{K}$ &  electron kinetic energy \\

$E_{vac}$ & vacuum level \\

$E_{vac,ML}$ & vacuum level upon the formation of first monolayer  \\

$E_{Vs}$ & valence band edge on the surface, \\

$E_{seco}$ & secondary electrons cut-off \\

$h$ & Planck's constant, 6.626$\times $10$^{ - 34}$ Js \\

$I_{0}$ & (photoemission) intensity of a bare substrate \\

$I_{film}$ & (photoemission) film-related intensity  \\

$I_{subs}$ & (photoemission) substrate-related intensity \\

$I_{\infty}$ & (photoemission) intensity of a semiinfinite film \\

$m/e$ & mass-to-charge ratio \\

$Q_{D}$ & depletion zone charge \\

$Q_{ss}$ & charge in surface states \\

$Q_{it}$ & charge in interface states \\

$S$ & slope parameter  \\

$t$ & film thickness \\

$U$ & bias \\

$U_{an}$ & voltage applied to the analyzer to detect an emitted electron \\

$U_{bi}$ & build-in potential \\

$w$ & depletion zone width \\

$\delta_{i}$ & interface region thickness \\

$\Delta \phi _{L \to U}$ & work function change induced by the orientational 
transition (of molecules) \\

$\Delta \phi _{ML}$ & work function change upon growth of 1$^{\rm st}$ 
monolayer (ML) \\

$\Delta \phi _{ML(L)}$ & work function change upon growth of 1$^{\rm st}$ 
ML formed by lying molecules \\

$\Delta \phi _{ML(U)}$ & work function change upon growth of 1$^{\rm st}$ 
ML formed by upright-oriented molecules \\

$\Delta _{BB}$ & band-bending magnitude \\

$\lambda$ &  electron inelastic mean free path \\

$\mu$ &  electrochemical potential \\

$\nu$ &  frequency \\

$\phi$ &  work function \\

$\phi_{an}$ & work function of energy analyzer \\

$\phi_{film}$ & work function of a film \\

$\phi_{ID}$ & interface dipole potential \\

$\phi_{int}$ & intrinsic work function \\

$\phi_{int,L}$ & intrinsic work function of a film formed by lying molecules \\

$\phi_{int,U}$ & intrinsic work function of a film formed by upright-oriented molecules \\

$\phi_{s}$ & semiconductor work function \\

$\phi_{SD}$ & surface-dipole potential \\

$\phi_{subs}$ & work function of a substrate \\

$\varphi_{e}$ & injection barrier for electrons \\ 

$\varphi_{h}$ & injection barrier for holes \\
\end{tabular}

\newpage
\section*{List of acronyms/abbreviations}
\fancyhead[LO]{List of acronyms/abbreviations}
\addcontentsline{toc}{section}{List of acronyms/abbreviations}

{\it\small\qquad\ \  (Please, refer to Tab.~1.1 for the chemical formulae)}
\bigskip

\begin{tabular}{ll}
BA &  band alignment \\

BB &  band bending \\

ELA &  energy level alignment \\

HOMO &  highest occupied molecular orbital \\

ID &  interfacial dipole \\

FET &  field-effect transistor \\

LNT &  liquid nitrogen temperature \\

L$ \to $U &  (orientational transition) from lying- towards upright-oriented molecules \\

LUMO &  lowest unoccupied molecular orbital \\

ML &  monolayer \\

OFET &  organic field-effect transistor \\

OLED &  organic light-emitting diode \\

OPC &  organic photovoltaic cell \\

PCS &  photoionization cross section \\

POD &  preferentially-oriented diffusion \\

RT &  room temperature \\

SAM &  self-assembled monolayer \\

UHV &  ultra-high vacuum \\

UPS &  ultra-violet photoemission spectroscopy \\

VB &  valence band \\

VL &  vacuum level \\

VLA &  vacuum-level alignment \\

wrt &  with respect to \\

XPS &  x-ray photoemission spectroscopy \\
\end{tabular}

\setcounter{equation}{0} \setcounter{figure}{0} \setcounter{table}{0}\newpage
\fancyhead[LE,RO]{\thepage} \fancyhead[RE]{\shorttitle}
\fancyhead[LO]{\rightmark}
\section{Introduction}

After some revived interest on conductivity of organic materials in early 
1960's [1, 2] and on photovoltaic effect in 1975 [3], it was the breakthrough 
and pivotal work on transformation of transpolyacetylene into an electrical 
conductor with the conductivity controlled over several orders emerged in 
1977 [4, 5], which installed a fresh oxymoron, namely `organic 
electronics'. Nowadays, the term epitomizes both the class of devices based 
on organic electro- and opto-active materials, and such materials 
themselves. The latter are also denoted as conjugated materials or organic 
semiconductors owing to their resemblance to inorganic semiconductors in 
terms of the electronic and optical properties.

Since then, the research in the field has expanded considerably: the 
published works related to conjugated materials (the figure covers the 
research both on fundamental properties and organic devices, batteries, 
etc.) has inflated to as much as about several thousand studies to this day. 
In 2000, Heeger, Shirakawa, and MacDiarmid, who had triggered the new field, 
were awarded by Nobel Prize in chemistry `\textit{For the discovery and development of conductive polymers}'.

Upon this, maiden demonstrations of devices based on the $\pi $-conjugated 
oligomer ($i.e.$ small organic molecules with the fixed $m/e$ ratio) films had 
followed, such as the single- [6, 7] and double-layer [8] organic 
photovoltaic cell (OPC), organic light-emitting diode (OLED) [9], and the 
organic field-effect transistor (OFET) [10, 11]. Recent trend has been 
oriented towards the replacement of oligomer by polymers with their 
pioneered employment in OLED [12], single-layer [13] and 
heterojunction-based [14] OPC, and OFET [15]. Particularly optical organic 
devices have been progressed far from their nascence and nowadays OLED$s$ and 
OPC$s$ have reached a commercial rank.

Currently, the organic devices are second to traditional inorganic ones in 
terms of output parameters; nevertheless, the organic electronic materials 
offer various advantages. Besides the low-temperature (and thereby cheap) 
processing, the molecular design allows tuning of their fundamental 
electronic and optical properties, leading to a plethora of distinct 
materials with a tremendous potential for research and applications. In 
contrast to the class of inorganic semiconductors, where practically only 
several kind of semiconductors have been employable, hundreds new 
`semiconducting' molecules produced and screened out up to date suggest 
the high figure of applicable materials. Along the molecular design 
determining fundamental electronic and optical properties of active 
molecular films, equally important are related interfaces, such as the 
interfaces between molecular films and contacting materials, or between 
distinct organic materials themselves. Therefore the interface engineering, 
hence the control of interfacial electronic properties, is the necessary 
prerequisite of the further progress in organic electronics. Even though the 
operating devices are routinely attained, the physics behind interfacial 
electronics is still not completely understood precluding an exploitation of 
the potential of organic electronics.

\subsection{Energy levels in organic device}

Figure~1.1 presents the energy diagram of a simple organic device under 
applied bias, $U$: an organic film is sandwiched between two metal electrodes 
with distinct work functions. Of relevance for transport properties are the 
injection barriers for electrons $\varphi_e$ and holes $\varphi_h$ tantamount 
to the energy level offset between the lowest unoccupied molecular orbital 
(LUMO) or highest occupied molecular orbital (HOMO), respectively, and the 
Fermi level of the contacting metal. The interfacial barrier for charge 
carriers can be indirectly evaluated from transport properties of final 
devices, or, alternatively, by photoemission. Particularly the latter 
approach is attractive, as the barrier can be determined directly by probing 
relevant energy levels, during the interface formation (the bottom-up 
manner) along the physical and chemical structure examined on atomic level.

\begin{figure}[tb]
\begin{center}
\includegraphics[width=8cm,clip]{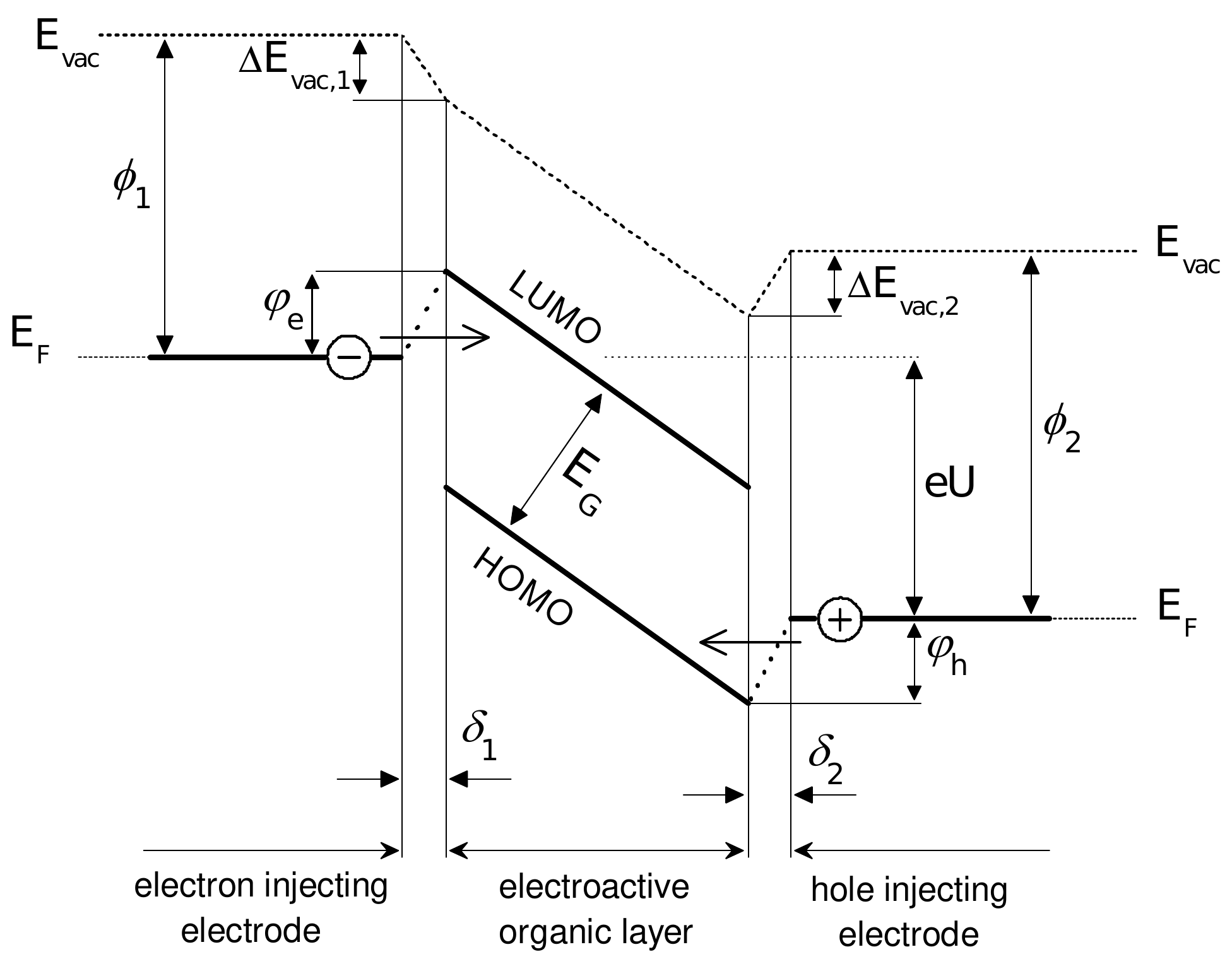}
\caption{A schematic energy diagram of an electroactive organic layer sandwiched
between electrodes under bias, $U$. The $\phi_i$ ($i=1,2$) are the respective work
functions of the metal electrodes $1$ and $2$, and $\Delta E_{vac,i}$ ($i = 1,2$)
are the vacuum level changes over the respective interface region thickness,
$\delta_i$. The $\phi_e$ and $\phi_h$ are the injection barriers for electrons and
holes. The $E_G$ is the energy gap, $E_F$ is the Fermi level.}
\end{center}
\vspace*{-0.5cm}
\end{figure}

\subsection{Photoemission characterization of electronic structure of molecular 
films}

Photoemission spectroscopy has been among the most traditional and potent 
techniques for the characterization of solid surfaces. Both chemical and 
electronic properties of solid surfaces and interfaces can be deduced. As 
for the mechanism, an electron from an occupied state is excited above the 
vacuum level by an impinging photon and can thus escape from the sample. The 
photoelectron kinetic energy of the ejected electron, $E_{K}$, allows to 
determine electron binding energy, $E_{B}$, via the Einstein relation:
\begin{equation}
\label{eq1}
E_B = h\nu - E_K 
\end{equation}
where $h\nu$ is the photon energy of the incidence beam. 

The standard laboratory photoemission facilities operate with x-ray sources 
thereby referred to x-ray photoemission spectroscopy (XPS). The most 
frequently used source based on the Al K$\alpha$ excitation gives $h\nu=1486.6$ eV,
the  energy sufficient to probe at least some core levels of majority of 
elements. Aside the core-level spectroscopy commonly linked with the 
characterization of the chemical structure, the characterization of the 
valence band (VB) determining the electronic structure is of key importance 
for electronically active materials. Even though electronic structure can be 
studied by means of XPS too, the more convenient approach is the employment 
of the ultra-violet photoemission spectroscopy (UPS). The typical light 
sources in UPS are rare gas discharges such as the He lamp with the photon 
energy of 21.22 eV. Compared to the $\sim 1$ keV-photon source, the He lamp 
admittedly covers only a portion of the electron binding energy range, yet 
the photoemission cross section at low incident photon energy is several 
orders higher; for example, the shallow C $2p$, C $2s$ orbitals, which are 
particularly relevant at examination of organic materials, have 
photoionization cross section higher by factor about $10^{5}$ and $2\times 
10^{3}$, respectively, when excited by $h\nu=21.22$ eV compared to 
$h\nu=1486.6$ eV [16]. 

If a homogenous film over a substrate is grown, the substrate-related 
photoemission intensity, $I_{subs}$, is attenuated approximately according to 
the equation $I_{subs}\propto I_{0}\exp(-t/\lambda)$ for overlayer thickness 
exceeding one monolayer, where $I_{0}$ is the intensity of the bare 
substrate, $t$ is the overlayer thickness, and \textit{$\lambda $} states for the electron 
inelastic mean free path. Thus, the overlayer thickness of $ \approx _{ 
}$3\textit{$\lambda $} results in a drop of the substrate signal to about 5{\%}. On the other 
side, the signal of a homogenous film, $I_{film}$, increases approximately 
according to formula $I_{film}$~$\alpha $~$I_{\infty }[1-\exp(-t/\lambda)]$, where 
$I_{\infty }$ states for the signal of a semi-infinite film. Again, the 
formula works for the film thickness exceeding one monolayer. While the 
photon penetration depth is of the order of microns, \textit{$\lambda $} is of the order of 
nanometres for XPS and as small as several angstroms for UPS making it 
extremely surface sensitive. This allows routine investigations of 
properties of adsorbents at deep-submonolayer coverages; since the substrate 
signal is fully eliminated by an overlayer with the thickness exceeding a 
few \textit{$\lambda $}, comparable and smaller thicknesses of homogeneous overlayers can be 
determined with a high relative accuracy.

Regarding the interfaces, photoemission characterization permits the 
assessment of relative positions of energy levels, thereby the determination 
of interfacial electronic structure. Since the photoemission probing depth 
appears much smaller than the usually employed film thicknesses of the 
film/substrate systems, the interface properties are conveniently 
investigated by means of so-called surface-science approach [17, 18]: the 
film is grown in a stepwise manner under ultra-high vacuum (UHV) conditions 
at pressures $< 10^{ - 9}$ mbar onto atomically clean and controlled 
substrates, while the evolutions of electronic, chemical, structural, and 
morphological properties are simultaneously examined \textit{in situ} at the onset of the 
interface formation. This allows the determination of factors governing the 
interface properties such as \textit{$\varphi $}$_{h}$ (Fig.~1.1). The surface-science approach 
benefits from the well-described and defined film/substrate systems under 
studies at atomic level free from ambient influence. 

\begin{figure}[tb]
\begin{center}
\includegraphics[width=8cm,clip]{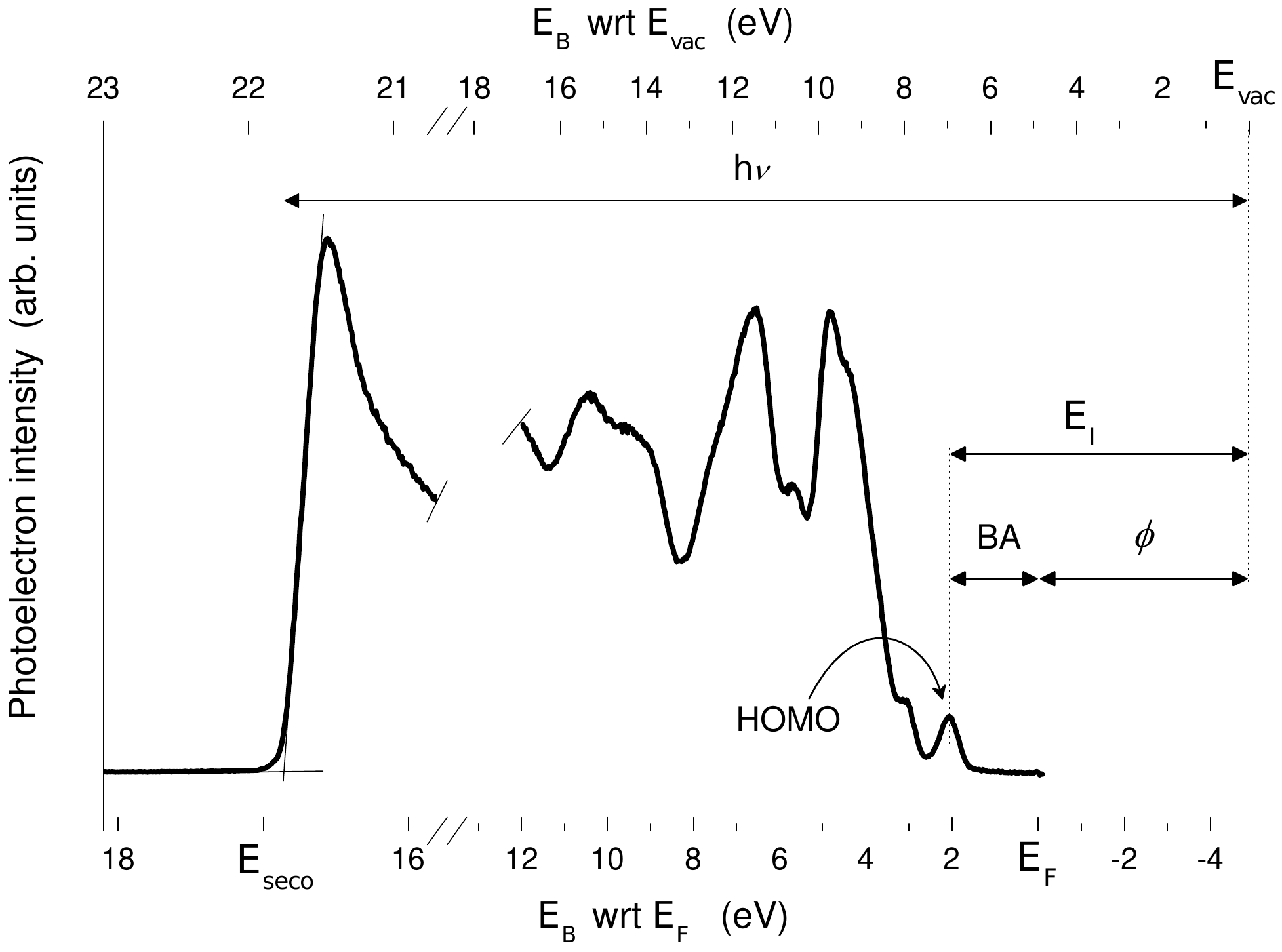}
\caption{The valence band of copper phthalocyanine (CuPc) film with indication
of fundamental electronic parameters such as the band alignment (BA),
ionization energy ($E_I$), and work function ($\phi$); the latter makes the
difference between the energy scales referred with respect to (wrt) the Fermi
level (bottom) and the vacuum level (upper), $h\nu$ corresponds to photon energy
of the incident beam, $E_{seco}$ is the secondary electrons cut-off.}
\end{center}
\vspace*{-0.5cm}
\end{figure}

Figure~1.2 illustrates the valence band (VB) of a molecular film, here of 
copper phthalocyanine (CuPc), grown on a substrate. The VB of a molecular 
film comprises of a number of shallow nearby molecular orbitals. Their 
relative positions are weakly affected by the intermolecular interaction 
occurring in condensed films [19]. Since the shallowest orbitals are shaped 
by $\pi $ bonds of a molecule, the VB region near the Fermi level is 
referred to as the $\pi $ band. Of orbitals constituting the VB, the HOMO is 
considered as the cardinal one; in terms of the molecular film, the energy 
difference between the HOMO and LUMO is termed the HOMO-LUMO gap, which 
determines the optical and transport band gap. As for a contact between a 
molecular film and a metal, the HOMO referred to the Fermi level is termed 
the band alignment (BA), or energy level alignment (ELA), and it corresponds 
to the transport barrier for holes \textit{$\varphi $}$_{h}$ injected from a metallic substrate 
to organics (Fig.~1.1). We note that \textit{$\varphi $}$_{h}$ is ascertainable by the (direct) 
photoemission, \textit{$\varphi $}$_{e}$ can be probed by an inverse photoemission. The HOMO 
referred to the vacuum level, $E_{vac}$, identifies with the ionization 
energy, $E_{I}$. Obviously,
\begin{equation}
\label{eq2}
E_I = BA + \phi ,
\end{equation}
where \textit{$\phi $} states for the electron work function. The essential
problem of the 
interface engineering would be the understanding a mechanism controlling the 
BA. That has motivated numerous photoemission studies focused on the 
examination of HOMO binding energies for various molecular films grown on a 
variety of substrates to reveal a correlation between electronic properties 
of isolated materials, hence prior to the contact formation, and the band 
alignment.

The edge at the high-$E_{B}$ side of the valence band in Fig.~1.2 manifests 
the onset of the secondary electrons cut-off ($E_{seco})$. Its position 
determines \textit{$\phi $}, which is defined as the minimal energy necessary to transfer an 
electron from the material to the nearby vacuum with its final kinetic 
energy being zero. The work function is obtained as \textit{$\phi $}$_{ }=_{ }$\textit{h$\nu  - $E}$_{seco}$. 

Note that the HOMO position is assigned to its maximum energy in Fig.~1.2. 
This is in variance with a frequent approach identifying the HOMO position 
with the low-$E_{B}$ onset of the HOMO approximated by its leading edge. The 
use of the onset had been substantiated in relation with optical properties, 
since the energy gap determined by optical absorption better suits the 
position of the HOMO and the LUMO onsets (see $e.g.$ Refs. [20] and [21]). Yet, 
the peak shape is affected by experimental factors such as instrumental 
resolution or sample temperature. Moreover---unlike $e.g.$ the onsets of the 
secondary electrons and the Fermi edge---the HOMO peak shows no leading 
edge, accordingly no linear portion. Particularly at low intensities, the 
`leading edge' of the HOMO can result in an uncertain determination of its 
binding energy. We therefore advocate the former approach eventually amended 
by a correction. Nevertheless, the experimentally determined HOMO values 
addressed in this study were prevailingly dug out in the
literature---where the low-$E_{B}$ onset had been commonly preferred
for the HOMO determination---and thereby adopted for our purposes as they were.

\begin{table}[!bp]
\begin{center}
\vspace*{-0.6cm}
\caption{Overview of molecular films tackled in this work.}
\begin{tabular}{lll}
\hline
\textbf{Acronym} & \textbf{Chemical} & \textbf{Name} \\
& \textbf{\ formula} & \\
\hline
6P &  C$_{36}$H$_{22 }$ & sexiphenyl \\
6T &  C$_{24}$H$_{14}$S$_{6 }$ & sexithiophene \\
$\alpha - $NPD, NPB &  C$_{44}$H$_{32}$N$_{2}$ & $N,N'$-diphenyl-$N,N'$-bis(1-naphthyl)-1,1$'$-biphenyl-4,\\&&4$'$-diamine \\
Alq$_{3 }$ & C$_{27}$H$_{18}$AlN$_{3}$O$_{3 }$ & tris(8-hydroxyquinoline)aluminum \\
BCP &  C$_{26}$H$_{20}$N$_{2 }$ & bathocuproine \\
BP2T &  C$_{32}$H$_{22}$S$_{2 }$ & 2,5-bis(4-biphenylyl) bithiophene \\
CBP &  C$_{36}$H$_{24}$N$_{2 }$ & 4,4$'$-$N,N'$-dicarbazolyl-biphenyl  \\
CuPc &  C$_{32}$H$_{16}$CuN$_{8 }$ & copper phthalocyanine \\
DH4T &  C$_{20}$H$_{26}$S$_{4 }$ & $\alpha {\rm s}\omega 
$-dihexyl-quaterthiophene \\
DH6T &  C$_{36}$H$_{38}$S$_{6 }$ & $\alpha {\rm s}\omega 
$-dihexyl-sexithiophene \\
DiMe-PTCDI &  C$_{26}$H$_{14}$O$_{4}$N$_{2}$ & $N,N'$-dimethyl-3,4,9,10-perylenetetracarboxylic diimide  \\
F$_{4}$CuPc &  C$_{32}$H$_{12}$F$_{4}$CuN$_{8 }$ & tetrafluoro copper 
phthalocyanine \\
F$_{4}$TCNQ &  C$_{12}$F$_{4}$N$_{4}$ & 2,3,5,6-tetrafluoro-7,7,8,8-tetracyanoquinodimethane  \\
F$_{16}$CuPc &  C$_{32}$F$_{16}$CuN &  hexadecafluoro copper phthalocyanine  \\
H$_{2}$Pc &  C$_{32}$H$_{18}$N$_{8 }$ & metal-free phthalocyanine \\
HBC &  C$_{42}$H$_{18 }$ & hexa-\textit{peri}-hexabenzocoronene \\
HATCN &  C$_{18}$N$_{12 }$ & hexaazatriphenylene-hexacarbonitrile  \\
NADPO &  C$_{20}$H$_{32}$O$_{4}$N$_{4 }$ & amphiphilic substituted 
2,5-diphenyl-1,3,4-oxadiazole  \\
NiPc &  C$_{32}$H$_{16}$NiN$_{8 }$ & nickel phthalocyanine \\
NTCDA &  C$_{14}$H$_{4}$O$_{6 }$ & 1,4,5,8-naphthalene tetracarboxylic dianhydride\\
PEN &  C$_{22}$H$_{14 }$ & pentacene \\
PFP &  C$_{22}$F$_{14 }$ & perfluoro-pentacene \\
PTCBI &  C$_{36}$H$_{16}$O$_{2}$N$_{4 }$ & 3,4,9,10-perylenetetracarboxylic 
bisbenzimidazole \\
PTCDA &  C$_{24}$H$_{8}$O$_{6 }$ & 3,4,9,10-perylenetetracarboxylic dianhydride \\
PTCDI &  C$_{24}$H$_{10}$O$_{4}$N$_{2 }$ & 3,4,9,10-perylenetetracarboxylic 
diimide  \\
TCNQ &  C$_{12}$H$_{4}$N$_{4 }$ & tetracyanoquinodimethane \\
ZnPc &  C$_{32}$H$_{16}$ZnN$_{8 }$ & zinc phthalocyanine \\
ZnTPP &  C$_{44}$H$_{28}$N$_{4}$Zn &  5, 10, 15, 20-zinctetraphenylporphyrin \\
\hline
\end{tabular}
\end{center}
\end{table}

\vspace*{-0.25cm}
\subsection{Oligomer molecular films}

Even though the polymers dominate in the role of active layers particularly 
in nowadays commercial organic devices, a great deal of knowledge on organic 
films was achieved by means of studies on model oligomer molecules. In 
contrast to the polymers, the oligomers can be sublimed in UHV making them 
employable for the surface-science approach. That---although not intended 
to be technologically competitive for the production of organic
devices---oriented towards small molecules permits studies on fundamental chemical 
and electronic issues. Table 1.1 lists the oligomers addressed in this 
study. Their chemical structures are collected in Fig.~1.3. 

\begin{figure}[!htb]
\begin{center}
\includegraphics[width=10cm,clip]{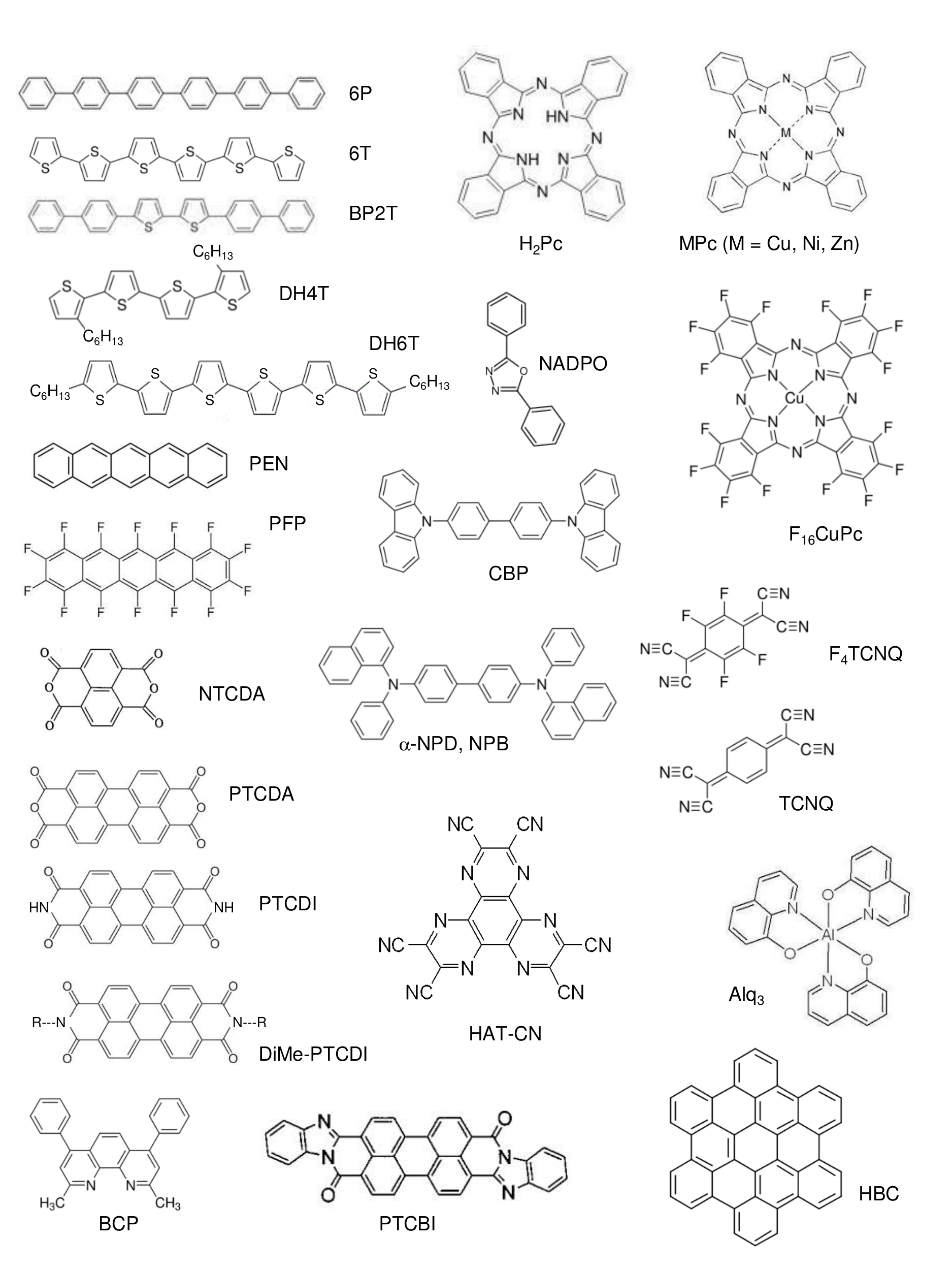}
\caption{The structural formulae of oligomer molecules listed in Tab.1.1.}
\end{center}
\end{figure}

\subsection{Molecular orientation}

Figure~1.4a shows the $\pi $ orbitals of benzene ring oriented along the 
normal of the ring plane. The overlapped orbitals form a conjugated system 
with the delocalized electrons. Molecular crystals are noted for anisotropic 
electronic properties: for example, transport properties of an organic 
device are dramatically affected by the molecular orientation relative to 
the current flow [22]. The most efficient current flow in a molecular film 
occurs along the direction of the stacked $\pi $ bonds being coincident with 
the interaction between the $\pi $ bonds of neighbouring molecules. This is 
illustrated in Fig.~1.4b schematically presenting the side view of an 
ordered assembly of molecules, $e.g.$ of pentacene. 

\begin{figure}[!tb]
\begin{center}
\includegraphics[width=10cm,clip]{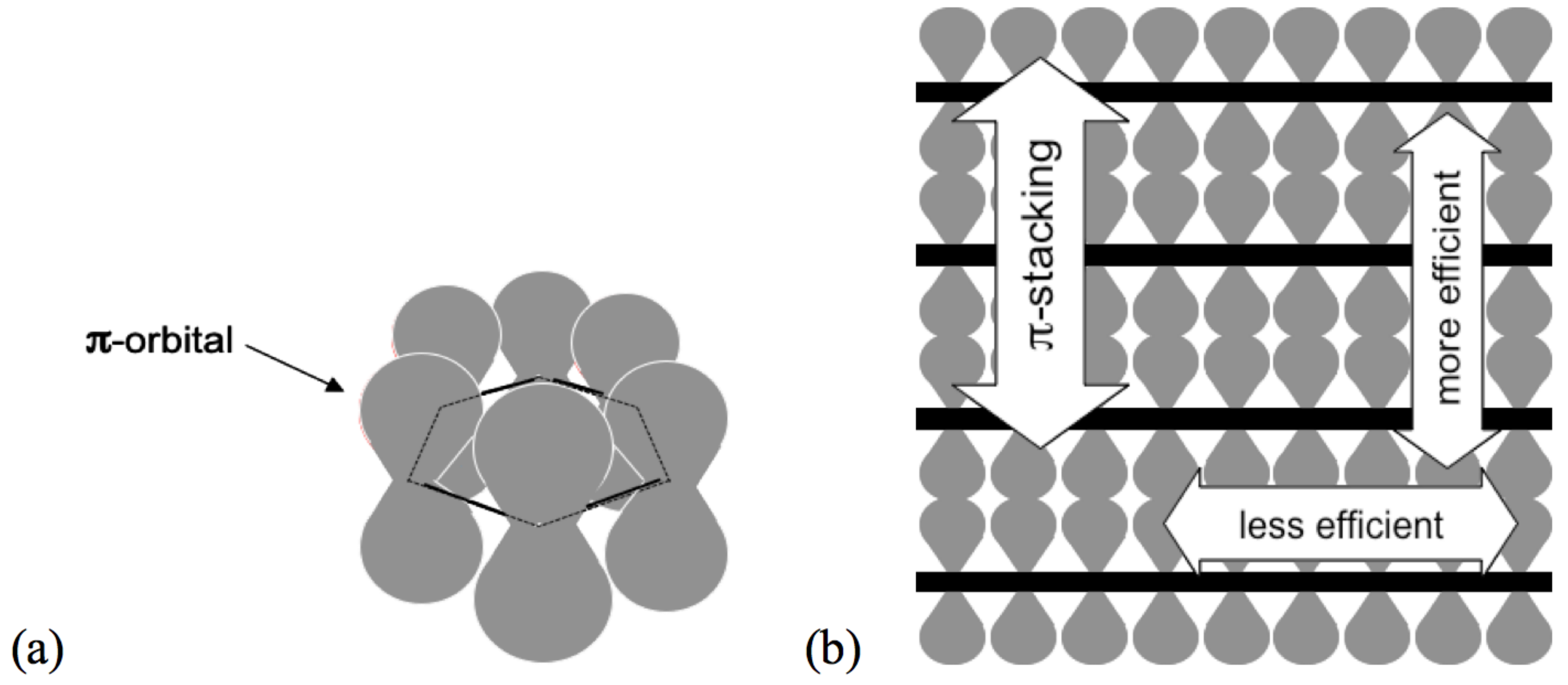}
\caption{The schematic views on (a) the spatial distribution of $\pi$ orbitals
of a benzene ring represented by a hexagon, and (b) the assembly of ordered
molecules viewed along their ring planes with indicated $\pi$-stacking direction.
The thick lines represent backbones of the molecules. The directions with the
more and less efficient charge carrier transport are shown.}
\end{center}
\end{figure}
\begin{figure}[!tbp]
\begin{center}
\includegraphics[width=8cm,clip]{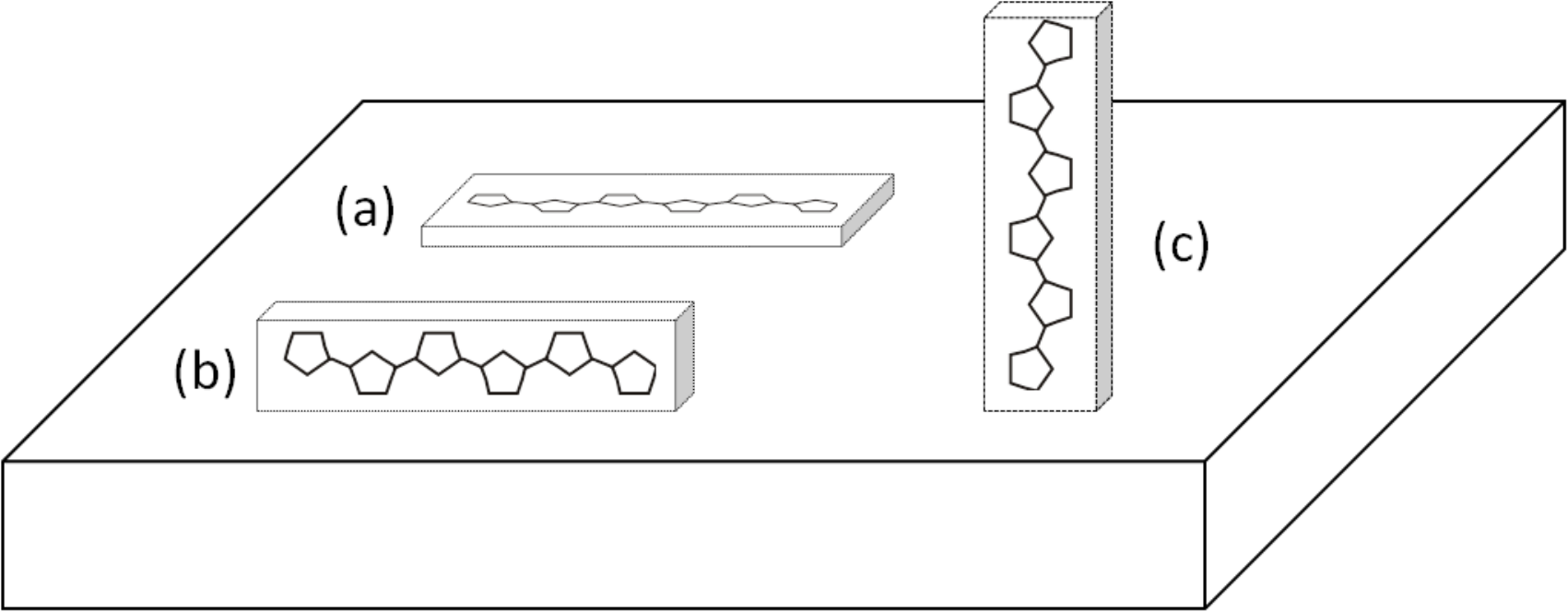}
\caption{Characteristic orientations of a molecule (here, exemplified by
sexithiophene) with respect to the substrate surface: the lying molecule
oriented flat-on (a) and edge-on (b), and the upright-oriented (end-on)
molecule (c).}
\end{center}
\end{figure}

Herringbone structure of crystalline molecular films implicates the 
different orientations of particular molecules within the unit cell, yet, an 
average molecular orientation pertinent to a film can be addressed; thus, 
the term `molecular orientation' describes the average orientation of 
molecules in a molecular film in terms of the orientation of the molecular 
plane and/or its backbone with respect to the surface normal. The molecules 
with their backbones parallel or near parallel to the substrate \textit{surface}
are referred to the lying ones (Fig.~1.5a,b). The molecules with their molecular 
planes and/or backbones parallel or near parallel with the surface \textit{normal}
have upright orientation (Fig.~1.5c). Obviously, the term molecular orientation 
refers to the orientation of molecules embedded in the crystalline film. 

\begin{figure}[tb]
\begin{center}
\includegraphics[width=8cm,clip]{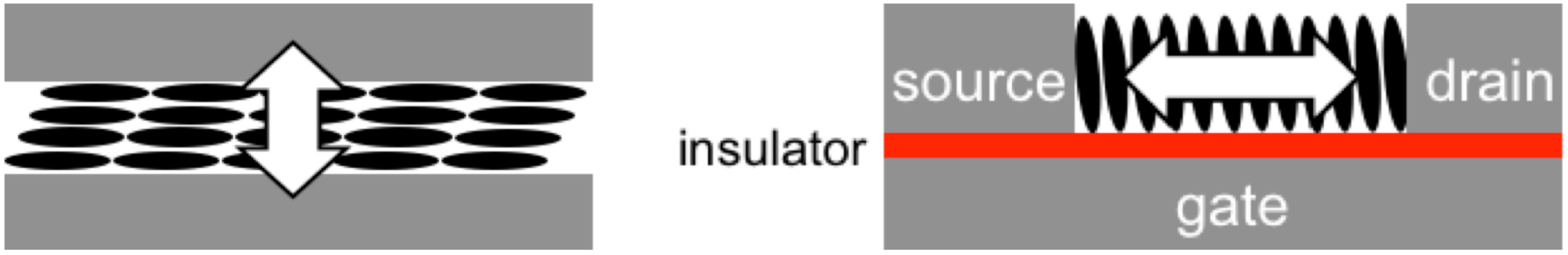}
\caption{Examples on organic devices with distinct targeted molecular orientation
(the molecules are emblemized by ovals): the lying orientation for the OLED (left)
and the upright orientation for the OFET (right). Molecular layer are sandwiched
between contacting layers. The arrows indicate the current flow directions.}
\end{center}
\end{figure}

Depending on the targeted planar device---typical contrasting examples 
being the OLED and the OFET---the current flows in the device parallel 
either to the substrate surface normal or to the substrate surface (Fig.~
1.6). Likewise, the molecular orientation determines interfacial energetics 
and by controlling the relative molecular orientation at the donor/acceptor 
interfaces in organic heterostructures the charge transfer [23] and charge 
dissociation [24] can be affected. It is therefore essential to control both 
the crystallinity of grown films and the molecular orientation.

\setcounter{equation}{0} \setcounter{figure}{0} \setcounter{table}{0}\newpage
\section{Control of molecular orientation}

To control the molecular orientation, the mechanism of the organic film 
growth has been widely studied via the molecule deposition on a variety of 
substrates; typical examples include reconstructed surfaces of 
single-crystal metallic substrates, inorganic semiconductors, and oxidized 
surfaces. The profound influence of the substrate nature on the film 
structure and packing were frequently demonstrated. Yet, this mostly 
concerned of thin films, as further growth towards thick films resulted in 
the different film structure and morphology weakly or not dependent on the 
substrate.

Commonly, the rod-like molecules, such as $e.g.$ 6P, 6T, pentacene, adsorb onto 
metallic substrates with the lying geometry. According to the conventional 
wisdom, the principal factor determining the lying molecular orientation had 
been assumed to be the high interaction strength between molecular $\pi $ 
bonds and the surface of metallic substrate, while the upright-oriented 
molecules in films grown on oxidized surfaces such as SiO$_{2}$, 
Al$_{2}$O$_{3}$, glass, etc., were rationalized by the weak(er) interaction 
of molecules to the surfaces passivated by oxygen [25-32]. 

Exceptions from the interaction-strength model, such as the upright 
molecular orientation in films grown onto polycrystalline and non-oxidized 
metallic surfaces (Au, Ag, and Cu) were substantiated by the higher 
roughness of the polycrystalline surfaces compared to the single-crystal 
surfaces [33-35]. Though, the interaction-strength model implicates further 
inconsistencies. For example, it does not rationalize the lying orientation 
of molecules beyond the first monolayer of the molecular film; whereas the 
first monolayer acts as a substrate for the second monolayer, the weak 
molecule-molecule interaction should result to universal upright orientation 
by the second layer upon the closing the first layer. Such
scenario---although reported occasionally---is in contrast with frequent 
observations. 

According to Ref. [36], the molecular orientation is controlled by the 
substrate surface topology such as the presence of the long-range surface 
order or by lack of it, instead of the interaction strength between the 
molecule and the substrate. Specifically, a diffusion of molecules over the 
substrate surface either along preferred directions or in azimuthally random 
directions was proposed to be the principal factor determining the molecular 
orientation [37]; the adsorbed molecules, which do not chemically react with 
the substrate, diffuse over the substrate surface till they are embedded 
into the formed organic crystal. The necessary requirement for getting the 
lying orientation in the film is the presence of a diffusion director on the 
substrate surface. Preferential azimuthal directions present on 
reconstructed surfaces manifested by a lower energy for the molecular 
diffusion [38], or unioriented steps on vicinal surfaces [39], or directed 
scratches [40], can act as directors. The molecules diffusing over preferred 
azimuthal direction result in the growth governed by attractive 
intermolecular interaction for the probability for a molecule of being 
involved in a crystal growth seemingly increases if diffusing molecules 
encounter aligned; this maximizes the overlap of $\pi$ bonds between the 
neighbouring molecules and thereby preserves the lying orientation upon 
their embedding in the film [36, 41]. The deficient or intentionally marred 
surface reconstruction ($e.g.$ by sputtering) has a dramatic effect on the film 
growth; instead of the lying molecular orientation, the molecules orient 
upright, thus sharply disproving the interaction strength model [36, 42]. 
This is due to the diffusion of the molecules over an unordered surface, 
thus in random directions, which results in the formation of energetically 
more favourable crystals made of upright-oriented molecules. Obviously, a 
single molecule adsorbs with the lying geometry and preserves it while 
diffusing over both ordered and disordered surface; the lying and upright 
molecular orientations are meant in terms of the final orientation of 
embedded molecules. 

It should be noted that the lying geometry, if achieved on proper 
substrates, usually applies for thin and up to moderately thick films only, 
and the molecular orientation changes towards the upright with the further 
film growth (Fig.~2.1). In other words, such molecular films are 
inhomogeneous in terms of the molecular orientation. The orientational 
change with the increasing film thickness from lying to upright is termed 
the L$ \to $U orientational transition. The orientational transition can 
occur either abruptly already by the second layer [43-45], or gradually 
within the nominal film thickness ranging from several nanometres up to 
several tenths of nanometre [27, 39, 46-49]. The upright molecular orientation 
was reported also to be owing to the higher growth temperature [43, 50]. The 
gradual orientational transition with the film thickness was reported for 
polymers too [51] and it may be a general growth phenomenon, although it has 
been often unnoticed. Importantly, the orientational transition affects the 
film electronic properties. Indeed, in terms of electronic properties, the 
molecular film with the orientational transition is a heterostructure. This 
will be discussed in the next chapters.

\begin{figure}[tb]
\begin{center}
\includegraphics[width=10cm,clip]{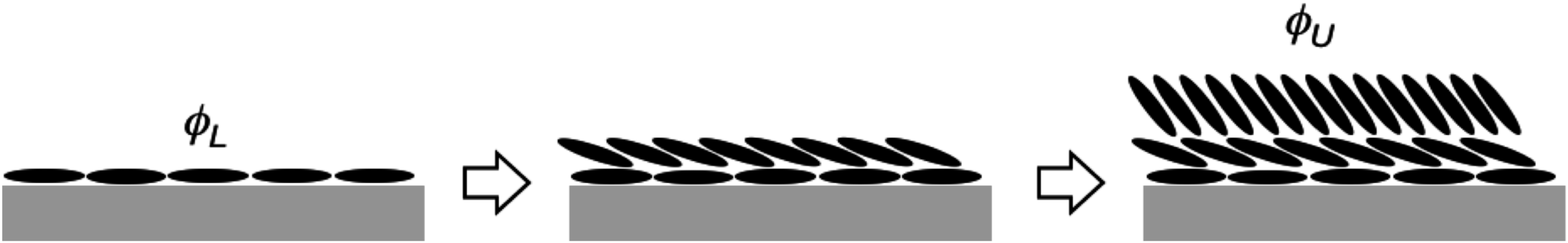}
\caption{Sketched illustration of the gradual L$ \to $U orientational transition
(from lying- towards upright-oriented molecules): the molecular tilt angle
gradually varies with the increased thickness of a molecular film (from left
to right). The $\phi_L$ and $\phi_U$ state for the work function corresponding
the molecular film surface formed by lying- and upright-oriented molecules,
respectively.}
\end{center}
\vspace*{-0.4cm}
\end{figure}

In general, the upright orientation is easier to achieve and it particularly 
takes place ($i)$ at elevated growth temperatures, owing to the high kinetic 
energy sufficient to free the diffused molecules confined by the director, 
(\textit{ii}) on unordered or poorly ordered surfaces lacking the directors. 

By elevating the substrate surface morphology to the primary factor 
determining the molecular orientation in the film, the 
preferentially-oriented diffusion (POD) model justifies distinct molecular 
orientations obtained on chemically identical but topologically distinct 
surfaces, such as the reconstructed and polycrystalline metallic surfaces 
[34-35]; the lack of the POD on the latter results in upright oriented 
molecules in the film, while the lying molecular orientation takes place on 
the reconstructed/ordered surface. The chemical origin of the substrate 
plays a minor role unless the adsorbed molecule is immobilized by chemical 
reaction with the substrate. 

The POD model rationalizes the frequently observed and undesirable gradual 
orientational transition occurring in thicker film (Fig.~2.1); the surface 
order of molecular films deteriorates with the increasing film thickness due 
to growth imperfectness and thereby the directors $[n$th layer on the film acts 
as the substrate for the $(n+1)$th layer$]$ gradually deteriorate and vanish. This 
results in the gradual orientational transition affecting the electronic 
properties (see the discussion on growth fashions and the band bending 
below). On the other hand, distinct molecular orientation affects the 
crystal orientation with respect to the substrate, yet the crystal structure 
can remain preserved as it was suggested by the vibronic progressions 
insensitive to the molecular orientation [52].

\setcounter{equation}{0} \setcounter{figure}{0} \setcounter{table}{0}\newpage
\section{Electronic properties}

\subsection{Ionization energy, $\boldsymbol{E_{I}}$}

Equivalent to the first ionization energy of elements, the ionization energy 
of a molecule is the minimal energy necessary for the removal of an electron 
from the HOMO and move it to infinity. The binding energies of an isolated 
molecule (in the gas phase) have to be referred to the vacuum level and the 
first ionization energy is a material constant. In photoemission 
characterization of solid surfaces, the Fermi level reference has been the 
convenient reference choice, since the Fermi level of grounded sample 
coincides with the ground of spectrometer. Thereby, the energy scale is 
shifted by the surface work function, $\phi $, from the absolute energy 
scale referred to the vacuum level (recall both energy scales in Fig.~1.2). 

The ionization energy of a condensed molecule constituting the molecular 
film is lower typically by about 1 eV in comparison to the ionization energy 
of that but isolated molecule due to the extra-molecular screening present 
in condensed phase [53]. Molecules embedded in the bulk of the molecular 
film and those terminating its surface experience distinct surroundings 
which is reflected in their photoemission spectra [54]. 

The ionization energies of the oligomer molecular films listed in Tab.~1.1 
range from about 4.4 eV to almost 10 eV reported for 6T [27] and HAT-CN 
[55], respectively. In a first approximation, the constituting elements with 
the high ionization energy, such as N, O, and/or F, tend to increase the 
ionization energy of the molecule compared to the hydrocarbon molecule 
comprised exclusively of C and H; for example. the substitution of hydrogen 
in the C-H bonds by fluorine leads to substantial increase of the ionization 
energy, as seen for and fluorine-substituted TCNQ [56], phthalocyanines 
[20, 57], PEN [58], and alkanethiols with different end groups [59]. 

Unlike an isolated molecule, which has a unique ionization energy, the 
ionization energy of a condensed molecular \textit{film} may show distinct values; the 
difference of 0.35 eV has been initially reported for differently prepared 
sexiphenyl films [60]. This is illustrated in Fig.~3.1, where the upper $\pi 
$ band of the 6P films grown at RT and 395 K are contrasted. For the spectra 
are referenced to the vacuum level, the HOMO offset indicates difference in 
the ionization energies. The varying ionization energy suggests that it is 
not a material constant and the growth conditions such as $e.g.$ the substrate 
surface properties and temperature can lead to changes in $\pi $-band 
electronic structures. Importantly, this may extend the possibility to 
control the band alignment and thereby the transport barrier at the contact.

\begin{figure}[tb]
\begin{center}
\includegraphics[width=5.5cm,clip]{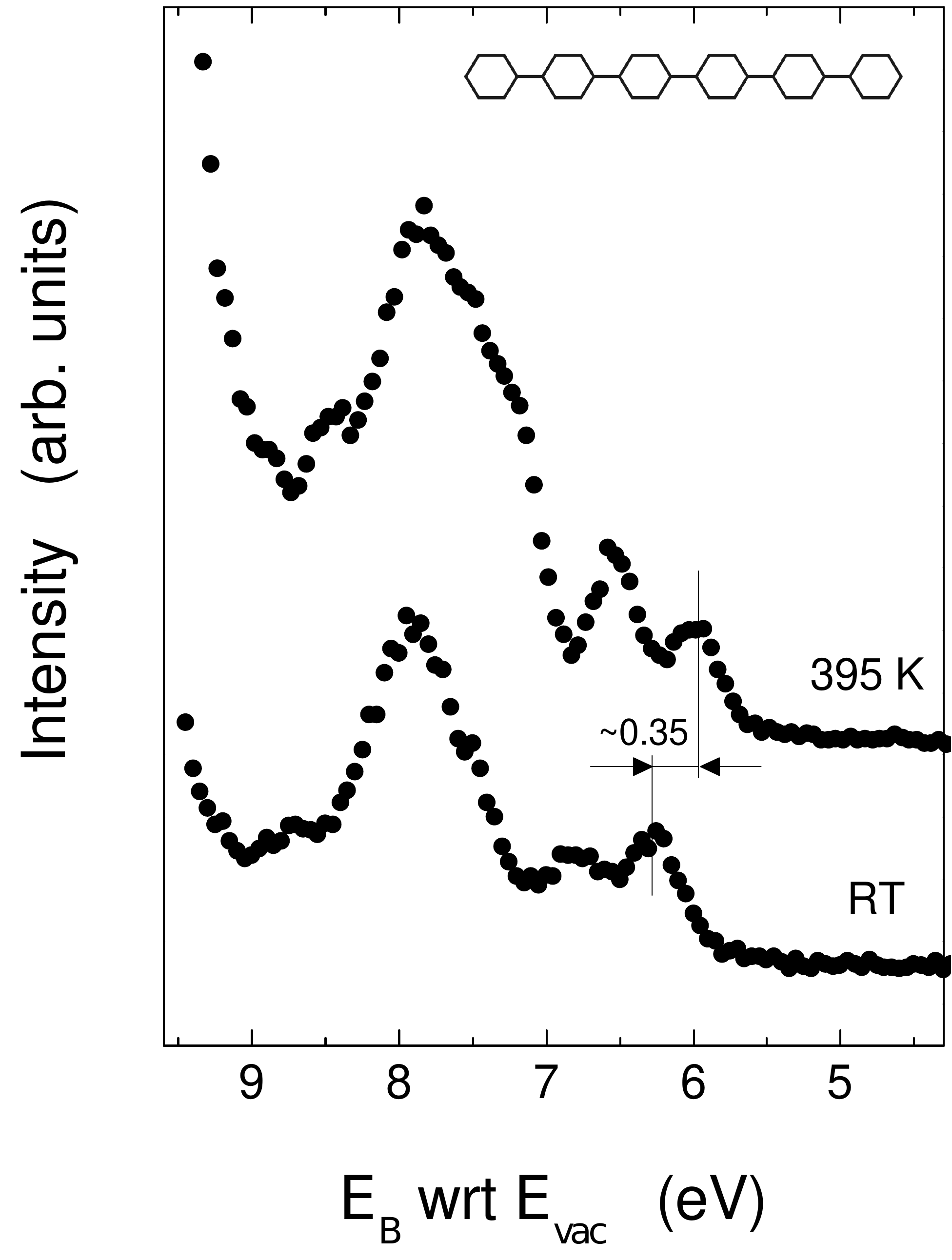}
\caption{The upper $\pi$ bands of two sexiphenyl films with distinct ionization
energies; the HOMO positions referred to the vacuum level are indicated by
vertical ticks, their relative position identifies with the difference in
ionization energies.}
\end{center}
\end{figure} 
\begin{figure}[!htb]
\begin{center}
\includegraphics[width=6cm,clip]{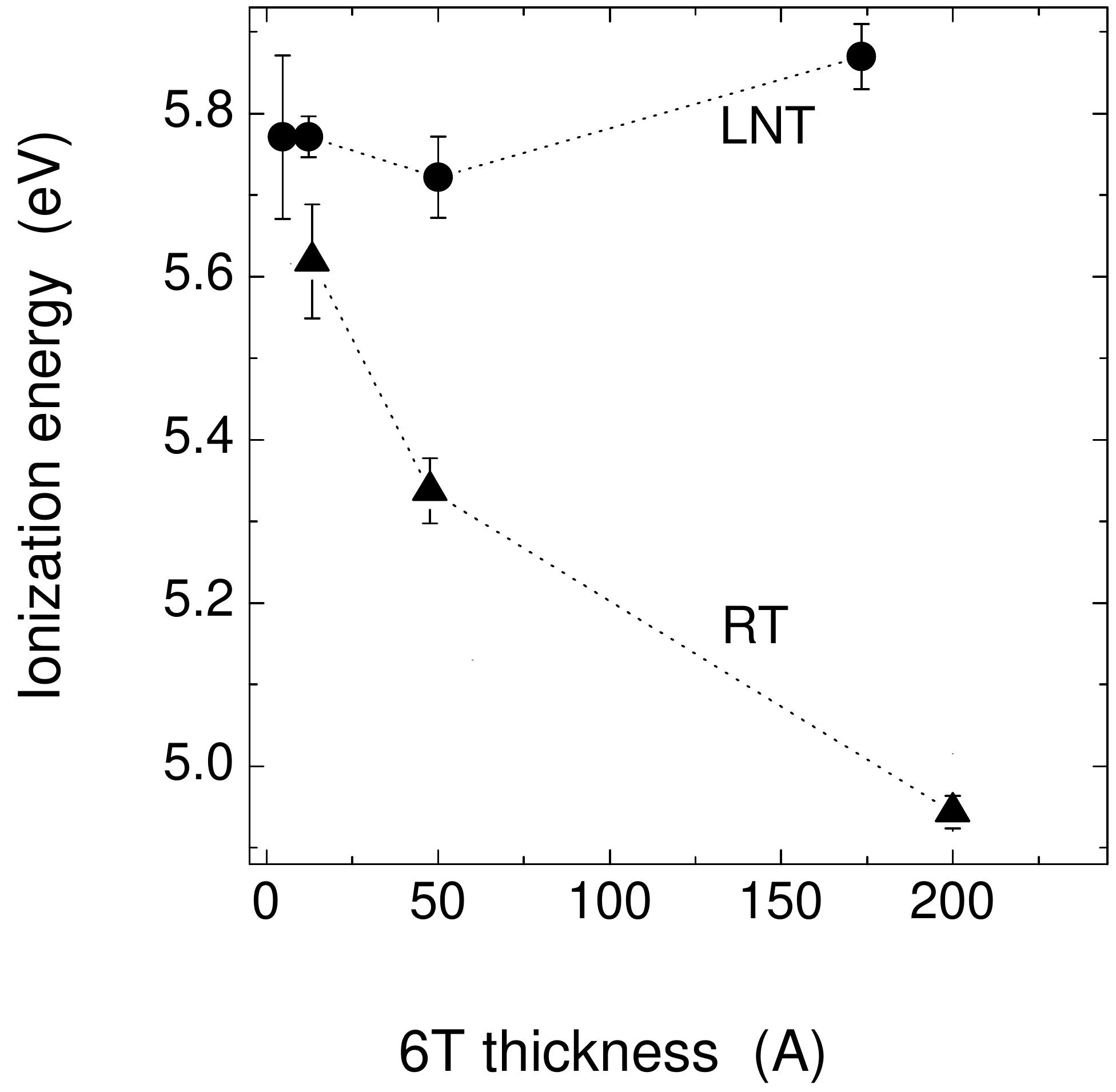}
\caption{The thickness dependence of the ionization energy of sexithiophene films grown on SiO$_x$
at liquid nitrogen (LNT) and room temperature (RT).}
\end{center}
\vspace*{-0.7cm}
\end{figure}

The observed $\Delta E_{I}$ was assumed to be the result of initial and/or 
final state differences. The former can occur due to differences in $\pi $ 
conjugation as a result of varying interring (torsional) angle between 
neighbouring benzene rings induced by the spherical hinderance of hydrogen 
atom. The latter can arise from differences in the screening of the 
photo-hole. Both differences can be affected by the packing in the molecular 
solid and are to a certain extent interrelated. 

Figure~3.2 shows the evolution of $E_{I}$ with the 6T film thickness for 
films grown at liquid nitrogen temperature (LNT) and RT on SiO$_{x}$ 
substrates. For the LNT growth, $E_{I} \approx 5.8$ eV results whether 
measured at LNT or after warming to RT. This value matches reasonably well 
the value of 5.9 eV derived from $E_{I}$ of the gas phase being of 7.0 eV 
[61] and corrected for extra-molecular screening of 1.1 eV, which is common 
in organic solids [53] and suggests a `frozen' gas phase. The ionization 
energy of the LNT-grown film is nearly independent on the film thickness, 
while that of the RT-grown film dramatically drops with the film thickness 
by about 0.8 eV suggesting gradual major changes in the electronic 
structure. The differences were presumably owing to the molecular order and 
the lack of it, as it occurs in the RT- and LNT-grown films (Ref. [27]). 
Similar trends were reported for the NADPO [62].

The variable $E_{I}$ implies its importance for the characterization of 
molecular films, thus the necessity of adhering to the vacuum level 
reference, and consequently---besides the BA evaluation---the need of 
the work function examination. If the characterization of the work function 
is neglected, the electronic distinctness of the molecular film can be 
unnoticed. In contrast to the habitually employed Fermi level reference, the 
vacuum-level reference eliminates rigid energy shift arising from the vacuum 
level changes and thereby enables the comparison of films grown on different 
substrates.

Further studies reported the ionization energy changes for 6T [27, 63], DH6T 
[63], and for rigid molecules, such as CuPc [64-66], F$_{16}$CuPc [65], PEN 
[67], and PFP [58], too. Provided that the molecular orientation of grown 
film was examined, an evident relation between the ionization energy and the 
molecular orientation was revealed. Specifically, the following films and 
the corresponding $\Delta E_{I}$---when going from the lying towards the 
(nearly) upright molecular orientation---were reported: CuPc: -0.4 eV 
[65, 68]; 6T: -0.4 eV [63]; PEN: -0.55 eV [69]; DH6T: -0.6 eV [63]; 
F$_{16}$CuPc: +0.7 eV [65, 68]; and PFP: +0.7 eV [70], +0.85 eV [69]. Note 
that both sign of changes were observed.

\subsection{Work function, $\boldsymbol{\phi}$}

The work function of a free surface, \textit{$\phi $}, has two contributions:
\begin{equation}
\label{eq3}
\phi = \mu _{bulk} + \phi _{SD} ,
\end{equation}
where \textit{$\mu $}$_{bulk}$ states for the internal (bulk) chemical potential and \textit{$\phi $}$_{SD}$ 
denotes the surface-dipole potential.

A particular material manifests its work function provided that the 
sufficient amount is probed. This is known for investigated inorganic 
($e.g.$ elemental) surfaces of films prepared on supported substrates, where the 
electronic properties of the substrate have to be eliminated and those of 
the overlayer developed. This is attained for film thicknesses exceeding a 
critical thickness [71]. The transition thickness range, accordingly the 
film thickness lower than the critical thickness, can span from a monolayer 
to multilayers and it was attributed to a violation of the local charge 
neutrality in films [72]. 

As for molecular films, the work function had been studied and mostly 
examined in the relation with the interfacial properties of the 
film/substrate interface. Specifically, the work function \textit{change}, $\Delta $\textit{$\phi $}$_{subs}$, 
upon the molecular film growth has been commonly examined: 
\begin{equation}
\label{eq4w}
\Delta \phi _{subs} = \phi _{subs} - \phi _{film} = \Delta E_{vac} ,
\end{equation}
for the vacuum level shift, $\Delta E_{vac}$, had been identified with a 
dipole formed at the film/substrate interface, the interfacial dipole (ID):
\begin{equation}
\label{eq5}
\Delta E_{vac} \equiv ID,
\end{equation}
where \textit{$\phi $}$_{subs}$ is the work function of the pristine substrate surface, 
which changes to \textit{$\phi $}$_{film}$ upon the film growth. The ID affects the 
transport barrier for charge carriers injected across the junction. Figure~3.3
shows a typical $\Delta $\textit{$\phi $}$_{subs}$ upon the molecular film growth, 
namely of the Al(111) surface upon sexiphenyl growth with $\Delta $\textit{$\phi $}$_{subs}$ 
$ \approx $ 0.45 eV. This \textit{abrupt} change of the vacuum level---occurring within 
the completion of about one monolayer here---is considered to be a 
manifestation of the ID. 

\begin{figure}[tb]
\begin{center}
\includegraphics[width=6.5cm,clip]{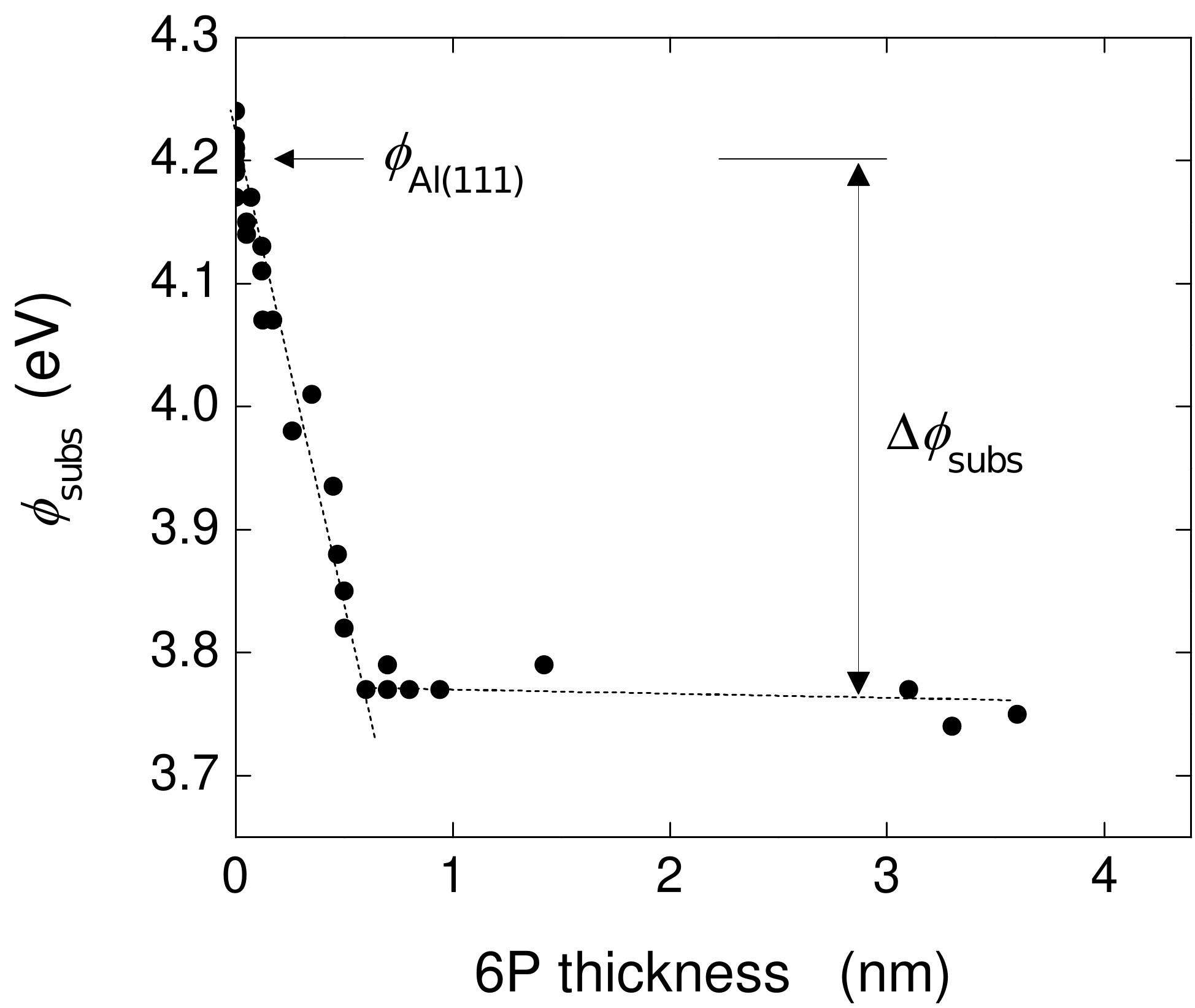}
\caption{The substrate work function change $\Delta\phi_{subs}$ upon the sexiphenyl (6P)
film growth.}
\end{center}
\vspace*{-0.7cm}
\end{figure}

In terms of the work function evolution upon the film growth, the research 
has been focused on the work function \textit{change}, and not the work function \textit{per se}; hence, 
$\Delta $\textit{$\phi $} have been conceptually perceived as $\Delta $\textit{$\phi $}$_{subs}$ in Eq. 
(\ref{eq4w}), accordingly, the modified work function pertinent to the substrate. 
Yet, a sufficiently thick molecular film should display the work function 
relevant to the film itself likewise to films formed by inorganic electronic 
materials. Such work function is referred to as an intrinsic work function, 
\textit{$\phi $}$_{int}$, throughout this study. It will be shown in the following that 
\textit{$\phi $}$_{int}$ both characteristic of the particular film and independent of the 
substrate can be detected. Similar to the work function of inorganic 
materials, the term `intrinsic' is apparently abundant, yet it is employed 
in this study in relation with the molecular films for it would distinguish 
\textit{$\phi $}$_{int}$ being the material constant from the work function of \textit{thin} films, 
$i.e.$ \textit{$\phi $}$_{film}$ in Eq. (\ref{eq4w}), which can differ from \textit{$\phi $}$_{int}$ due to subcritical 
thickness of the molecular film. In spite of ample photoemission studies on 
the organic film growth, the usable data in the literature for the 
distinguishing the intrinsic work function of various molecular films are 
relatively scarce as studies on thinner films have been favoured owing to a 
focus on examination of the interfacial properties, such as the BA and ID 
determinations.

Conclusive examples supporting the notion on the intrinsic work function of 
molecular films based on compiled data are collected in Tab.~3.1. The data
demonstrate that the film work function (\textit{$\phi $}$_{film})$ of a 
particular sufficiently thick ($t_{film})$ film tends to converge to the 
specific value irrespective the substrate characterized by its \textit{$\phi $}$_{subs}$. 
Thereby, \textit{$\phi $}$_{subs}$ has no effect on the resulting work function of the thick 
molecular film. For example, the 10 nm-thick NiPc films displayed the same 
work function of 4.0 eV, whether grown on Au or Ag polycrystalline 
substrates, accordingly in spite of the marked difference of 0.8 eV between 
\textit{$\phi $}$_{sub}$'$s$ being of 5.2 eV and 4.4 eV, respectively [73]. This suggests that 
\textit{$\phi $}$_{int}$ of NiPc is of about 4 eV. Likewise, the work function of 
12.8-nm-thick PTCBI films displayed the same magnitude of about 4.5 eV 
irrespective the employed substrates, such as Au, Ag, or Mg [74]. A next 
example would be the 6T films with almost the same work function (spanning 
from 4.3 to 4.5 eV) in spite of the growth onto three distinct substrates 
with their work function ranging from 3.9 to 5.4 eV [75]. Further, \textit{$\phi $}$_{int}$ 
of PTCDA is assumed to be of 4.7 eV, as this was the work function of 12 
nm-thick films grown on GaAs (100) substrates with different work functions 
ranging from 4.52 to 5.39 eV attained via specific surface modifications 
[76]. 

The above mentioned and further examples are summarized in Tabs.~3.1 and 
3.2, whilst Tab.~3.2 contains solitaire measurements, thus of thick 
particular films prepared on one kind of substrate. Admittedly, the latter 
examples do not rigorously prove that the particular film ends up with 
\textit{$\phi $}$_{int}$ irrespective of the substrate. Yet, the film thickness and the 
growth conditions suggest that \textit{$\phi $}$_{int}$ was attained. Rather large span of 
work functions can be noticed when various thick films of a particular 
molecule are compared, $e.g.$ of 0.47 eV among the F$_{16}$CuPc$s$ and about 0.6 eV 
among 6T$s$; the data are apparently scattered mostly owing to the varying 
molecular orientation. The issue on the molecular orientation-resolved 
\textit{$\phi $}$_{int}$ will be discussed below. 

Of the numerous reported molecular films, eligibility of a particular film 
for the compilation resumed in Tabs.~3.1 and 3.2 was determined by the 
following concerns: 
\begin{itemize}
\item[($i$)] The sufficient film thickness: the molecular films with the thickness
of 5-10 nm were adopted. Thinner films, but still moderately thick ones, were 
included in the compilation provided that work function was similar to the 
work function of the same thick film grown however on distinct substrates 
and substrate-related features in the spectra were absent. 
\item[($ii$)] The growth fashion: nominally thick molecular films, yet with the
detectable  substrate features, were excluded, since this suggested electronically
and  morphologically markedly inhomogeneous films [58, 70, 77, 78]. 
\end{itemize}
The mentioned selection concerns applied on the variety of reported 
photoemission studies naturally limit the number of eligible data, which 
eventuates to the list of rather frequently investigated molecules. Note 
that the employed substrates include metallic, semiconducting, and molecular 
surfaces.
\renewcommand{\arraystretch}{1.1}
\begin{table}[!htbp]
\vspace*{-0.2cm}
\begin{center}
\caption{The ionization energy, $E_{I,film}$, the work function, $\phi_{film}$, 
of thick molecular films with the thickness $t_{film}$ grown on a variety of 
substrates described by their work function, \textit{$\phi $}$_{subs}$. The
indices $L$ and $U$ indicate the values relevant to the lying and upright geometry, 
respectively, provided that the molecular orientation was probed. The work 
function values, which were not explicitly reported in the literature, were 
calculated using the formula $\phi = E_{I}-BA$. The HOMO positions necessary for 
the $E_{I}$ evaluation were determined according to their low-$E_{B}$ onsets. 
The UHV-SiO$_{2}$ states for the SiO$_{2}$ prepared under UHV conditions 
[27]. Further employed acronyms state as follows: PEDOT:PSS 
[Poly(3,4-ethylenedioxythiophene):poly(styrenesulfonate)], SAM 
(self-assembled monolayer), and HOPG (highly-ordered pyrolitic graphite).}
\medskip
\begin{tabular}
{|p{60pt}|p{25pt}|p{25pt}|p{30pt}|p{88pt}|p{24pt}|p{28pt}|}
\hline
\multicolumn{4}{|l|}{\qquad\qquad\quad Molecular film} & 
\multicolumn{2}{l|}{\qquad\quad\ \  Substrate} & 
\raisebox{-3.00ex}[0cm][0cm]{Ref.} \\
\cline{1-6} 
\raisebox{-1.50ex}[0cm][0cm]{Acronym}& 
$t_{film}$& 
\textit{$\phi $}$_{film}$ & 
$E_{I,film}$ & 
\raisebox{-1.50ex}[0cm][0cm]{Acronym}& 
\textit{$\phi $}$_{subs}$ & 
  \\
& 
(nm)& 
(eV)& 
(eV)& 
& 
(eV)& 
  \\
\hline
\raisebox{-7.00ex}[0cm][0cm]{6T }& 
10& 
3.95$^{U}$& 
4.4$^{U}$& 
\raisebox{-1.50ex}[0cm][0cm]{Native SiO$_{2}$}& 
4.26& 
\raisebox{-7.50ex}[0cm][0cm]{[27]} \\
\cline{2-4} \cline{6-6} 
 & 
23& 
3.88$^{U}$& 
4.6$^{U}$& 
 & 
4.06& 
  \\
\cline{2-6} 
 & 
5& 
4.35$^{L}$& 
5.35$^{L}$& 
SiO$_{2}$ (90K)& 
4.83& 
  \\
\cline{2-6} 
 & 
20& 
3.96$^{U}$& 
4.65$^{U}$& 
Si(111)-7$\times $7& 
4.6& 
  \\
\cline{2-6} 
 & 
20& 
3.94$^{U}$& 
4.55$^{U}$& 
\raisebox{-1.50ex}[0cm][0cm]{UHV-SiO$_{2}$}& 
5.05& 
  \\
\cline{2-4} \cline{6-6} 
 & 
30& 
3.97$^{U}$& 
4.7$^{U}$& 
 & 
5.15& 
  \\
\hline
\end{tabular}
\end{center}
\end{table}

\begin{table}[!htbp]
{\small \ \ \ \ Tab.~3.1 {\it continued}}\vspace{-0.2cm}
\begin{center}
\begin{tabular}
{|p{60pt}|p{25pt}|p{25pt}|p{30pt}|p{88pt}|p{24pt}|p{28pt}|}
\hline
\cline{2-7} 
\raisebox{-15.00ex}[0cm][0cm]{6T } & 
2.8& 
4.3& 
5.3& 
Au& 
5.1& 
\raisebox{-3.00ex}[0cm][0cm]{[75]} \\
\cline{2-6} 
 & 
5.2& 
4.5$^{L}$& 
5.15$^{L}$& 
Au/PFDT& 
5.4& 
  \\
\cline{2-6} 
 & 
5.3& 
4.3& 
5.2& 
Au/ODT& 
3.9& 
  \\
\cline{2-7} 
 & 
6.4& 
3.5& 
5.3& 
Ag& 
4.2& 
\raisebox{-10.50ex}[0cm][0cm]{[79]} \\
\cline{2-6} 
 & 
2-3& 
3.9& 
4.6& 
Pd& 
5.1& 
  \\
\cline{2-6} 
 & 
2-3& 
4.0& 
5.2& 
Au& 
5.2& 
  \\
\cline{2-6} 
 & 
2-3& 
4.0& 
4.6& 
Pt& 
5.5& 
  \\
\cline{2-6} 
 & 
2-3& 
3.8& 
4.5& 
Contaminated Ag& 
4.0& 
  \\
\cline{2-6} 
 & 
2-3& 
3.9& 
4.4& 
Contaminated Pd& 
4.4& 
  \\
\cline{2-6} 
 & 
2-3 & 
3.9& 
4.4& 
Contaminated Au& 
4.3& 
  \\
\cline{2-6} 
 & 
2-3& 
4.0& 
4.5& 
Contaminated Pt& 
4.1& 
  \\
\hline
\raisebox{-4.50ex}[0cm][0cm]{$\alpha - $NPD, NPB}& 
5.5& 
3.8& 
5.2& 
Au/Alq$_{3}$& 
4.0& 
[80] \\
\cline{2-7} 
 & 
10& 
3.95& 
5.35& 
Au& 
5.1& 
[81] \\
\cline{2-7} 
 & 
10& 
4.18& 
5.4& 
Au& 
5.14& 
\raisebox{-6.00ex}[0cm][0cm]{[82]} \\
\cline{2-6} 
 & 
10& 
3.4& 
5.5& 
Mg& 
3.65& 
  \\
\cline{1-6} 
\raisebox{-3.00ex}[0cm][0cm]{BCP}& 
10& 
4.18& 
6.4& 
Au/$\alpha $-NPD& 
4.18& 
  \\
\cline{2-6} 
 & 
n.a.& 
4.14& 
6.4& 
Au/doped $\alpha $-NPD& 
4.74& 
  \\
\cline{2-6} 
 & 
n.a.& 
3.4& 
6.4& 
Mg/$\alpha $-NPD& 
3.4& 
  \\
\hline
\raisebox{-1.50ex}[0cm][0cm]{BP2T}& 
25.6& 
4.47& 
5.3& 
Au/F$_{16}$CuPc& 
4.8& 
\raisebox{-1.50ex}[0cm][0cm]{[83]} \\
\cline{2-6} 
 & 
20& 
4.71& 
5.27& 
Au& 
5.3& 
  \\
\hline
\raisebox{-1.50ex}[0cm][0cm]{CBP}& 
10& 
4.45& 
6.21& 
Au/ZnPc& 
4.45& 
\raisebox{-1.50ex}[0cm][0cm]{[82]} \\
\cline{2-6} 
 & 
10& 
4.54& 
6.16& 
Au/doped ZnPc& 
5.04& 
  \\
\hline
\raisebox{-12.00ex}[0cm][0cm]{CuPc}& 
5-10& 
3.9& 
5.0& 
Au& 
5.3& 
[84] \\
\cline{2-7} 
 & 
9.4& 
4.1& 
5.0& 
Au(100), Au& 
5.3& 
[85,86] \\
\cline{2-7} 
 & 
11& 
4.1& 
5.0& 
GeS(100)& 
4.6& 
[86] \\
\cline{2-7} 
 & 
15& 
3.87& 
4.82& 
$p$-Si(111)& 
4.22& 
[20] \\
\cline{2-7} 
 & 
5& 
4.35$^{L}$& 
5.2$^{L}$& 
HOPG& 
n.a.& 
\raisebox{-1.50ex}[0cm][0cm]{[65]} \\
\cline{2-6} 
 & 
5& 
3.95$^{U}$& 
4.8$^{U}$& 
SiO$_{2}$& 
n.a.& 
  \\
\cline{2-7} 
 & 
20& 
4.24& 
4.82& 
Native SiO$_{2}$/F$_{16}$CuPc& 
5.3& 
[87] \\
\cline{2-7} 
 & 
3.6& 
4.0& 
4.6& 
H-Si(111)& 
4.28& 
[39] \\
\cline{2-7} 
 & 
9& 
3.9& 
5.0& 
Au& 
5.3& 
\raisebox{-3.00ex}[0cm][0cm]{[84]} \\
\cline{1-6} 
\raisebox{-4.50ex}[0cm][0cm]{F$_{4}$CuPc}& 
5-10& 
4.7& 
5.7& 
Au& 
5.3& 
  \\
\cline{2-6} 
 & 
5-10& 
4.3& 
5.7& 
ITO& 
4.2& 
  \\
\cline{2-7} 
 & 
8.5& 
4.7& 
5.7& 
Au(100)& 
5.3& 
[85] \\
\cline{2-7} 
 & 
15& 
4.7& 
5.55& 
$p$-Si(111)& 
4.22& 
[20] \\
\hline
\raisebox{-9.00ex}[0cm][0cm]{F$_{16}$CuPc}& 
9& 
4.95& 
6.1& 
Au& 
5.3& 
[84] \\
\cline{2-7} 
 & 
15& 
5.42& 
6.32& 
$p$-Si(111)& 
4.22& 
[20] \\
\cline{2-7} 
 & 
25.6& 
5.13& 
6.44& 
Au/BP2T& 
4.71& 
[83] \\
\cline{2-7} 
 & 
20& 
5.3& 
6.66& 
Native SiO$_{2}$/CuPc& 
4.24& 
[87] \\
\cline{2-7} 
 & 
10& 
5.1& 
6.3& 
Au& 
5.0& 
[88] \\
\cline{2-7} 
 & 
4& 
4.7$^{L}$& 
5.9$^{L}$& 
HOPG& 
4.3& 
\raisebox{-1.50ex}[0cm][0cm]{[65]} \\
\cline{2-6} 
 & 
4& 
5.3$^{U}$& 
6.6$^{U}$& 
Au(111)/C$_{8}$-SAM& 
n.a.& 
  \\
\hline
\end{tabular}
\end{center}
\end{table}

\begin{table}[!htbp]
{\small \ \ \ \ Tab.~3.1 {\it continued}}\vspace{-0.2cm}
\begin{center}
\begin{tabular}
{|p{60pt}|p{25pt}|p{25pt}|p{30pt}|p{88pt}|p{24pt}|p{28pt}|}
\hline

\raisebox{-1.50ex}[0cm][0cm]{HAT-CN}& 
90& 
5.95& 
9.75& 
ITO& 
4.46& 
[89] \\
\cline{2-7} 
 & 
80& 
6.1& 
9.9& 
Au& 
5.2& 
[90] \\
\hline
\raisebox{-1.50ex}[0cm][0cm]{NiPc}& 
10& 
4.0& 
5.0& 
Au& 
5.2& 
\raisebox{-1.50ex}[0cm][0cm]{[73]} \\
\cline{2-6} 
 & 
10& 
4.0& 
5.0& 
Ag& 
4.4& 
  \\
\hline
\raisebox{-6.00ex}[0cm][0cm]{PEN}& 
6.4& 
4.33& 
4.91& 
SiO$_{2}$& 
4.33& 
[91] \\
\cline{2-7} 
 & 
15& 
4.35& 
5.2& 
Au& 
5.4& 
[81] \\
\cline{2-7} 
 & 
12.8& 
4.5& 
4.85& 
PEDOT:PSS& 
5.3& 
[92] \\
\cline{2-7} 
 & 
20& 
4.32$^{U}$& 
4.77$^{U}$& 
SiO$_{2}$& 
n.a.& 
\raisebox{-1.50ex}[0cm][0cm]{[67]} \\
\cline{2-6} 
 & 
3.2& 
4.47$^{U}$& 
4.73$^{U}$& 
HOPG/CuPc& 
n.a.& 
  \\
\hline
\raisebox{-1.50ex}[0cm][0cm]{PFP}& 
7& 
4.95& 
5.6$^{L}$& 
Au(111)& 
5.45& 
\raisebox{-1.50ex}[0cm][0cm]{[70]} \\
\cline{2-6} 
 & 
7& 
4.95 & 
$\sim$$6.3^{U}$& 
Au(111)& 
5.45& 
  \\
\hline
\raisebox{-3.00ex}[0cm][0cm]{PTCBI}& 
12.8& 
4.6& 
6.2& 
Au& 
5.0& 
\raisebox{-3.00ex}[0cm][0cm]{[74]} \\
\cline{2-6} 
 & 
6.3& 
4.5& 
6.2& 
Ag& 
4.3& 
  \\
\cline{2-6} 
 & 
6.3& 
4.5& 
6.0& 
Mg& 
3.8& 
  \\
\hline
\raisebox{-12.00ex}[0cm][0cm]{PTCDA}& 
12& 
4.77& 
6.67& 
GaAs(100)-c(4$\times $4)& 
4.63& 
\raisebox{-4.50ex}[0cm][0cm]{[76]} \\
\cline{2-6} 
 & 
12& 
4.56& 
6.6& 
S-GaAs(100)& 
4.52& 
  \\
\cline{2-6} 
 & 
12& 
4.55& 
6.58& 
S-GaAs(100)& 
4.68& 
  \\
\cline{2-6} 
 & 
12& 
4.65& 
6.56& 
S-GaAs(100)& 
4.98& 
  \\
\cline{2-7} 
 & 
15& 
4.94& 
6.95& 
S-GaAs(100)& 
5.4& 
[20] \\
\cline{2-7} 
 & 
12& 
4.8& 
6.65& 
Se-GaAs(100)-(2$\times $1)& 
5.39& 
[76] \\
\cline{2-7} 
 & 
10& 
5.0& 
6.8& 
Au& 
5.2& 
[93] \\
\cline{2-7} 
 & 
8& 
4.8& 
6.55& 
HOPG/F$_{16}$CuPc$^{L}$& 
4.7& 
\raisebox{-1.50ex}[0cm][0cm]{[65]} \\
\cline{2-6} 
 & 
8& 
4.95& 
6.4& 
C$_{8}$-SAM/F$_{16}$CuPc$^{U}$& 
5.3& 
  \\
\hline
\end{tabular}
\end{center}
\end{table}
\begin{table}[!htbp]
\vspace*{-0.2cm}
\caption{The continuation of Tab.~3.1 but for molecular films with only 
solitary reported growth on a particular substrate.}
\begin{center}
\begin{tabular}
{|p{60pt}|p{25pt}|p{25pt}|p{30pt}|p{88pt}|p{24pt}|p{28pt}|}
\hline
\multicolumn{4}{|l|}{\qquad\qquad\quad Molecular film} & 
\multicolumn{2}{l|}{\qquad\quad\ \  Substrate} & 
\raisebox{-3.00ex}[0cm][0cm]{Ref.} \\
\cline{1-6} 
\raisebox{-1.50ex}[0cm][0cm]{Acronym}& 
$t_{film}$& 
\textit{$\phi $}$_{film}$ & 
$E_{I,film}$ & 
\raisebox{-1.50ex}[0cm][0cm]{Acronym}& 
\textit{$\phi $}$_{subs}$ & 
  \\
& 
(nm)& 
(eV)& 
(eV)& 
& 
(eV)& 
  \\
\hline
6P& 
20& 
4.3& 
6.1& 
Au& 
5.1& 
[81] \\
\hline
Alq$_{3}$& 
15& 
4.0& 
5.7& 
Au& 
n.a.& 
[80] \\
\hline
DH4T& 
10.4& 
4.05$^{U}$& 
4.9& 
Au& 
5.25& 
[94] \\
\hline
DH6T& 
6& 
$\sim 3.6$& 
4.85& 
Ag(111)& 
4.5& 
[45] \\
\hline
DiMe-PTCDI& 
15& 
4.55& 
6.58& 
S-GaAs(100)& 
5.17& 
[20] \\
\hline
H$_{2}$Pc& 
15& 
4.04& 
4.96& 
$p$-Si(111)& 
4.22& 
[20] \\
\hline
PTCDI& 
15& 
4.67& 
6.42& 
S-GaAs(100)& 
5.28& 
[20] \\
\hline
ZnPc& 
10& 
4.45& 
5.29& 
Au& 
5.14& 
[82] \\
\hline
\end{tabular}
\end{center}
\end{table}

\subsubsection{Molecular orientation-resolved intrinsic work function.}

The work function comprises of two contributions, namely of the chemical 
potential and the surface dipole [Eq. (\ref{eq3})]. The latter depends on the 
molecular orientation, since different molecular orientations imply the 
distinct surface terminations leading to the different surface dipole. 
Therefore, it is reasonable to presume that the intrinsic work function of a 
particular molecular film depends on the molecular orientation in the 
surface region. Two extreme situations in terms of the molecular orientation 
can occur: the film formed by lying- and upright-oriented molecules with the 
corresponding intrinsic work function, namely \textit{$\phi $}$_{int,L}$ and \textit{$\phi $}$_{int,U}$. 
Thus, the molecular orientation-resolved work function of a particular 
molecular film would represent the work functions of the corresponding faces 
of a molecular crystal. 

It can be noted that various surface reconstruction of single-crystal 
elemental surfaces can manifest the work function differing as much as 
several tenths of electronvolts. An extreme example would be the work 
functions of W(111) and W(110) being of 4.47 and 5.25 eV, respectively, 
thereby differing by 0.78 eV [95]. 

Given the data presented in Tab.~3.1 and 3.2, estimations on intrinsic 
work functions of selected molecular films are compiled in Tab.~3.3. 
Admittedly, the available data present a small
\begin{table}[!htbp]
\vspace*{-0.2cm}
\renewcommand{\arraystretch}{1.0}
\caption{The estimated intrinsic work function $\phi_{int}$ of selected 
molecular films discriminated by the molecular orientation. The values typed 
in bold were adopted from studies, where both work-function and 
molecular-orientation characterizations were performed. The remaining values 
were reported in works with lacking characterization of the molecular 
orientation; yet both the film thickness and the growth conditions suggest 
the upright-oriented molecules.}
\begin{center}
\begin{tabular}
{|p{60pt}|c|c|}
\hline
Molecular film& 
\textit{$\phi $}$_{int,L}$ \par (eV)& 
\textit{$\phi $}$_{int,U}$ \par (eV) \\
\hline
6T& 
\textbf{$\sim $4.35}& 
\textbf{$\sim $3.95} \\
\hline
$\alpha - $NPD, NPB& 
--& 
3.8-4.2  \\
\hline
Alq3& 
--& 
$\sim 4.0$ \\
\hline
BCP& 
--& 
$\sim 4.15$ \\
\hline
BP2T& 
--& 
4.5-4.7 \\
\hline
CBP& 
--& 
$\sim 4.5$ \\
\hline
CuPc& 
\textbf{$\sim $4.4}& 
\textbf{$\sim $3.95} \\
\hline
DH4T& 
--& 
\textbf{$\sim $4.05} \\
\hline
DiMe-PTCDI& 
--& 
$\sim 4.55$ \\
\hline
F$_{4}$CuPc& 
--& 
4.7 \\
\hline
F$_{16}$CuPc& 
\textbf{4.7}& 
\textbf{5.3} \\
\hline
H$_{2}$Pc& 
--& 
$\sim 4.0$ \\
\hline
NiPc& 
--& 
4.0 \\
\hline
PEN& 
--& 
\textbf{4.3-4.5} \\
\hline
PFP& 
$\sim 4.95$& 
$\sim 4.95$ \\
\hline
PTCBI& 
--& 
4.5-4.6 \\
\hline
PTCDA& 
\textbf{4.5-4.94}& 
-- \\
\hline
PTCDI& 
--& 
$\sim 4.7$ \\
\hline
\end{tabular}
\end{center}
\end{table}
statistical ensemble not 
sufficient for reliable estimates; nevertheless, the presented values are 
intended to provide an aid for their further verifications. Note that the 
following molecular films show about the same $\phi_{int}\sim 4.0$ eV for 
upright geometry: 6T [27, 79], 6P [96], H$_{2}$Pc [20], CuPc [20, 39, 84], NiPc 
[73], Alq$_{3}$ [91], $\alpha $-NPD [82], and BCP [82]; we assume that the 
similar \textit{$\phi $}$_{int}$ are owing to similar surface terminations,
namely carbon  atoms with bonded hydrogen suggesting the similar surface dipole. 

As it was mentioned above, thick films can undergone the L$ \to $U 
orientational transition (Fig.~2.1), which depends on the growth conditions 
such as the substrate surface morphology, growth rate, and growth 
temperature. This implies that the work function switches from $\phi_{int,L}$ to 
$\phi_{int,U}$ with the film thickness and the orientational transition-related 
work function change is
\begin{equation}
\Delta \phi _{L \to U} = \phi _{int,L} - \phi _{int,U}.
\end{equation}

The orientational transition can be abrupt with the first lying monolayer 
followed by upright-oriented next monolayers [43-45, 98, 99]. Yet, the gradual 
transition accomplished up to thickness of several nanometres had been more 
frequently encountered [27, 39, 46-49, 100]. As the molecular orientation had 
been rarely determined in reported studies, the compilation presented in 
Tabs.~3.1 and 3.2 focussed to the molecular films with the thickness of 5-10 nm, 
where the L$ \to $U orientational transitions were presumably accomplished. 
Admittedly, the minimal thickness required for attainment of $\phi_{int}$ may be 
substantially lower provided that the particular molecular orientation is 
preserved through the entire film. 

In fact, the $\Delta\phi_{L \to U}$ can be inferred from several studies, 
where thicker films were examined, yet with lacking molecular-orientation 
characterization: with the increasing film thickness, the work function 
suggests its saturation after the ID formation, but the work function 
continues to gradually drop by several tenths of electronvolts with the 
further growth. Such evolution was observed for 6T [27], NiPc [73], CuPc 
[39, 66, 86, 101, 102], and thiophene [75], and DH4T [94]. Yamane \textit{et al.}
suggested that the gradual work function change is due to the summation of
incremental  dipoles, which were arising due to the gradually changed molecular
tilt angle [101]. 

Chen \textit{et al.} measured simultaneously both the work function and the
molecular  orientation of CuPc films and observed the work function of 4.35 eV
and 3.95 
eV, $i.e.,$ differing by -0.4 eV, for the films built from lying- and 
upright-oriented molecules [103]. Interestingly, the $\Delta\phi_{L \to U}
\approx -0.4$ eV, when going from lying- towards upright-oriented 
molecules, had been frequently observed for unsubstituted (hydrogen 
terminated) molecules, such as 6T [27, 63], NiPc [73], CuPc 
[66, 86, 101, 103-104], DH4T [94], HBC [49], DH6T [45]. In contrast,
$\Delta\phi_{L \to U} > 0$ was observed for fluorine-substituted molecular films such 
as F$_{16}$CuPc (+0.85 eV) [103] (+0.6 eV) [65], and PFP [58]. Since 
fluorine has a high $E_{I}$, which in turn suggests the high work function of 
fluorine-terminated surface (see Section 3.4), the inequality $\phi_{L}<\phi_{U}$ 
for the fluorine-substituted molecules can be qualitatively explained via 
the increased density of fluorine atoms terminating the film comprising 
upright oriented molecules in comparison to the film terminated by lying 
molecules.

\subsection{Molecular orientation-resolved growth modes}

Growth fashions of grown films are commonly categorized into three basic 
modes according to the morphology evolution, namely the Frank-Van der Merwe, 
Stranski-Krastanov, and Volmer-Weber growth modes referring to the laminar 
growth, the islanding on a wetting layer, and the island growth, 
respectively. The growth fashion determines the evolution of the probed 
surface electronic properties and, conversely, the evolution of electronic 
properties can be employed for the growth mode determination. For the 
surface electronic properties of molecular films are dramatically affected 
by the molecular orientation that has to be involved in the discernment of 
the growth fashions. Figure~3.4 illustrates the often observed growth 
fashions of molecular films with the consideration of the molecular 
orientation: 
\begin{itemize}
\item[($a$)] the laminar growth of lying molecules; 
\item[($b$)] the Stranski-Krastanov growth of lying molecules; 
\item[($c$)] the laminar growth of upright-oriented molecules; 
\item[($d$)] the laminar growth with the abrupt L$ \to $U orientational transition 
beyond the first lying monolayer; 
\item[($e$)] the laminar growth with the gradual L$ \to $U orientational transition 
beyond the first lying monolayer; and 
\item[($f$)] the laminar growth of a multilayer formed by lying molecules followed by 
the gradual L$ \to $U orientational transition. 
\end{itemize}

\begin{figure}[tbp]
\begin{center}
\includegraphics[width=6.5cm,clip]{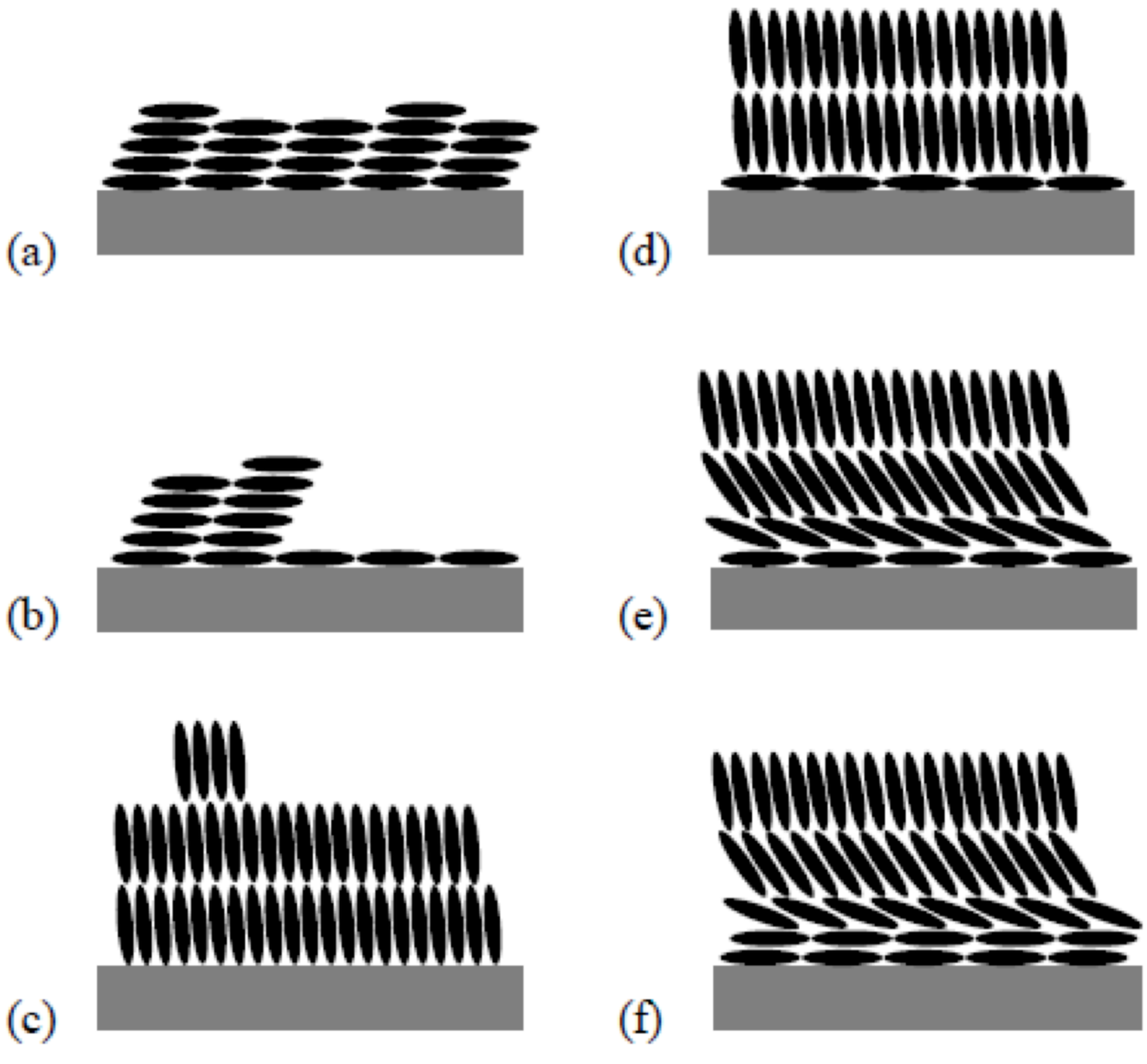}
\caption{The sketches of the often observed growth fashions of molecular films
with the consideration of the molecular orientation. Except the panel ($b$) indicating
the Stranski-Krastanov growth mode with the lying molecular orientation, all
panels depict the laminar growth, albeit with distinct evolutions of the molecular
orientational transition. Reprinted with permission from J. Ivanco, Thin Solid
Films {\bf 520} (2012) 3975. Copyright 2012 Elsevier. }
\end{center}
\end{figure}

Except the ($b$) representing the islanding, the further examples pertain to 
the laminar growth. The ($a-c$) presume the molecular orientation preserved 
during the growth, the ($d-f$) consider the orientational transition. Note that 
the growth fashions illustrated in Fig.~3.4 are rather elementary ones 
chosen to illustrate possible scenarios. In reality, the structure evolution 
with the film thickness may be more complex particularly at a 
heterostructure growth [105]. 

As for the work function change, $\Delta\phi$, induced by the molecular 
film growth, two contributions can be distinguished; namely due to ($1$) the 
formation of the 1$^{\rm st}$ monolayer, $\Delta\phi_{ML}$, accounted for the ID 
formation, and ($2$) the L$ \to $U orientational transition, $\Delta\phi_{L \to 
U}$ [Eq. (3.4)]: 
\begin{equation}
\Delta \phi = \Delta \phi _{ML} + \Delta \phi _{L \to U} ,
\end{equation}

\begin{figure}[tb]
\begin{center}
\includegraphics[width=7.0cm,clip]{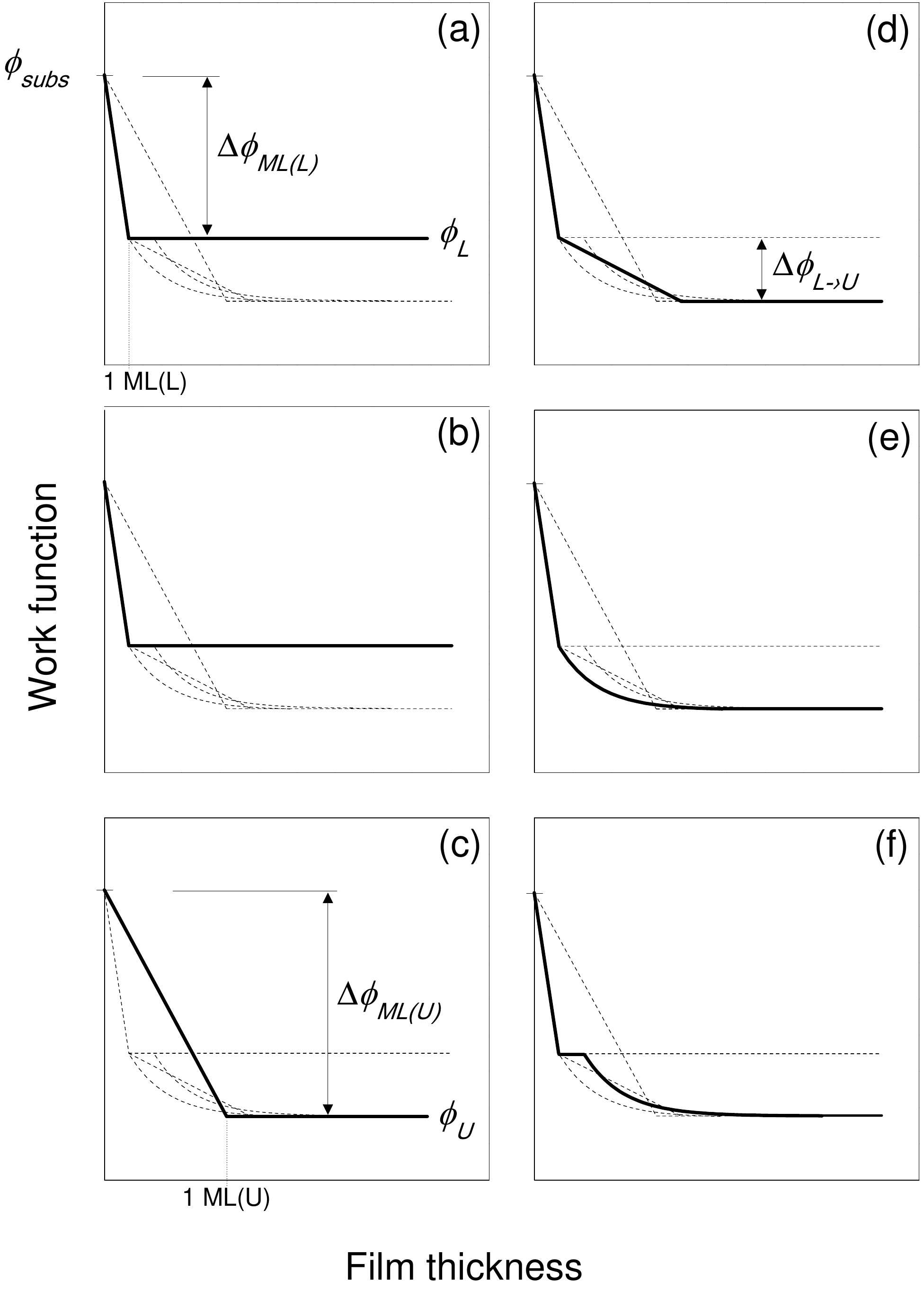}
\caption{The sketched work function evolutions---and thereby the energy level
shift---with the film thickness corresponding to the distinct growth scenarios
shown in Fig.~3.4 are indicated by thick solid lines. Besides, all discussed
$\phi$-evolutions are repeatedly drawn by dashed lines in each panel to allow
the comparison. The more frequent encountered inequality $\phi_L > \phi_U$ was
assumed here. The $\Delta\phi_{ML(L)}$ and $\Delta\phi_{ML(U)}$ refer to the
$\Delta\phi$ examined upon the growth of 1$^{\rm st}$ monolayer formed by
lying- and upright-oriented molecules, respectively. Reprinted with permission
from J. Ivanco, Thin Solid Films {\bf 520} (2012) 3975. Copyright 2012 Elsevier.}
\vspace*{-0.5cm}
\end{center}
\end{figure}

Panel $f$ shows a scenario with the $\Delta\phi_{L \to U}$ occurring at higher 
film thickness, which is manifested by an intermediate flattening of the 
work function at small thickness. The plateaued work function may suggest 
further examinations at higher thickness redundant. The work-function 
plateau can be absent provided that molecules are exclusively oriented 
upright through the entire film (panel $c$), $i.e.$ including the first monolayer, or 
the abrupt change from the lying towards upright occurs when going from the 
first do second monolayer (panel $d$) [45, 98, 99]. Note, that only the scenarios 
($a$-$c$) correspond to the expected work-function change illustrated in
Fig.~3.3. 
The situations ($d$-$f$) representing the orientational transitions had frequently 
implied erroneous conclusions on the presence of a band bending ({\it e.g.} Refs. 
[98, 99]), the situation ($e$) was rationalized as the mixing the interfacial 
dipole and a band bending [106]; the issue will be discussed in more details 
in Section 3.5. 

\subsection{The ionization energy-work function correlation}

In a first approximation, the first ionization energy of elements and the 
work function of the corresponding surfaces are correlated [107]: the 
correlation plot is shown in Fig.~3.6 with $E_{I}$ and $\phi $ values adopted from 
Refs. [95] and [108]. So far as distinct surface reconstructions for a 
particular element are reported, the magnitude for the unreconstructed 
surface was assumed. 

\begin{figure}[tb]
\begin{center}
\includegraphics[width=9.0cm,clip]{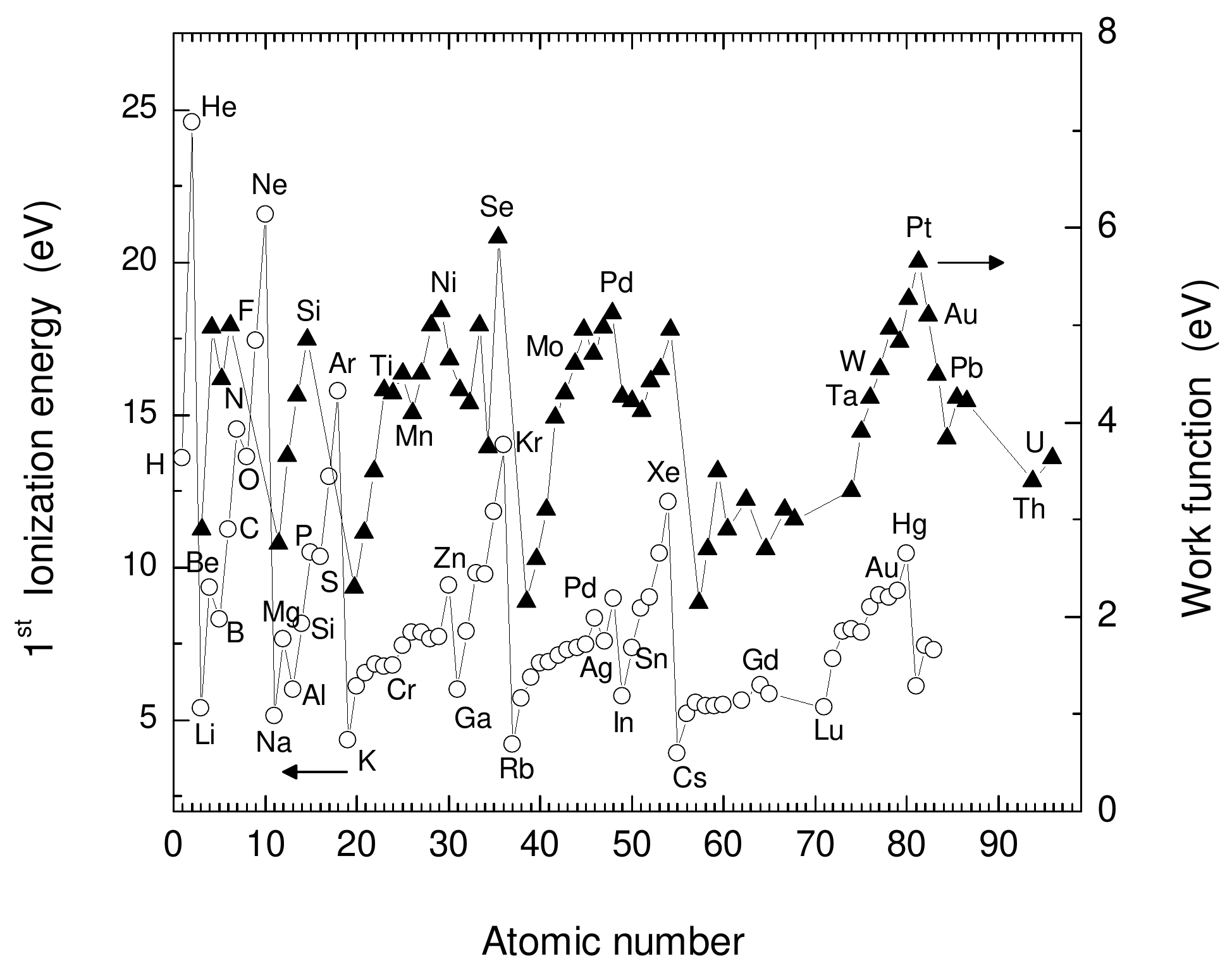}
\caption{The first ionization energy of elements (empty circles, [95]) suggests the
correlation with the work function (full triangles, [108]) of corresponding surfaces.
The solid line connecting the data points are to guide the eye. The arrows indicate
the relevant coordinate.}
\vspace*{-0.5cm}
\end{center}
\end{figure}

Analogous to the $\phi$\hspace*{0.04cm}-$E_{I}$ correlation for elements (Fig.~3.6), the thick 
molecular films show the $\phi_{int}$-$E_{I}$ correlation too. This is shown in 
Fig.~3.7a, where the $\phi_{int}$-$E_{I}$ points of molecular films (empty 
circles, adopted from Tab.~3.1) and the replotted correlation graph for 
elements introduced in Fig.~3.6 (full circles) are juxtaposed. Figure~3.7b 
repeats the $\phi_{int}$-$E_{I}$ correlation for molecular films though with 
their assignment. The dashed lines represent the corresponding linear fits, 
whilst As, C, and Hg---the merely labelled elements in the graph---were 
retracted from the fitting due to their large deviation. The linear fits 
follow the equations:
\begin{eqnarray}
&&\phi _{el} = 0.08 + 0.56\times E_{I,el} \\
&&\phi _{int} = 2.1 + 0.41\times E_{I,org} 
\end{eqnarray}
for the elements and for the molecular films addressed in this work. Note 
that the dispersion of the \textit{$\phi $}$_{int}$-$E_{I}$ dependences for both molecular 
films and elements are comparable. The correlation plot for molecular films 
does not discriminate the molecular orientation as there is a shortage of 
data explicitly relating the work function and the molecular orientation for 
specific organics. Note that the correlation suggests the commensurably high 
\textit{$\phi $}$_{int}$ of the films with high $E_{I}$, such as TCNQ, F$_{4}$TCNQ, 
F$_{4}$CuPc, F$_{16}$CuPc, NTCDA, and HAT-CN. 

\begin{figure}[tb]
\begin{center}
\includegraphics[width=11.0cm,clip]{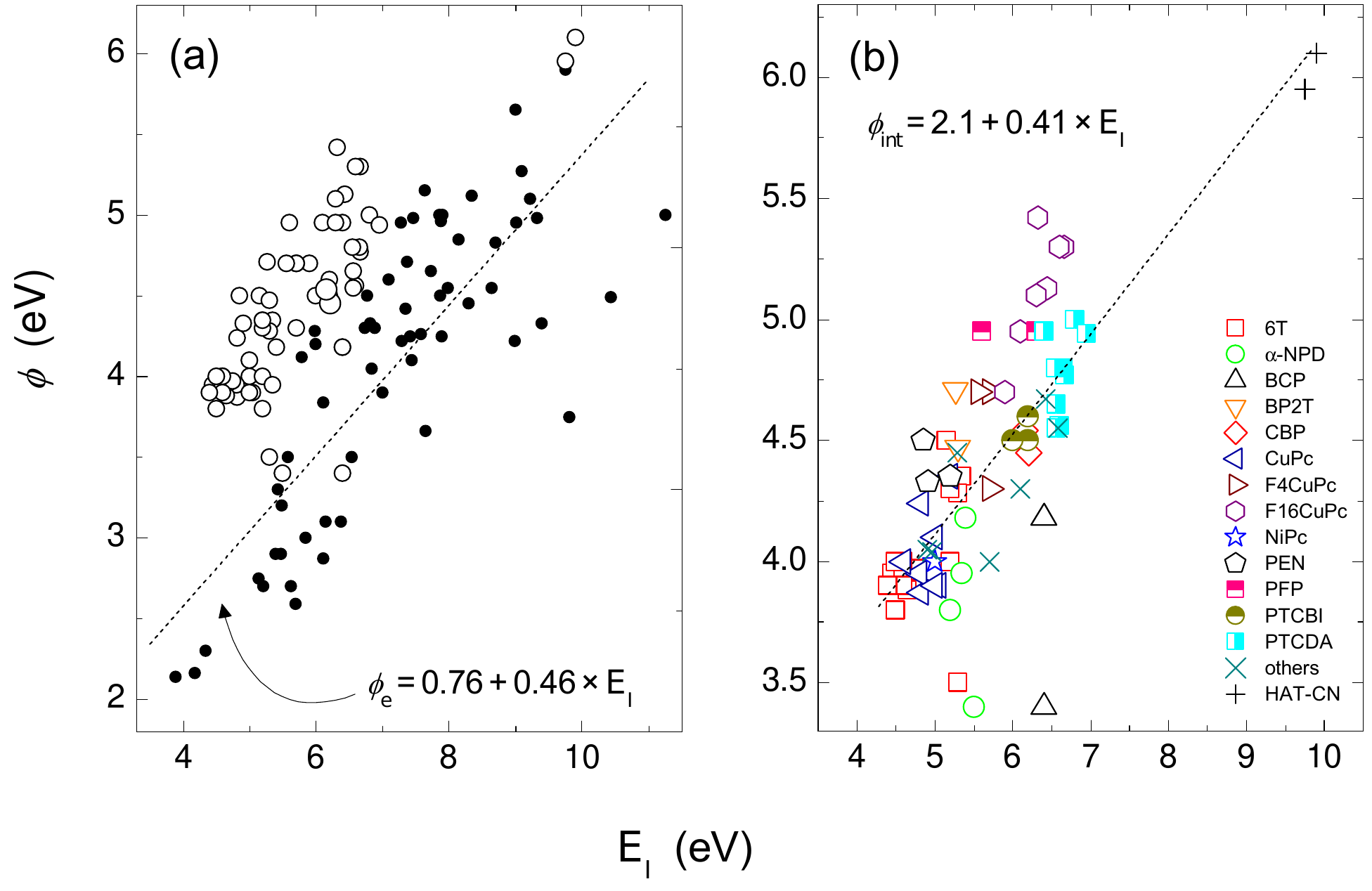}
\caption{(a) The $\phi$\hspace*{0.04cm}-$E_I$ correlation for elemental surfaces (full circles, [95, 107, 108])
and for surfaces of oligomeric molecular films addressed in this work (empty circles),
(b) the $\phi_{int}$-$E_I$ correlation for molecular films replotted from panel (a) with
indications of particular molecules.}
\vspace*{-0.5cm}
\end{center}
\end{figure}

\subsection{Band bending deduction from photoemission}

The free charge carrier concentration in the subsurface region of a 
semiconductor can be controlled by a field effect, $i.e.$ altering the electron 
energy levels by an external field. The effect is also termed the band 
bending as the energy bands change their relative position with respect to 
the Fermi level in the subsurface region. Both the upward and downward band 
bending can occur, the former being illustrated in Fig.~3.8. In contrast to 
$e.g.$ metals, where the binding energies of core levels $E_{B}$ are constant, 
$E_{B}$ of semiconductors core levels depend on the doping level, as that 
determines the position of the Fermi level in the energy gap. The band 
bending implies the shift of photoemission features for those are determined 
with respect to the analyser ground, so to the Fermi level. Conversely, the 
energy difference between the core levels at a semiconductor surface and in 
the bulk assigns the band bending. The upward-bent bands result in the 
high-$E_{B}$ rigid shift of all levels with the depth in semiconductor, 
accordingly away from the Fermi level; this is shown in Fig.~3.8 for the 
valence band edge. The band bending is caused by an external field or by a 
surface/interface charge and---conversely---the latter can be 
determined, $e.g.$, by examination the core-level shift upon an external bias 
[109].

\begin{figure}[tb]
\begin{center}
\includegraphics[width=6.7cm,clip]{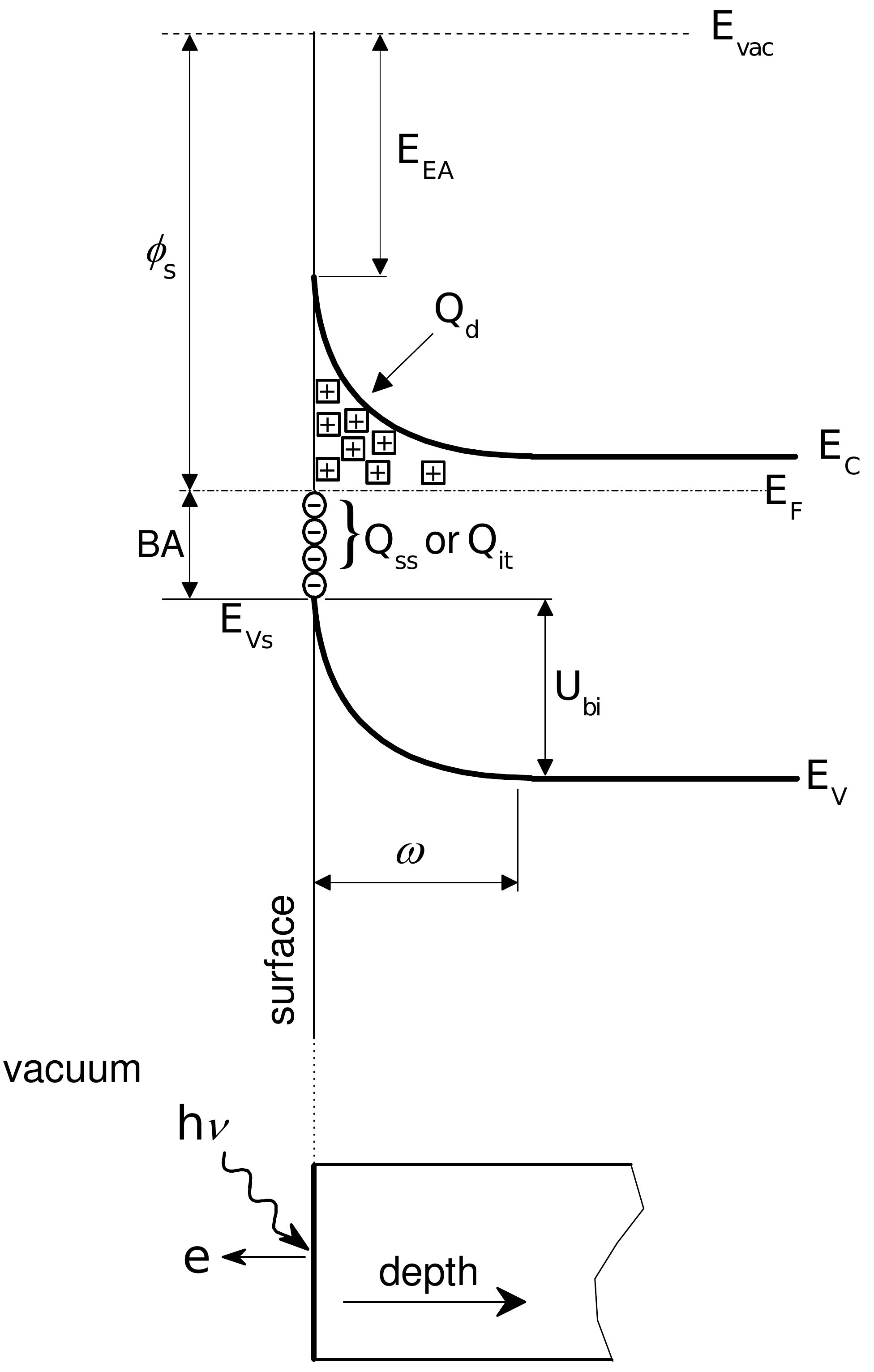}
\caption{The energy diagram of $n$-type semiconductor with the upward band bending;
$\phi_s$ is the semiconductor work function, $E_{EA}$ is the electron affinity at
the surface, $Q_d$ is the depletion zone charge, $w$ is the depletion zone width,
$U_{bi}$ is the build-in potential identified with the band bending magnitude, and
the BA is the band alignment, i.e. the binding energy of the valence band edge on
the surface, $E_{V_s}$. $Q_{ss}$ and $Q_{it}$ are the charge in surface and/or
interface states, respectively. The bottom pictogram sketches the experimental set up.}
\vspace*{-0.5cm}
\end{center}
\end{figure}

In organic electronics, some concepts well established in inorganic 
semiconductors, $e.g.$ the band bending, had been adopted for the molecular films. 
The presence of the band bending in molecular films would be of great 
importance for organic FETs; it would mean that the underlying operational 
mechanism of OFET mimics that of the traditional (inorganic) FET. Therefore, 
the issue had been carefully followed in photoemission characterizations of 
molecular films; the conclusion on upward band bending had been habitually 
allured by observing the high-$E_{B}$ shift of photoemission features with 
increasing film thickness [73, 74, 77, 80, 83, 86, 87, 106, 110-115]. The downward 
band bending was reported in F$_{16}$CuPc [116]. One should recognize that 
the high-$E_{B}$ shift associated with the band bending is an additional, 
further shift to that induced by the interfacial dipole (ID); the ID-related 
shift is abrupt; it is manifested by the work function drop, which commonly 
saturates within the completion of about a monolayer (Fig.~3.3). On the 
other side, the band-bending-associated shift is the gradual one in the 
range of several nanometres. The band bending in organic materials 
supposedly occurs owing to a light doping induced by impurities derivable 
from synthesis.

\begin{figure}[tb]
\begin{center}
\includegraphics[width=6.2cm,clip]{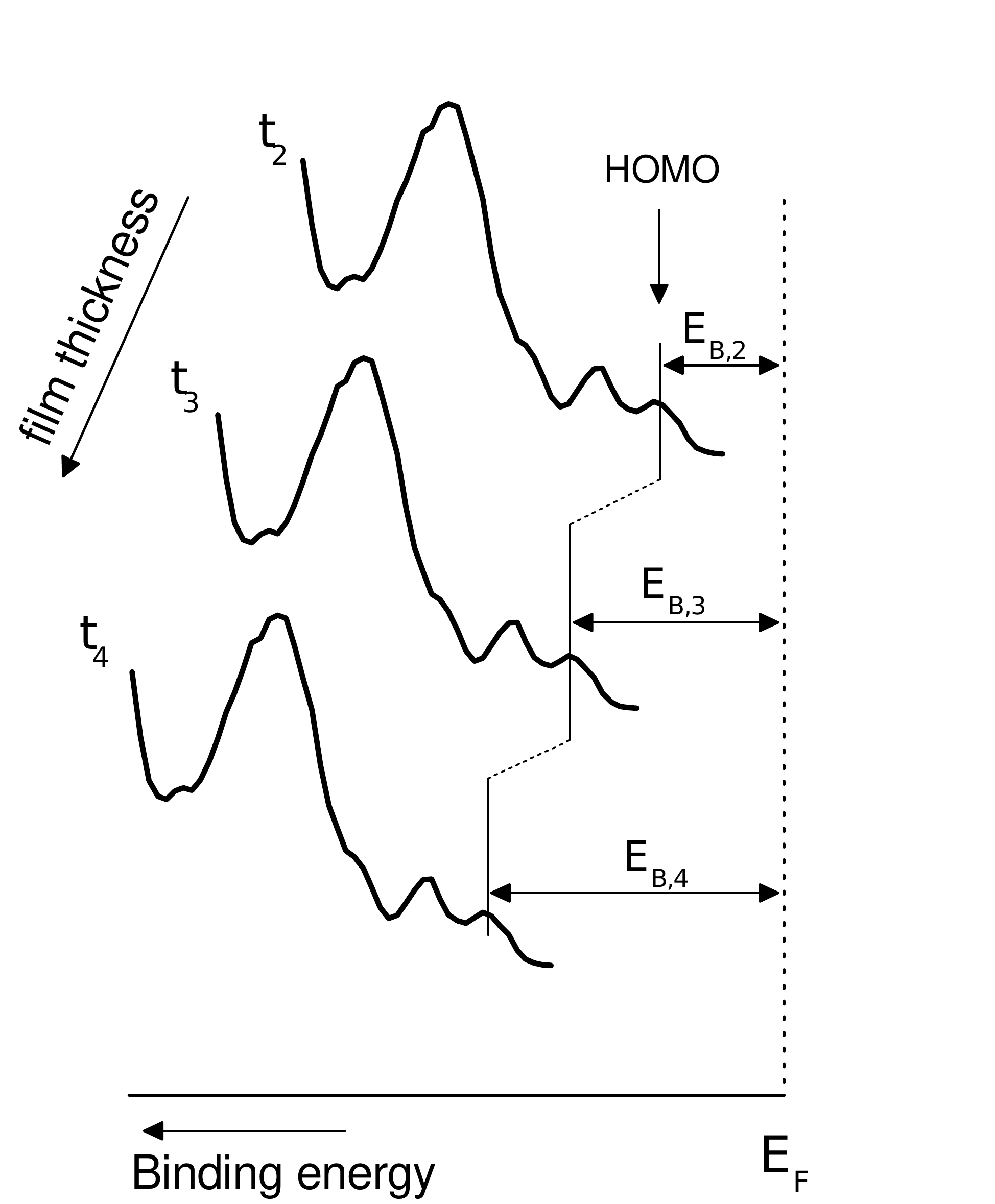}
\includegraphics[width=6.2cm,clip]{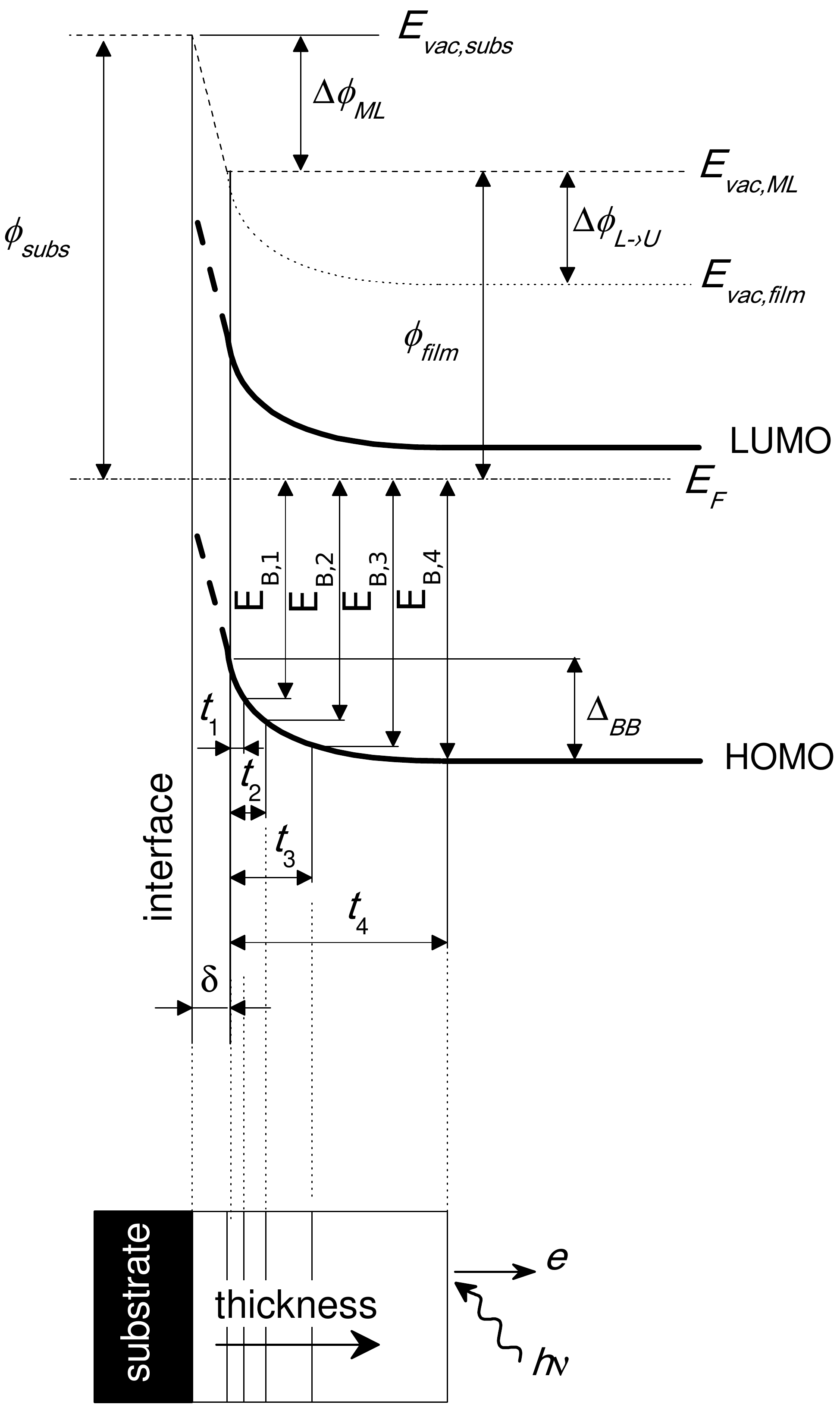}
\caption{The high-binding energy shift of energy levels with the increasing film
thickness, $t_2<t_3<t_4$ (left panel) and the corresponding notion in terms of the
band bending in the grown molecular film (right panel). The $t_1$ is the thickness
of the first monolayer. The pictogram sketches the experimental set up.}
\vspace*{-0.5cm}
\end{center}
\end{figure}

The prevailing notion on the (upward) band bending in a molecular film is 
explained in Fig.~3.9; the left panel shows the high-$E_{B}$ shift of VBs of 
a molecular film with the increasing film thickness, 
$t_{2}<t_{3}<t_{4}$. The right panel transforms the varying particular HOMO 
positions, $E_{B,2}<E_{B,3}<E_{B,4}$, to the corresponding points in the 
energy band diagram suggesting the band bending. The $E_{vac,ML}$ is the 
vacuum level upon the formation of the first monolayer with the thickness 
$\delta$, $\Delta\phi_{ML}$ is the corresponding vacuum-level change identified with 
the interfacial dipole. The $\Delta_{BB}$ is the total shift of energy 
levels of the film beyond the first monolayer interpreted as the band 
bending. 

The above interpretation framework presumes that the situation introduced in 
Fig.~3.9 is equivalent to the detection of the band bending at the surface 
or interface of inorganic semiconductors (Fig.~3.8). Yet, there is an 
essential difference as far as the respective photoemission 
characterizations are concerned; whereas the band bending in inorganic 
semiconductors is examined by a single measurement, the band bending in 
molecular films is deduced from a series of sequentially performed 
measurements during the film growth (see sketches at the bottom of the
Fig.~3.8 and Fig.~3.9). The most importantly, the successive characterizations of 
the molecular film are performed at distinct conditions, namely, at varying 
work function, which itself affects the shift of binding energies employed 
for the conclusions on the band bending. In other words, the interpretation 
framework for the band bending tacitly presumes the constant vacuum level 
after the first ML formation, $E_{vac,ML}$, with the increasing film 
thickness (Fig.~3.9 right panel, dashed line). Instead, the vacuum level 
varies according to $E_{vac,film}$ (dotted line) owing to the L$ \to $U 
orientational transition, thereby differing by $\Delta\phi_{L \to U }(t)$ from 
$E_{vac,ML}$. In the next paragraph, it will be shown that the work function 
change itself of the probed surface results in the equal rigid shift of 
photoemission spectra.

\begin{figure}[tb]
\begin{center}
\vspace*{-0.8cm}
\includegraphics[width=8.0cm,clip]{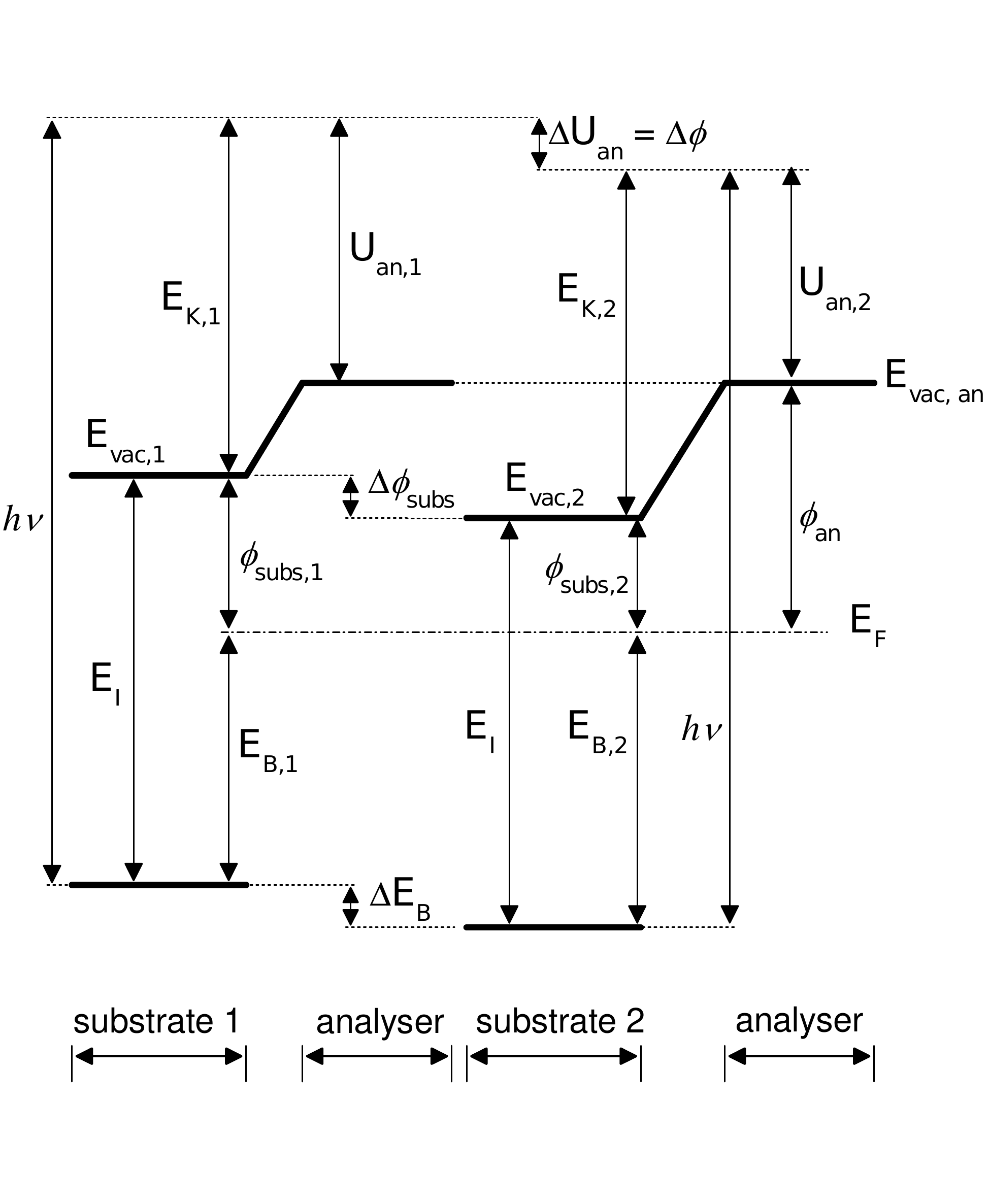}
\vspace*{-0.9cm}
\caption{The principal energy diagram of the photoemission measurement contrasting
an adsorbate characterized by its ionization energy on surfaces with work function
$\phi_{subs,1}$ and $\phi_{subs,2}$ (see text).}
\vspace*{-0.7cm}
\end{center}
\end{figure}

In probing the photoemission process, the binding energy, $E_{B}$, of a core 
level is given by the equation [117]: 
\begin{equation}
\label{eiiq1}
E_B = h\nu - U_{an} - \phi _{an} ,
\end{equation}
where $h\nu$ is the photon energy, $U_{an}$ is the voltage applied to the 
analyzer to detect an emitted electron, and $\phi_{an}$ is the work function of 
energy analyzer. $E_{B}$ is referred to the Fermi level. Even though Eq.~(\ref{eiiq1}) does not explicitly involve the work function of the substrate, this 
affects the binding energy $E_{B}$ of an adsorbate. This is shown in
Fig.~3.10, where the principal energy diagrams for the photoemission 
characterization of a particular atom, yet adsorbed on two electronically 
distinct substrates, are contrasted: the substrates are described by their 
work functions, $\phi_{subs,1}$ and $\phi_{subs,2}$, the adsorbate is detected via 
its core level with the binding energy $E_{B}$. Since the ionization energy 
of the adsorbate $E_{I }$---being a sum of $E_{B}$ and substrate work 
function $\phi_{subs}$---is constant on both substrates, the analyzer detects 
$\Delta E_{B}=E_{B,2} - E_{B,1}$ equal to the substrate work function 
difference, $\Delta\phi_{subs} = \phi_{subs,1} - \phi_{subs,2}$ [118]. 

The high-$E_{B}$ shift in thicker films (which is regarded as the band 
bending) and the work function drop are yoked effects and they virtually 
occur upon the first monolayer formation too: the molecule photoemission 
features shift towards high $E_{B}$ simultaneously with the work function 
drop with the film thickness at submonolayer coverages. Consequently, the 
shift of photoemission spectra, either upon the growth of first monolayer or 
with further increase of the film thickness, has a common nature, namely the 
work function change of the probed surface. 

Ref. [119] provides plausible examples demonstrating the influence of the 
substrate work function on the photoemission spectra; photoemission spectra 
of ZnTPP grown on various metal (Mg, Ag, Au, Al) substrates shifted upon the 
oxygen exposure by amount equal to the expected work function changes due to 
the oxidation of the corresponding substrate. Likewise, the photoemission 
spectra of bithiophene were aligned according to local substrate work 
function on nanoscopically patterned surfaces [120]. 

As for the origin of the work-function change inducing the shift of the 
photoemission spectra, this is commonly caused by the L$ \to $U 
orientational transition occurring in thick molecular films (Fig.~3.5), 
though the oxidation can be at play too [115]. Upon the growth, the adsorbed 
molecules successively build particular monolayers, while the $n$th monolayer 
represents a substrate for the ($n+1$)th monolayer. Due to molecular 
orientation-resolved work function, the gradual L$ \to $U orientational 
transition with the film thickness implies the concomitant and incremental 
work function change. Accordingly, the photoemission spectra of ($n+1$)th 
molecular layer will be rigidly shifted relative to the spectra of the $n$th 
layer by their work function difference. The summed shift corresponds to the 
total work function change due to the L$ \to $U orientational transition 
$\Delta\phi_{L \to U}$ [Eq. (3.4)], which is identical to $\Delta _{BB}$ 
(see Fig.~3.9). Consequently, the shift of photoemission features with the 
film thickness considered to imply the band bending is, in fact, due to 
$\Delta\phi_{L \to U}$ induced by the gradual L$ \to $U orientational 
transition. In other words, the high-$E_{B}$ (low-$E_{B})$ shift accompanied 
by the same work function drop (increase) does not implicate any band 
bending. The spectra shift $\Delta E_{B}$ would indicate the factual band 
bending provided that the shift exceeds the work function change; in such a 
case, the difference would indicate the band bending:
\begin{equation}
\label{eq2}
\Delta E_B - \Delta \phi _{L \to U} = \Delta _{BB} .
\end{equation}

The spurious band bending can be recognized also at the organic-organic 
interfaces. This can be illustrated on CuPc/CuPcF$_{16}$ and 
F$_{16}$CuPc/CuPc heterostructures [87], so prepared by both deposition 
sequences; the F$_{16}$CuPc-related features shifted towards low $E_{B}$ with 
the F$_{16}$CuPc growth onto CuPc, while the growth of CuPc over 
CuPcF$_{16}$ resulted in the high-$E_{B}$ shift of the CuPc spectra. The low- 
and high-$E_{B}$ shifts were interpreted as the downward and upward band 
bending, respectively and the depletion/accumulation region of about 15 nm 
gave to an estimate of the free charge carrier concentration of about 
$10^{18}$ cm$^{ - 3}$. With reference to the discussion above, we presume 
that the would-be band bending in Ref. [87] is an artefact induced by the 
varying work function: The intrinsic work function of CuPc and F$_{16}$CuPc 
for lying (upright) films are 4.4 (3.95) and 4.7 (5.3) eV, respectively 
(Tab.~3.1). Therefore, the work function upon the growth of CuPc onto 
F$_{16}$CuPc converges to $\phi_{CuPc}$ (accordingly, it drops) and thereby 
implicates the high-$E_{B}$ shift. The opposite situation occurs for the 
reversed growth sequence, namely the work function increase towards the 
intrinsic value of F$_{16}$CuPc and thereby the low-$E_{B}$ shift. Depending 
on the molecular orientation of CuPc and F$_{16}$CuPc, and the growth 
sequence, the total work-function change ranges between 0.3 to 1.35 eV. The 
qualitatively same situation was observed for F$_{16}$CuPc/BP2T and 
BP2T/F$_{16}$CuPc organic heterostructures [83].

\subsection{$\boldsymbol{\pi}$ band}

Unlike the isolated and thereby well-identifiable core levels, the shallow 
levels determining the electronic structure of an organic molecule are 
markedly affected by the band dispersion and intramolecular interactions. 
Many vaguely identified, narrowly distributed and to various extents 
overlapped molecular orbitals populate the sub-Fermi region to eventually 
form the (valence) band (recall Fig.~1.2), also referred to the density of 
states (DOS). Apart from the band alignment determination, $i.e.$ the examinations 
of the HOMO positions with respect to the Fermi level, the rest of the 
valence band has been rather scarcely examined experimentally [121-123] 
thereby providing a weak feedback to \textit{ab initio} calculations of the molecular 
electronic structure [124, 125]. 

\begin{figure}[tb]
\begin{center}
\includegraphics[width=5.5cm,clip]{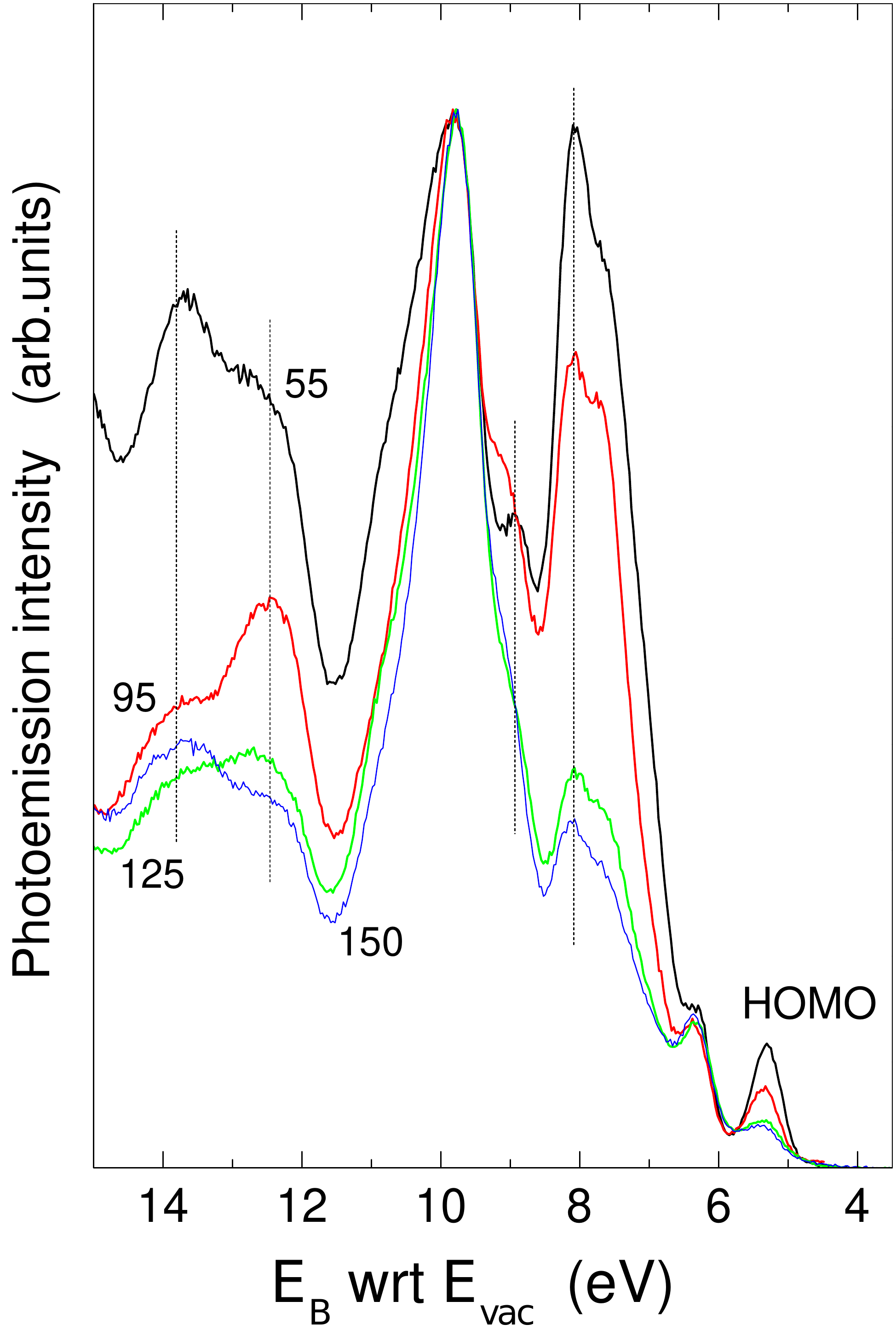}
\caption{The valence band spectra of the CuPc film examined at photon energies
of 55, 95, 125, and 150 eV and normalized with respect to the dominant peak.
Reprinted with permission from T. Toader {\it et al.}, Phys. Stat. Sol. (b)
{\bf 246} (2009) 1510. Copyright 2009, {\it John Wiley \& Sons, Inc.}}
\vspace*{-0.5cm}
\end{center}
\end{figure}

Employment of the tunable photon energy in the photoemission 
characterization allows in some extent to specify the character of 
particular molecular orbitals, accordingly their assignment to constituting 
atoms of a molecule. The approach is viable for assemblies comprising 
elements with distinct photoionization cross section (PCS) dependences on 
photon energy; the marked differences can be encountered between $e.g.$ carbon and 
metallic elements [16]. Thus, by monitoring intensities of particular 
molecular orbitals in dependence of the photon energy, the orbitals can be 
discriminated in terms of their character. 

Figure~3.11 shows normalized VB spectra of CuPc (C$_{32}$H$_{16}$CuN$_{8})$ 
probed with various photon energies. By knowing the ratio between the 
photoionization cross sections of Cu $3d$, C $2p$, and N $2p$ energy levels and relative 
intensities of individual orbitals in dependence of the photon energy, the 
character of particular orbitals can be deduced. Figure~3.12 shows 
individual molecular orbitals obtained by the deconvolution with the 
indication of their character spanning from the C $2p$- and N $2p$-derived (which 
are not distinguishable by this method owing to their very similar PCS 
photon energy dependences) to the Cu $3d$-derived orbitals. The notable result 
would be the experimental confirmation of the pure C $2p$-character of the HOMO, 
of which dissonant characters were suggested by theoretical investigations 
[124, 125]. 

\begin{figure}[tb]
\begin{center}
\includegraphics[width=9.0cm,clip]{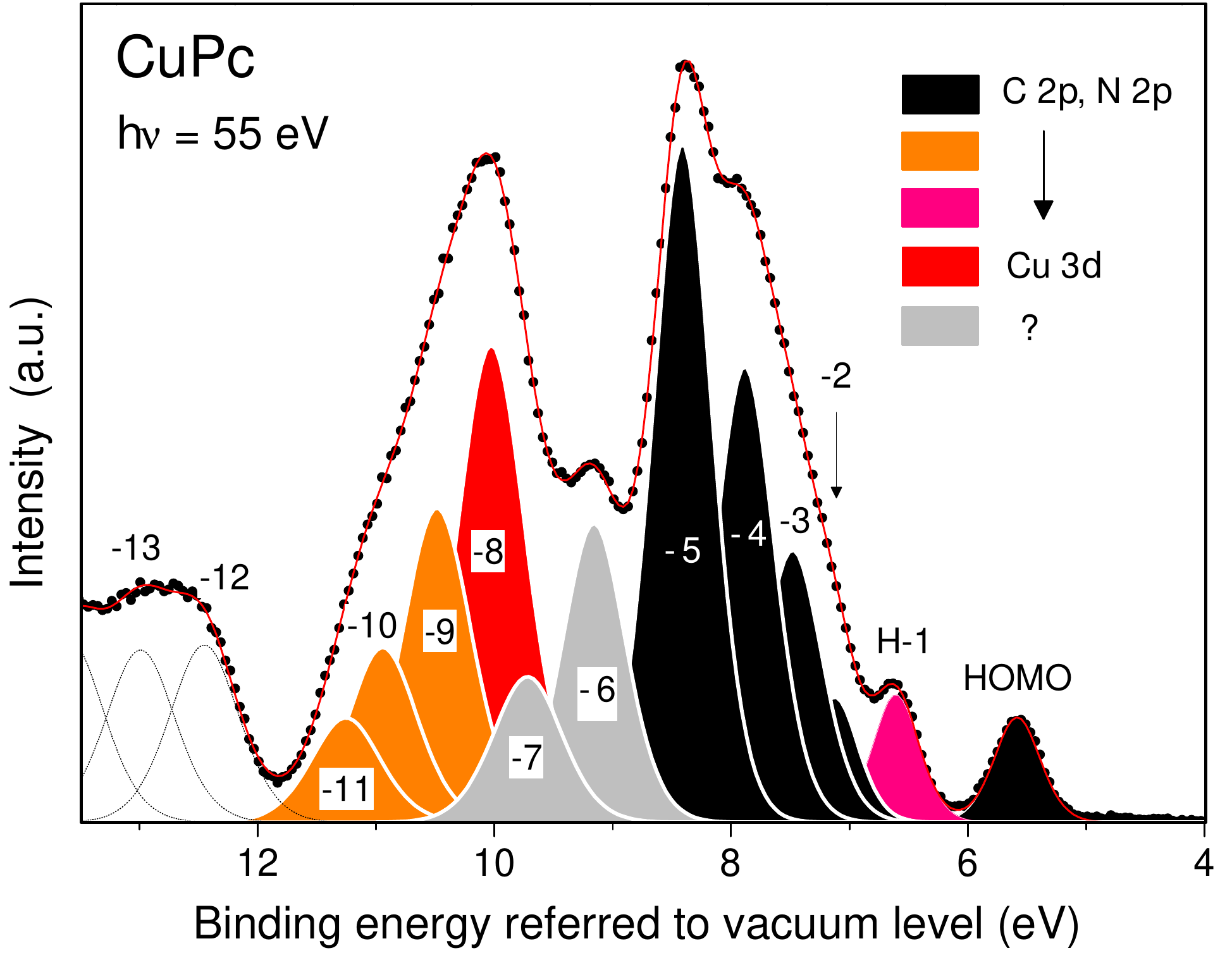}
\caption{The deconvoluted valence band of CuPc: The characters of particular
molecular orbitals ranging from entirely Cu $3d$-derived to entirely C $2p$ or
N $2p$-derived energy levels are indicated. The particular orbitals are numbered
with respect to the HOMO. Reprinted with permission from T. Toader {\it et al.},
Phys. Stat. Sol. (b) {\bf 246} (2009) 1510. Copyright 2009, {\it John Wiley \&
Sons, Inc.}}
\vspace*{-0.5cm}
\end{center}
\end{figure}

\setcounter{equation}{0} \setcounter{figure}{0} \setcounter{table}{0}\newpage
\section{Interfaces associated with molecular films}

The interfaces are---besides the active films---the functional parts of 
electronic devices. The interface engineering aims to control interfacial 
electronic properties (energy level alignment determining the 
injection/extraction barriers), chemistry, and morphology. In terms of 
electronic properties, organic electronics has, in fact, introduced a novel 
class of interfaces, namely these formed between molecular films and metals 
or inorganic semiconductors. In organic devices, matched pairs of materials 
are necessary to conduct away charges across junctions on metal/organic 
interfaces or to separate charges at organic acceptor/donor interfaces. As 
for the latter class of interfaces, the interest had been nourished due to 
organic heterostructures, which have become an integral part of organic 
devices based on a double layer or a bulk heterostructure. In spite of 
conclusive success of organic devices, the ability to predict and control 
the electronic properties of interfaces has been limited. 

Conductive contacts in organic devices have been regularly realized by means 
of metals. Among the most relevant ones were the coin metals such as $e.g.$ Ag, 
Al, Au, Cu, Ni, Pt, Sn, Zn. The metallic contact to the molecular film can 
be accomplished either by the growth of a molecular film on a metallic 
substrate/film or---reversely---by the evaporation of a metal onto a molecular 
film. In practise, the former interface has been studied via the growth of 
an organic layer prevailingly on both chemically and structurally 
well-controlled inorganic substrates, typically of single-crystal surfaces, 
whereas the latter interface (the reverse one) through the evaporation of a 
metal onto an organic film. 

The evaporated molecules have virtually no morphological effect on the 
metallic substrate and such interfaces are sharp with the topology 
determined by the substrate surface, and they can even manifest an epitaxial 
relation between the film and the substrate. Therefore, they allow more 
straightforward analysis and interpretation of interface properties, and 
particularly electronic interfacial properties have been conveniently 
studied via organic-on-metal contacts simply because of higher 
reproducibility, characterization, and overall control of the interface 
compared to the reversed interfaces. 

The metal evaporation onto organic surface may be more invasive to the 
organic substrate. The higher heat of adsorption of metal atoms can increase 
their reactivity and result in chemical reaction not present on the 
reversely prepared system. The metal atoms were reported to diffuse into the 
organic film forming a blurred interface. Moreover, the morphology of 
organic films acted as substrates for the metal evaporation is inferior to 
the surface morphology of metallic substrates. 

Consequently, the couple of materials forming the contact is an ordered pair 
since the succession of preparation steps markedly affects the interfacial 
structure and properties, and both types of interfaces have to be treated 
separately. Note that the Schottky barrier on traditional semiconductors had 
been studied via metal evaporation on semiconductor surface, organic 
semiconductors allow to accomplish both sequences, namely organic-on-metal 
and metal-on-organic. 

\subsection{Energy level alignment}

\begin{figure}[tb]
\begin{center}
\includegraphics[width=6.0cm,clip]{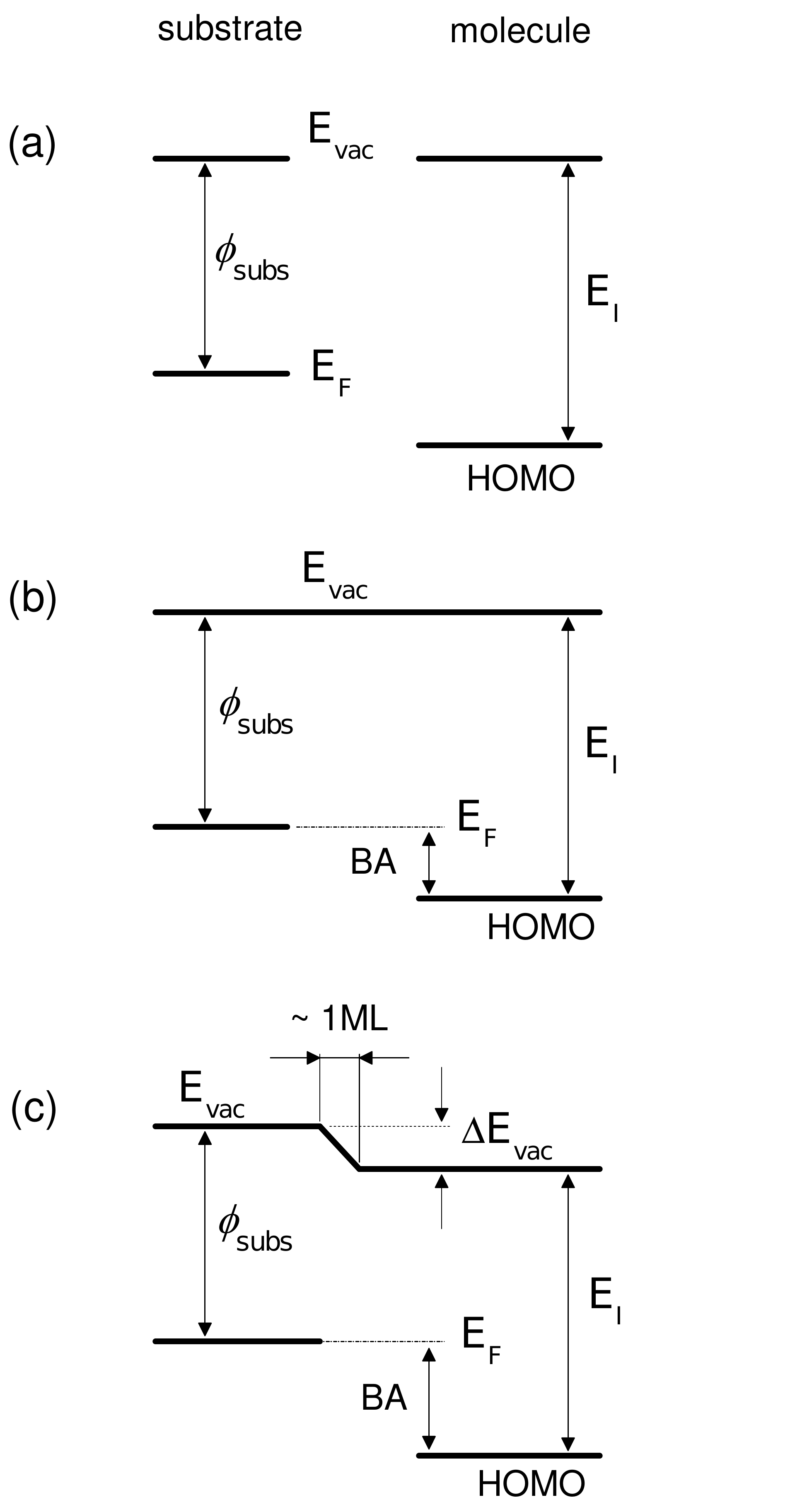}
\caption{Energy band diagram of a metal-organic pair prior the contact (a),
upon the contact with aligned vacuum levels (b), and with vacuum-level
offset (c).}
\vspace*{-0.5cm}
\end{center}
\end{figure}

If two electronically dissimilar materials make a contact, their distinct 
energy level positions result in an electronic discontinuity seen by charge 
carriers as an interfacial electronic barrier. Such interfacial barriers for 
electrons and holes, $\varphi_{e}$ and $\varphi_{h}$, respectively, established between 
the molecular film and metals are shown in Fig.~1.1. The fundamental 
question is the mechanism governing the alignment of energy levels at the 
interfaces between the materials forming the contact, and accordingly, the 
assignment of the initial electronic properties of isolated materials, 
$i.e.$ prior the contact, to the eventual energy level alignment (ELA). The answer 
would allow the design, control, and prediction of any kind of junction, 
metal/organic, organic/organic, and semiconductor/organic. As for a 
metal-organic pair, Figure~4.1a shows energy band diagram of a bare metal 
and a molecule characterized by $\phi$ and $E_{I}$, respectively, prior the contact 
formation. Two fundamental models on the ELA distinguished by (mis)alignment 
of the vacuum levels have been proposed: ($i$) the vacuum-level alignment model 
(Fig.~4.1b) and the interface dipole model (Fig.~4.1c). 

The vacuum-level alignment (VLA), which applies for weakly interacting 
(physisorbed) organic/inorganic systems, had also been referred to the 
Schottky-Mott rule (or Schottky-Mott limit), inasmuch as the assumption of 
the VLA was initially employed by Schottky to assess the barrier formed at 
metal-semiconductor contacts [126, 127]. Provided that the energy levels 
align according to the common vacuum level (Fig.~4.1b), the BA is
\begin{equation}
\label{eqo3}
BA = E_I - \phi _{subs} \,.
\end{equation}

Equation (\ref{eqo3}) implies that the BA is a linear function of the $\phi_{subs}$ 
with the slope parameter $S_{BA}=\Delta$BA/$\Delta\phi_{subs}=-1$ 
for each molecular film characterized by its $E_{I}$. The BA insensitive to 
$\phi_{subs}$, $i.e.$, $S = 0$, is referred to Bardeen limit. The Schottky-Mott rule had 
been shown generally unworkable and only part of interfaces (mostly organic 
heterointerfaces) were reported to obey the VLA.

Disobedience to the VLA model was suggested to be due the vacuum-level 
offset by amount of $\Delta E_{vac}$ $[$Eq. (3.2)$]$, corresponding to a dipole 
formed at the organic/metal interface, the interfacial dipole (ID) [119] 
(Fig.~4.1c). The ID controls the barrier across the interface and it is 
saturated typically upon the completion of the first monolayer. Considering 
Fig.~4.1c, the relation between the relevant parameters can be written as 
follows: 
\begin{equation}
\label{eq4}
BA = E_I + ID - \phi _{subs} 
\end{equation}

Since the initial parameters $E_{I}$ and $\phi_{subs}$ are known, the knowledge 
on the mechanism of the ID formation would be essential for the 
determination of the BA. The dipole model had thereafter attracted a lot of 
attention and since it has been the most frequently used framework for the 
description of electronic interfacial properties [27, 93, 128-136]. Various 
origins of the interfacial charge redistribution were suggested, such as 
\begin{itemize}
\item[(\textit{i})] The charge redistribution across the interface due to the interaction 
between a molecule with the substrate surface and eventually chemical bonds 
formed at the reactive interface [74, 98, 119, 128, 129, 137]; 
\item[(\textit{ii})] Push-back (or pillow) effect [129] meaning that $\phi_{subs}$ is modified by 
pushing-back of the tail of the electronic cloud of metal surface by 
repulsion with the electron cloud of an adsorbate; 
\item[(\textit{iii})] The interfacial dipole is of quantum origin due to the exchangelike 
mechanism, not sensitive to electrostatic repulsion or interfacial chemistry 
[138, 139];
\item[(\textit{iv})] Weak chemical reaction between metal and a molecule results in interface 
states in the organics' gap describable by charge neutrality level, which 
aligns to the metal Fermi level [130, 140, 141]; 
\item[(\textit{v})] The permanent dipole moment of the adsorbed molecule, the dipole 
component along the interface normal will induce a potential drop across the 
interface [131, 136];
\item[(\textit{vi})] Charge transfer due to the alignment of the substrate Fermi level and the 
positive polaronic level of the polymer [142, 143]; 
\item[(\textit{vii})] The equalization of electron chemical potentials of components, $i.e.$ the 
metal work function and $E_{I}$ or $E_{EA}$ of the molecular film [144, 145].
\end{itemize}

There have been basically two approaches in an experimental search for 
pivotal parameters of contacting materials determining the ELA: namely, the 
growth of a particular molecular film on a variety of substrates and the 
growth of variety molecular films on a particular substrate. In particular, 
searches for correlations between the metal work function and the $E_{I}$ 
(HOMO) or $E_{EA}$ (LUMO) have been frequently pursued. For example, no ID$ - 
E_{I}$ correlation was found for various molecular films grown on gold 
foil, yet an apparent correlation was observed between the ID and the 
product $E_{I}-\raise.5ex\hbox{$\scriptstyle 1$}\kern-.1em/ 
\kern-.15em\lower.25ex\hbox{$\scriptstyle 2$} E_{t}$ [116], where 
$E_{t}$ states for the transport gap. In fact, the observed correlation 
suggested that the ID is determined--as far as the adsorbed molecule 
contribution is concerned--by the LUMO position of the molecular film 
referenced to the vacuum level. This was explicitly claimed in Refs. 
[20, 76], where the linear correlation between the ID and $E_{EA}$ for four 
various phthalocyanines, namely H$_{2}$Pc, CuPc, F$_{4}$CuPc, and 
F$_{16}$CuPc, and PTCDA, has been observed. Yan \textit{et al.} [146] reported the linear 
relations between the ID and the product [$\phi_{subs} - (E_{I } - 
E_{G}$/2)]. Such correlation would indicate that the ID is correlated with 
the \textit{difference} between $\phi_{subs}$ and the LUMO of the organic film,
$i.e.$ with the barrier to electron injection from the metal to the organic.

Several studies reported the linear dependences of the ID on the substrate 
work function [119, 129, 147-152]. However, the claimed linear correlations 
for specific molecule were typically based on up to three measured values 
only, with various slope parameters, $S_{ID}=\Delta E_{vac}/\Delta 
$\textit{$\phi $}, ranging from 0 to 1 depending on the molecular film. Furthermore, some 
molecules were reported to reveal two distinct linear dependences; for 
example, PEN showed the slope parameter either zero or of +1, if grown on 
substrates with their work function smaller or higher than $E_{I}$ of PEN, 
respectively [153], while quite reverse behaviour has been observed for 
BCP, $i.e.$, the slope parameters of +1 and zero for small and high substrate work 
function, respectively [150]. In contrast, no correlation between the ID and 
the substrate work function has been detected when growing 6T on eight 
various substrates with work function ranging from about 2 to 5 eV [134]. 

In terms of the permanent dipole model [131], substrate work function should 
not show a change upon the growth of a molecule with no permanent dipole, 
such as $e.g.$, CuPc. Yet, different $\Delta E_{vac}$'$s$ were reported for CuPc 
grown on various substrates with the work function ranging over 1~eV
(Tab.~3.2a); the examples rather suggest that $\Delta E_{vac}$ corresponds to the 
difference between the substrate work function and the CuPc intrinsic work 
function of ca 4.4 eV (corresponding to upright-oriented molecules).

\subsubsection{Equalization of electrochemical potentials}

The work function of a molecular film $\phi_{film}$ converges to $\phi_{int}$ with 
the film thickness (Section 3.2). Accordingly, the total vacuum level change 
upon the film growth equals to:
\begin{equation}
\label{eqt5}
\Delta E_{vac} = \phi _{subs} - \phi _{int} .
\end{equation}

It is reasonable to presume that the same mechanism for the ELA is valid 
universally for all film thicknesses (including the monolayer) and Eq. (\ref{eqt5}) 
then becomes: 
\begin{equation}
\label{eq6}
\Delta E_{vac} (t) = \phi _{subs} - \phi _{film} (t),
\end{equation}
where $\phi_{film }(t)$ is the function of the thickness $t$ and $\phi_{film}$ converges 
to \textit{$\phi $}$_{int}$ for thickness higher than the critical thickness. 

We want to stress that even though Eq. (\ref{eq6}) is formally identical to Eq. 
(3.2) employed for the $\Delta E_{vac}$ determination, Eq. (\ref{eq6}) expresses 
the driving force for the vacuum level change: it implies that the Fermi 
levels of contacting materials defined via their respective (intrinsic) work 
function, $i.e.$ electrochemical potential of electrons (\textit{$\mu $}), equalize, whilst the 
difference between their work functions is compensated by the vacuum level 
offset. Note that the absolute value of the molecular film's work function 
matters instead of the work function change. 

Knowing the rule for the $\Delta E_{vac}$ determination [Eq. (\ref{eq6})], the BA 
of a {\it thick} film determined by photoemission is:
\begin{equation}
\label{eq7}
BA = E_I - \phi _{int} 
\end{equation}

\begin{figure}[tb]
\begin{center}
\includegraphics[width=5.7cm,clip]{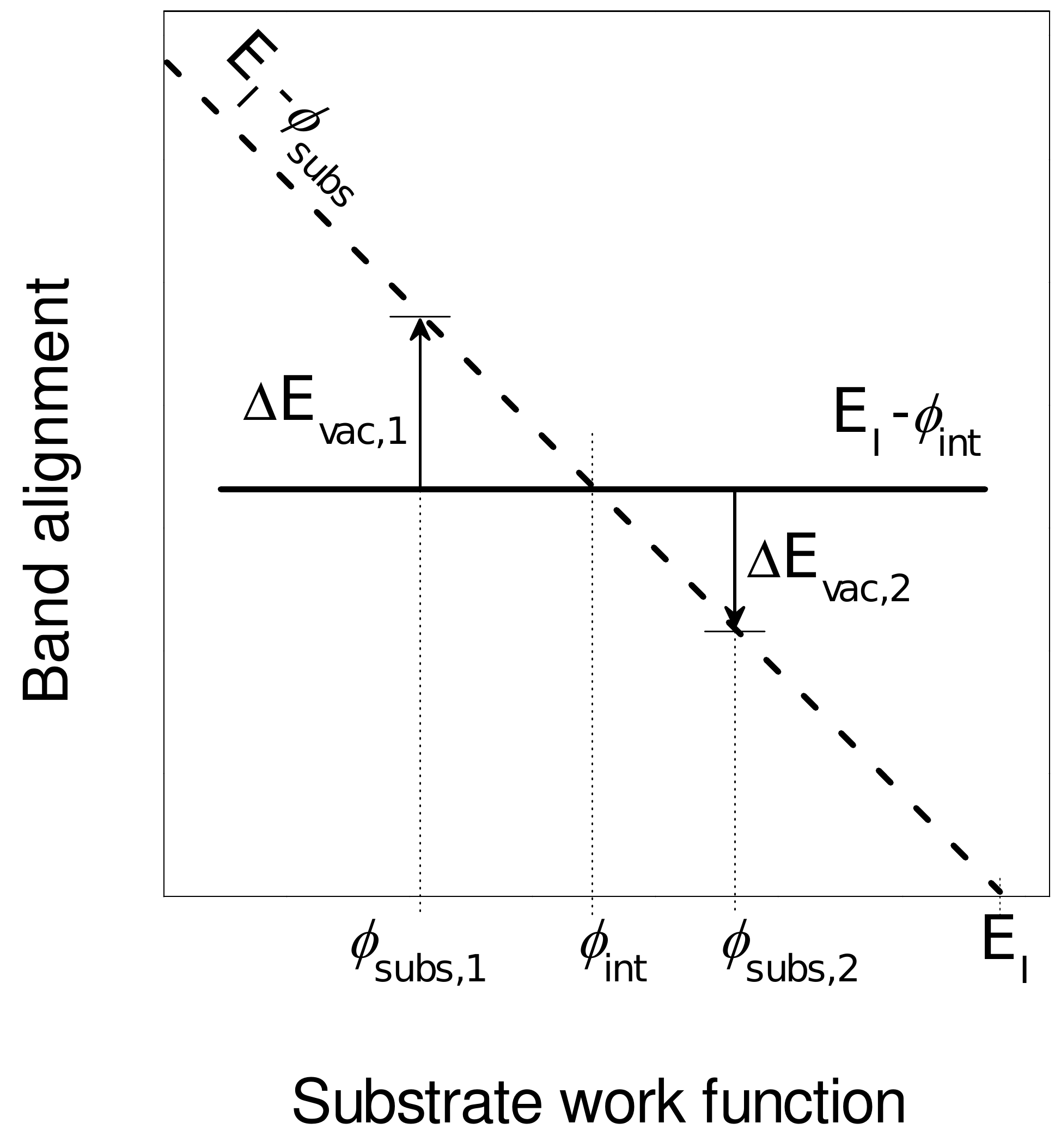}
\includegraphics[width=7.3cm,clip]{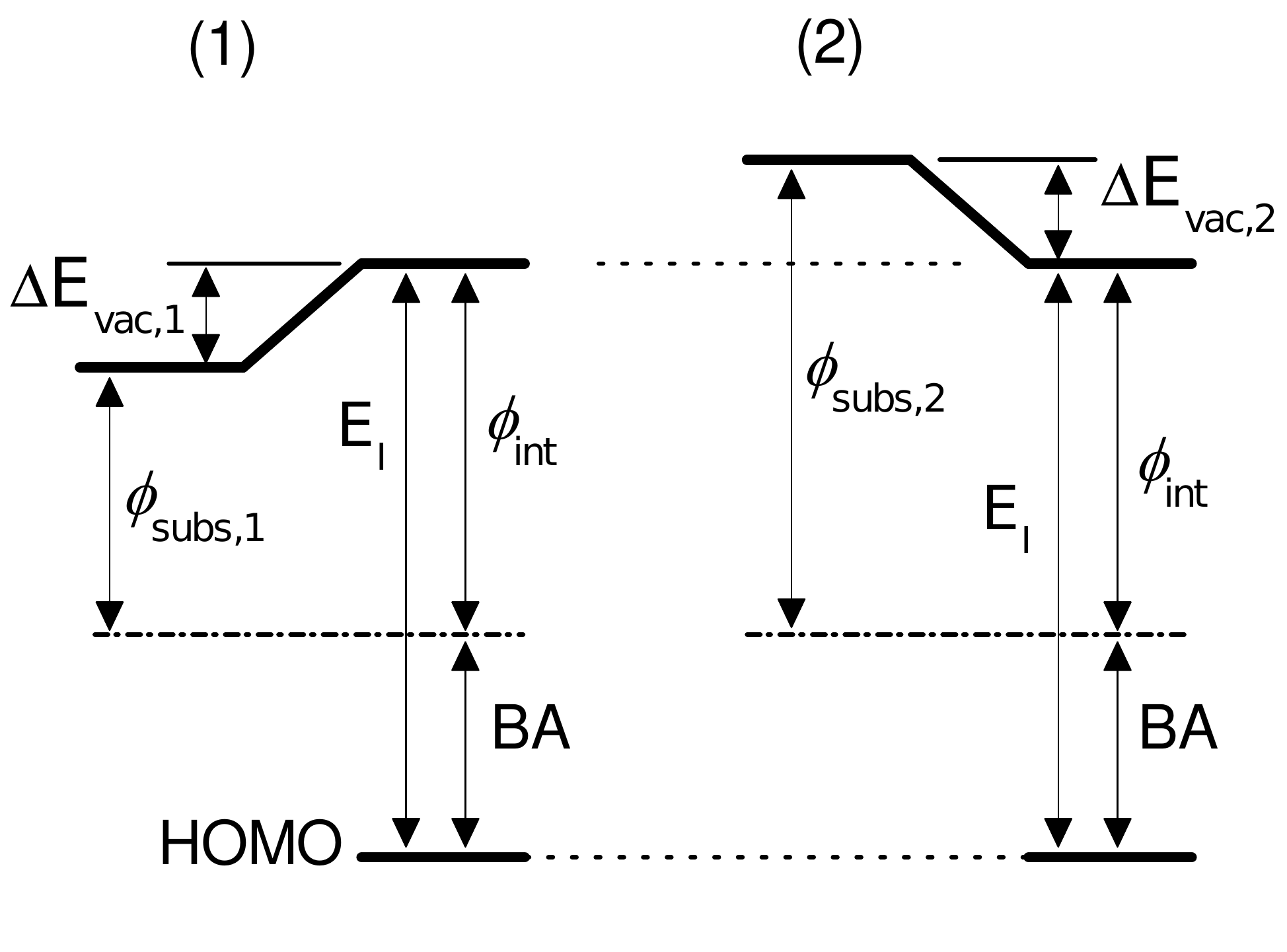}
\caption{The graphical representation of the BA versus φsubs in the framework of the
model proposing the electrochemical potentials equalization. The solid horizontal
line expresses ${\rm BA} = E_I-\phi_{int}$, i.e. the BA insensitive to the substrate
work function, the dashed line indicates the Schottky-Mott rule. They intersect at
$\phi_{subs}=\phi_{int}$. The arrows refer to particular vacuum level changes,
$\Delta E_{vac,1}$ and $\Delta E_{vac,2}$, upon the growth onto two substrates with
different work functions, $\phi_{subs,1}$ and $\phi_{subs,2}$. Their corresponding
band diagrams of the interfaces are sketched at right. Reprinted with permission
from J. Ivanco, Thin Solid Films {\bf 520} (2012) 3975. Copyright 2012, {\it Elsevier}.}
\vspace*{-0.5cm}
\end{center}
\end{figure}

Figure~4.2 presents a graphical representation of Eq.~(\ref{eq7}) for a particular 
molecular film characterized by the $E_{I}$ and grown on various substrates: 
the BA is independent of the substrate work function as shown by the 
straight solid line. The arrows (both their length and direction) indicate 
total vacuum-level changes upon the growth of a thick film onto two distinct 
substrates with $\phi_{subs,1}$ and $\phi_{subs,2}$. The vacuum-level changes 
$\Delta E_{vac,i}$ indicate actually the departure of the $\mu$-equalization 
model from the VLA model, the latter being drawn by a dashed line with the 
slope of -1. The $\mu$-equalization model gives the same result as the VLA 
model at the intersection of both dependences: the growth of a molecular 
film with $\phi_{int}$ onto a substrate with $\phi_{subs}=\phi_{int}$ results 
in zero $\Delta E_{vac}$ upon the contact formation, thereby suggesting the 
VLA. 

The band alignments both independent ($S=d$BA$/d\phi_{subs}=0$), $i.e.$ governed by Eq. 
(\ref{eq7}) and dependent ($-0.8 < S < 0$) on $\phi_{subs}$ were reported [56, 129]. In 
the latter case, we presume--forasmuch the reported experimental data often 
lack details such as the film morphology--that the disobedience to Eq. (\ref{eq7}) 
may be due to the insufficient film thickness (either nominally or due to 
the heavy islanding). In such situation, the substrate work function may 
still affect the measured work function of the film. 

Upon the film growth, both signs of the vacuum level change can occur. The 
downward change occurs provided that molecular films with low $\phi_{int}$, such 
as 6P, 6T, phthalocyanines, or $\alpha$-NPD are grown on substrates with 
high $\phi$, either inorganic ($e.g.$, Au, Ni, Pt) or organic ($e.g.$, F$_{16}$CuPc, 
F$_{4}$TCNQ, PTCDA) ones. In contrast, the upward shift of the vacuum level 
can be observed provided that molecular films with high $\phi_{int}$, such as 
F$_{16}$CuPc, F$_{4}$TCNQ, PTCDA, or HAT-CN, are grown on substrates with 
low and moderate work function, either inorganic ($e.g.$ Ag, Al, Mg) or organic 
($e.g.$ CuPc, Alq$_{3})$ ones. Notable, the downward shift of $E_{vac}$ can change 
upwardly upon the doping of the low-$\phi_{int}$ film by the high-$\phi_{int}$ 
molecular film, such as the doping of $\alpha $-NPD grown on Au by 
F$_{4}$TCNQ [56], or by choosing the low-$\phi_{int}\to$ high-$\phi_{int}$ 
growth sequence in the organic heterostructure [90]. The vastly frequent 
reports on downward changes of the vacuum level upon the film growth are 
obviously owing to the higher popularity of high-$\phi$ substrates employed for 
the surface studies in combination with frequent molecular films having the 
moderate of low $\phi_{int}$. The infrequent reports on upwards work-function 
changes had been due to relatively less investigated high-$\phi_{int}$ molecular 
films, $e.g.$ perfluorinated (fluorine substituted) ones, PTCDA, NTCDA, or HAT-CN, 
grown on substrates with low(er) work function.

Among the listed molecules (Tabs.~3.1 and 3.2), the F$_{4}$TCNQ and HAT-CN have the 
highest $\phi_{int}$ of $\approx 6.25$ eV [154] and $\approx 6$ eV [89]. The 
$\phi_{int}$ magnitudes rank to the highest work functions among the both 
elements and organics, or Mo-, Cr-, V-, and W-oxides ($\phi \approx 7$ eV, 
[145]). The very high $\phi_{int}$ of F$_{4}$TCNQ and HAT-CN suggests that the 
film would increase the work function of typical high-$\phi$ elemental surfaces 
too, such as Au; indeed, the work function of the Au substrate being beyond 
about 5 eV increased by 0.29 eV upon the growth of even ultrathin film of 
F$_{4}$TCNQ [155], The trend was confirmed in [156]. Similarly, HAT-CN grown 
on Au resulted in $\phi \approx 6.1$ eV [90]. Notable, such molecular 
film-induced metal work function \textit{increase} would disprove the push-back model [129], 
which accounts a lowering of the metal work function owing to the 
compression of the electron tail of metal substrate electrons by the 
electrons localized in the molecules. 

At organic-organic interfaces, no changes of the vacuum level have been 
presumed due to the charge confinement over a molecule. This indeed applies 
for heterointerfaces such as ZnPc/CBP, ZnPc/BCP, $\alpha$-NPD/BCP,
$\alpha$-NPD/CBP [157], and PEN/CuPc [158], 6T/6P [159], which showed no or 
negligible $\Delta E_{vac}$. Nevertheless, the vacuum-level changes $\Delta 
E_{vac} \approx 0.4$-$0.5$ eV were observed at the heterostructures such 
as PTCDA/CuPc, PTCDA/Alq$_{3}$ [160], PTCBI/BCP [160], C$_{60}$/CuPc [161], 
NTCDA/Alq$_{3}$, NTCDA/CuPc, NTCDA/BCP [157], and CuPc/CuPcF$_{16}$ [65, 87]. 
Both trends can be explained in the framework of the electrochemical 
potential-equalization model: the heterointerfaces in the former group are 
formed by molecular films with the same or similar $\phi_{int}$, while the 
latter set of examples represents interfaces between molecules with distinct 
$\phi_{int}$. Likewise, $\Delta E_{vac} \approx 0$ would arise provided 
that $\phi_{int}$ of organic films equals to $\phi$ of an inorganic substrate. Indeed, 
for example, the PEN on the highly oriented pyrolitic graphite (HOPG) shows 
the preserved vacuum level; this is apparently due to similar values of 
$\phi_{PEN}$ and $\phi_{HOPG}$ being of $\sim 4.4$ eV [158]. 

PFP grown on Ag shown non-monotonous evolution of the $\Delta E_{vac}$ with 
the film thickness [58]: the vacuum level abruptly dropped by 0.42 eV after 
the first monolayer and remained constant with further growth. Yet, 
$E_{vac}$ increased by 0.12 eV at the thickness of 5 nm. The F$_{16}$CuPc 
grown on Au showed a similar non-monotonous evolution [116]. Such curious 
behaviours of the work function can occur in the perfluorinated molecular 
film ($i.e.$ distinguished by the inequality $\phi_{int,L} < \phi_{int,U}$) provided 
that the L$ \to $U orientational transition takes place, and the inequality 
$\phi_{int,L} < \phi_{subs} < \phi_{int,U}$ holds. The reversed work function 
evolution, $i.e.$ the work function increase followed by the work function decrease 
upon the film growth, may occur for an unsubstituted molecular film provided 
that the inequality $\phi_{int,L} > \phi_{subs} > \phi_{int,U}$ holds. Indeed, such 
behaviour of the work function was reported for the growth of titanyl 
phthalocyanine film on HOPG [162]. 

The charge neutrality level (CNL) model [130, 132] by Vazquez \textit{et al.} introduces the 
CNL, the parameter being an intrinsic property of a molecule. The ELA at the 
molecule/metal interface is determined by the alignment of CNL with the 
Fermi level of a metal substrate. The calculated CNL$s$ for several molecules 
are given in Tab.~4.1. The values are similar to $\phi_{int}$`$s$ (refer to the 
next column), which suggests the common origin of both quantities. 
Admittedly, the comparison is rather preliminary as ($i$) the calculations were 
based on the one molecule positioned near the substrate surface suggesting 
that $\phi_{int}$ was not yet attained, and ($ii$) the statistically meagre 
ensemble of experimental data on $\phi_{int}$'$s$. 

\begin{table}[!bp]
\begin{center}
\vspace*{-0.5cm}
\caption{The charge-neutrality levels of several molecules [132, 133] and their 
intrinsic work functions, $\phi_{int}$, adopted from Tab.~3.2. Both quantities 
are referred to the vacuum level.}
\medskip
\begin{tabular}
{|c|c|c|c|c|}
\hline
& 
& 
& 
\multicolumn{2}{|c|}{$\phi_{int}$ [this work]}  \\
{Molecule}& 
\raisebox{1.50ex}[0cm][0cm]{CNL [132]}& 
\raisebox{1.50ex}[0cm][0cm]{CNL [133]}& 
\multicolumn{2}{|c|}{(eV)}  \\
\cline{4-5} 
 & 
\raisebox{1.50ex}[0cm][0cm]{(eV)} & 
\raisebox{1.50ex}[0cm][0cm]{(eV)} & 
lying& 
upright \\
\hline
$\alpha - $NPD& 
4.2& 
-& 
-& 
3.8-4.2 \\
\hline
Alq$_{3}$& 
3.8& 
-& 
-& 
$\sim 4.0$ \\
\hline
BCP& 
3.8& 
-& 
-& 
$\sim 4.15$ \\
\hline
CBP& 
4.2& 
4.05& 
-& 
$\sim 4.5$ \\
\hline
CuPc& 
4.0& 
3.8& 
-& 
$\sim 3.95$ \\
\hline
PTCBI& 
4.4& 
4.4& 
-& 
4.5-4.6 \\
\hline
PTCDA& 
4.8& 
4.8& 
4.5-4.94& -
 \\
\hline
\end{tabular}
\end{center}
\end{table}

The presented data and proposed ELA mechanism via the equalization of the 
Fermi levels (see also Ref. [163]) supports the previously reported concepts 
suggesting that the driving force for the ELA is an equalization of electron 
chemical potential of metal (corresponding to the metal work function) to 
that of molecular film [144, 145]; yet, both approaches related the 
electrochemical potential of the molecular film with its $E_{I}$ and 
$E_{EA}$ (referred to vacuum level). In this work, the electrochemical 
potential of a molecular film is---according to the definition and thereby 
preferable---identified with the intrinsic work function, which is 
experimentally accessible. The $\phi_{int }$-$E_{I}$ correlation claimed above 
(Fig.~3.7) would suggest that both parameters are equivalent as far as they 
change in parallel. Yet, this does not hold for doped semiconductor, where 
the doping affects the Fermi level and thereby the electrochemical 
potential, but has no effect on $E_{I}$ and $E_{EA}$.

\subsection{On equivalency between the interfacial dipole and the vacuum-level offset}

The interfacial dipole (ID) formed at the organic/metal interface affects 
the transport barrier for charge carriers injected across the junction. The 
current concept considers the equivalency between the ID and the measured 
vacuum-level offset, $\Delta E_{vac}$, [recall Eq. (3.3)] detected upon a 
film growth [119,164]. The concept can be described as follows: Referring 
back to the definition of the work function, Eq. (3.1) can be formally 
rewritten and the work function of a bare metal substrate $\phi_{subs}$
\begin{equation}
\label{eq8}
\phi _{subs} = \mu _{subs,bulk} + \phi _{subs,SD} 
\end{equation}
is modified upon the film growth to
\begin{equation}
\label{eq9}
\phi _{subs + film} = \mu _{subs,bulk} + \phi _{subs + film,SD} ,
\end{equation}
where the indices \textit{subs} and \textit{film} state for the substrate and the
film. It is supposed 
here that the surface-dipole potential of the bare substrate, $\phi_{subs,SD}$, 
is changed to $\phi_{subs + film,SD}$ upon the film growth. As a result, the 
substrate work function change $\Delta\phi$ amounts to 
\begin{equation}
\label{eq10}
\Delta \phi = \phi _{subs} - \phi _{subs + film} = \phi _{subs,SD} - \phi 
_{subs + film,SD} \equiv \phi _{ID} ,
\end{equation}
where $\phi_{ID}$ is the interface-dipole potential. Note that the ID is an 
additional dipole to that assigned to the bare substrate surface. We 
consider the concept inconsequent whereas it assumes a relation between 
$\phi_{ID}$ and the final vacuum level even for a thick film, which has to 
electronically screen out the substrate and the interface. In the following 
paragraph, an alternative view is proposed. 

The quantity $\Delta\phi$ is routinely yielded by the external photoemission, 
which means that the spectrometer detects a photoelectron escaping the 
probed solid. In comparison to the internal `transport electron' passing 
across the barrier created at the substrate/film interface, the 
photoelectron is further affected by the film, thus by its electrochemical 
potential and the surface dipole. It is thereby presumed here that---in 
variance with the notion expressed by Eq.~(\ref{eq9})---the work function of the 
substrate upon the film growth is instead described as follows:
\begin{eqnarray}
\nonumber
\phi _{subs + film}^\ast &=& (\mu _{subs,bulk} + \phi _{subs + film,SD} 
)f(1\buildrel {t \to \infty } \over \longrightarrow 0) \\
\label{eq11}
&+& (\mu _{film,bulk} + 
\phi _{film,SD} )\left[ {1 - f\left( {1\buildrel {t \to \infty } \over 
\longrightarrow 0} \right)} \right],
\end{eqnarray}
where $t$ states for the film thickness and $f(1\buildrel {t \to \infty } \over 
\longrightarrow 0)$ is an attenuation function with output converging from 
unity to zero with the film thickness ranging from near zero to high 
thickness. The first and second term on the right side of Eq. (\ref{eq11}) 
correspond to the substrate and the film, respectively, while $\phi_{subs + 
film,SD}$ states for the dipole potential of the substrate surface upon the 
film formation, thereby the interfacial dipole. Obviously, the work function 
change $\Delta\phi = \phi_{subs} - \phi_{subs + film}^\ast$ obtained by 
subtracting Eq. (\ref{eq8}) and Eq. (\ref{eq11}) is not equal to $\Delta \phi =
\phi_{subs} - \phi_{subs + film}$ [Eq. (\ref{eq10})] for the nonzero film term. 
Admittedly, the first term in Eq. (\ref{eq11}) can dominate for ultrathin films, 
$e.g.$ one-monolayer thick, yet, it has to converge to zero for thicker overlayers 
($t \to \infty )$, where both the substrate and the film/substrate 
interface are electronically shielded by the film. For a thick film, the 
second term in Eq. (\ref{eq11}), which represents the electrochemical potential of 
the bulk film and the film surface dipole, dominates and the measured work 
function is independent of both the substrate and the film/substrate 
interface. Consequently, the $\Delta\phi_{subs}$ (or $\Delta E_{vac})$ 
measured by photoemission upon the film growth converges to the difference 
between the intrinsic work functions of the substrate and film with the 
increasing film thickness.

\subsection{Metal-on-organic interface}

The metal growth on organic films may occur in distinct fashions depending 
on the nature of both the metal and the organic film; the evaporated metals 
were observed to form chemically reacted [113, 165-176] or inert 
[172, 177-179] interfaces with the film, to form clusters [167, 180-182] 
floating on the organic film surface, or the evaporated metal species were 
reported to diffuse into the bulk of the film [88, 166, 168-174, 183-185]. 

Compared to the organic-on-metal contacts, the metal-on-organic ones have 
been mostly focused to examinations of their chemical structure. 
Understanding the chemical reactions at the interface is a prerequisite for 
the interface barrier control and some contradicting results on the 
metal-on-organic systems point to its intricacy. For example, inert metals 
were reported to form abrupt interface in contrast to the reactive ones 
diffusing into the organics [172], while the opposite trend was observe in 
Ref. [168], namely, the in-diffusion of weakly interacting and formation of 
reacted interface on the organics' surface. 

The photoemission studies at the onset of the interface formation obviously 
deal with minute amounts of metals in the range of---nominally---monolayers.
The conclusions on the in-diffusion were often based on the 
observation that the photoemission signal of an evaporated element was 
incommensurate to the amount expected on the surface; specifically, the 
lower photoemission signal indicated a lesser amount of the evaporated metal 
and thereby suggesting its losses due to diffusion into a bulk, thus out of 
the probed depth. The chemical reactions are deduced via the `chemical 
shift' directly observed by photoemission, that is the (often 
high-$E_{B})$ shift of a core level corresponding to the reacted element. 

The next example illustrates how spurious conclusions, namely a metal 
diffusion into the organic layer with a concomitant chemical reaction, can 
be deduced [166]. Figure~4.3 shows the evolution of In $4d$ core level upon the 
incremental evaporation of indium onto CuPc film measured at two photon 
energies of 68 eV (a) and 335 eV (b) to set the `surface sensitive mode' 
and `bulk mode', as the corresponding electron inelastic mean free paths 
$\lambda$ were about 2 and 4 monolayers, respectively. A doublet emerging 
at about 19 eV (Fig.~4.3a), thus shifted by about 2 eV towards high 
$E_{B}$ from the In $4d$ detected for higher In coverage reveals a reacted 
component suggesting the reaction between In and CuPc. Yet, the comparison 
with the measurement at higher photon energy (Fig.~4.3b) suggests that the 
`reacted' component is more pronounced in the surface-sensitive mode, it 
therefore comes out rather from the In surface instead of the In/CuPc 
interface. 

\begin{figure}[tb]
\begin{center}
\includegraphics[width=9.0cm,clip]{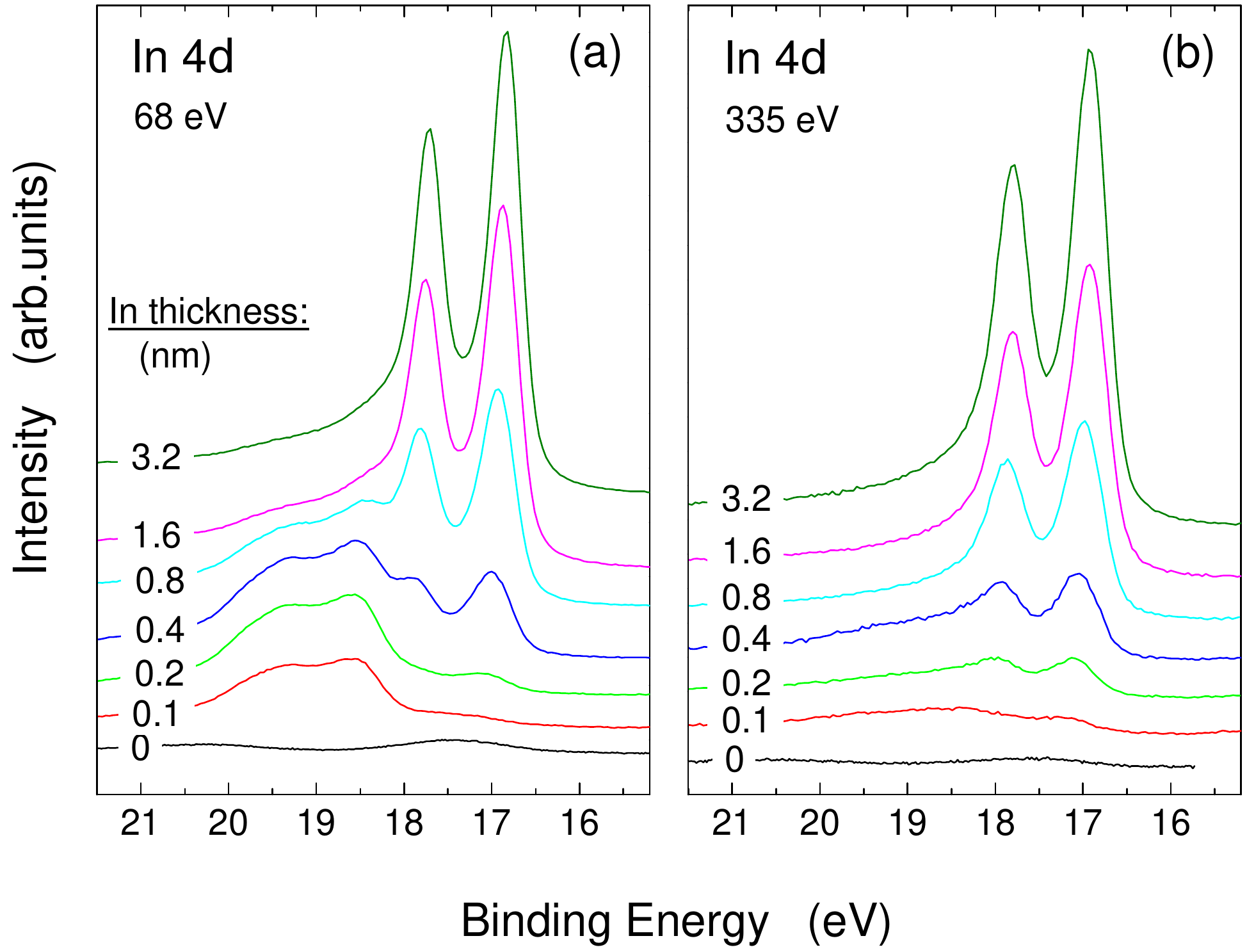}
\caption{Evolution of the In $4d$ core-level photoemission spectrum recorded
at the photon energy of (a) 68 eV and (b) 335 eV upon the incrementally
increased indium deposited on a thick CuPc layer. The nominal indium
thicknesses are indicated next to the spectra. Reprinted with permission
from J. Ivanco {\it et al.}, Phys. Rev. B {\bf 81} (2010) 115325. Copyright
2010, {\it American Physical Society}.}
\vspace*{-0.4cm}
\end{center}
\end{figure}
\begin{figure}[!htbp]
\begin{center}
\includegraphics[width=4.6cm,clip]{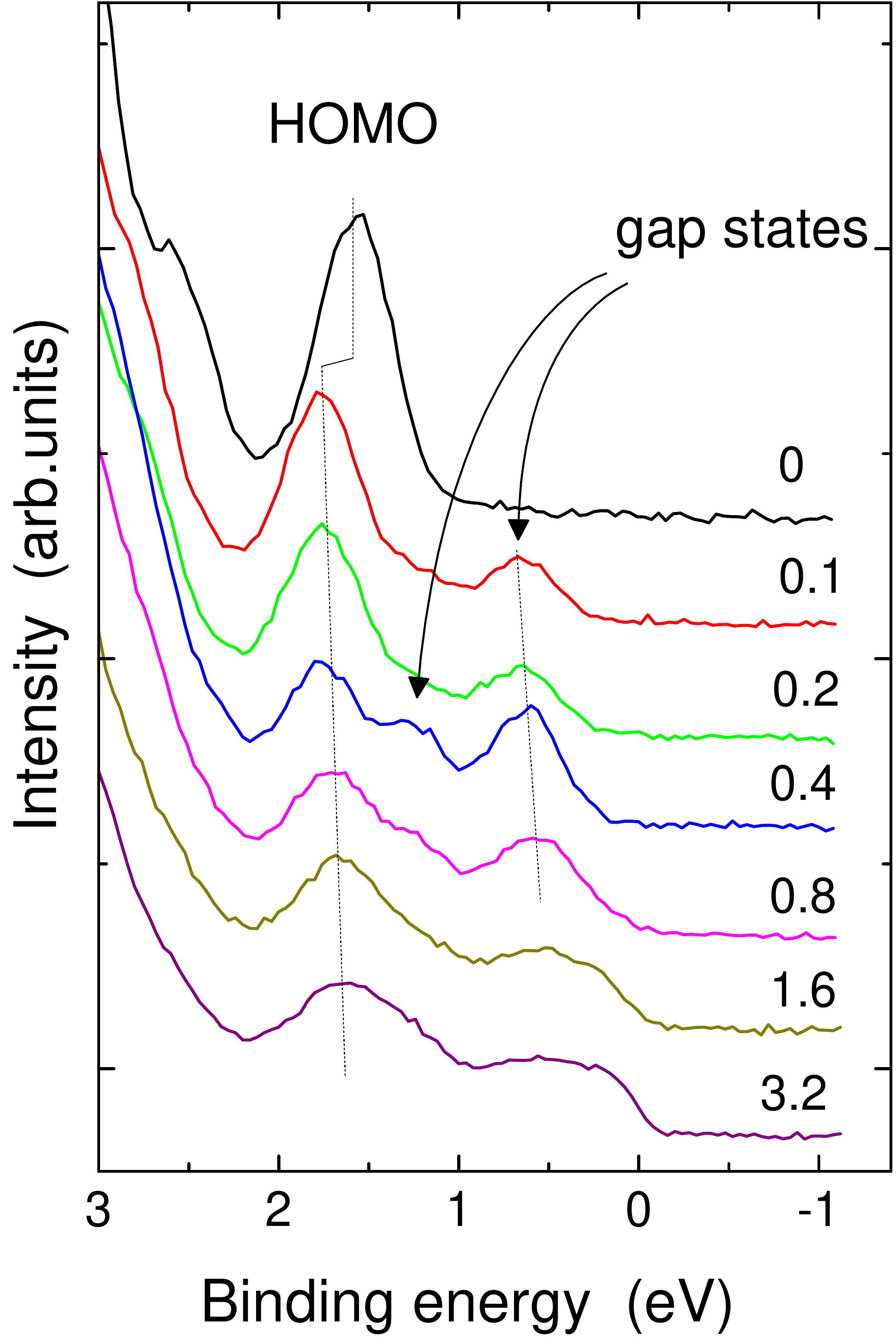}
\caption{Evolution of the valence band photoemission spectrum of a thick CuPc
layer upon the stepwise increased indium coverage. The nominal indium thicknesses
in nanometers are given next to the spectra. The arrows indicate new orbitals
regarded as gap states. Reprinted with permission from J. Ivanco {\it et al.},
Phys. Rev. B {\bf 81} (2010) 115325. Copyright 2010, {\it American Physical Society}.}
\vspace*{-0.4cm}
\end{center}
\end{figure}

Furthermore, an inspection of the CuPc valence band near the Fermi level 
(Fig.~4.4) reveals peaks emerging between the HOMO and the Fermi level. The 
peaks can be interpreted as gap states created in the energy gap of CuPc 
owing to the reaction between In and CuPc and thereby support the notion on 
the reactive In/CuPc interface [166]. 

Yet, photoemission spectra of nanoclusters such as Au and Ag (note that 
nanoclustering of In on CuPc was confirmed by the secondary electron 
microscopy [186]) manifest the increasing high-$E_{B}$ shift with the 
decreasing size due to the final-state effect [187-192]. Further, the Fermi 
level edge of such nanoclusters is not developed and the photoemission 
signal is high-$E_{B}$ shifted too. Consequently, an ensemble of even inert 
nanoclusters with varying size can produce superposed spectra with varying 
shift masquerading as the chemical shift and accordingly to allure a 
conclusion on the chemical reaction between metal and the substrate. 
Likewise the core levels, the Fermi edge of nanoclusters is high-$E_{B}$ 
shifted and simply superposed to the $\pi$ band of the underlying molecular 
film. Accordingly, the mere presence of a feature in the energy band does 
not prove a chemical reaction. Evidently, not all organic/metal systems are 
inert; for example, the chemical reaction can occur between metal with a 
high oxidation potential and molecule comprising oxygen [172]. 

\begin{figure}[tb]
\begin{center}
\includegraphics[width=6.0cm,clip]{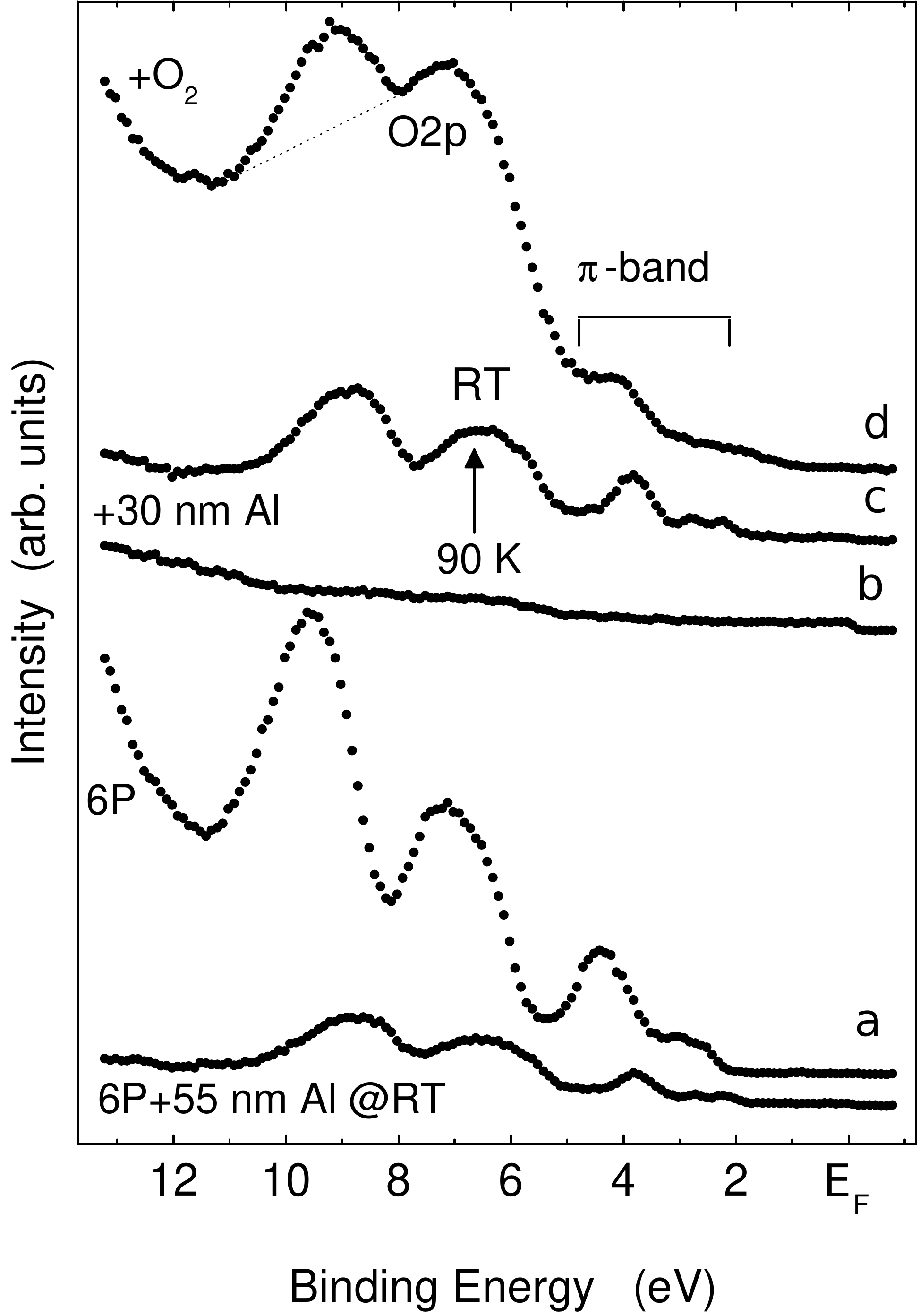}
\caption{Photoemission spectra of a sexiphenyl film upon 55-nm thick
RT-evaporated Al (bottom spectrum). Further spectra: (a) 15 nm sexiphenyl,
(b) after adding 30 nm of Al at 90 K, (c) after warming to RT, and (d) after
oxygen exposure of 4000 langmuirs. Reprinted with permission from J. Ivanco
{\it et al.}, Appl. Phys. Lett. {\bf 85} (2004) 585. Copyright 2004, {\it AIP
Publishing LLC}.}
\vspace*{-0.5cm}
\end{center}
\end{figure}

The In/CuPc system addressed above was prepared in UHV with the base 
pressure about $10^{-10}$ mbar. Likewise, aluminium evaporated over 
sexiphenyl surface in UHV showed marked departure from the laminar growth as 
the 6P spectra were visible even after nominally 55 nm-thick Al overlayer 
(bottom spectrum in Fig.~4.5). In contrast, evaporation of nominally only 3 
nm-thick Al layer, but at about liquid nitrogen temperature (curve $b$), led to 
the complete elimination of the substrate signal of 6P suggesting the 
laminar growth of Al. Eventually, the 6P spectrum emerged upon warming the 
Al/6P system to RT (curve $c$) suggesting either clustering or in-diffusion of 
Al species. Yet, an oxygen-related feature emerges upon the followed 
exposure to oxygen (curve $d$) implying that aluminium remained on the 6P 
surface as opposed to the in-diffusion. 

The clustering of Al on 6P surface (or $e.g.$ of In on CuPc discussed above) can 
be rationalized by means of surface-free energy arguments (Fig.~4.6). The minimization 
of the total surface free energy of the Al/6P system can be expressed by 
Young's equation [193]:
\begin{equation}
\label{eq12}
\gamma _{6P} = \gamma _{Al} \cos \theta + \gamma _{Al / 6P} ,
\end{equation}
where $\gamma_{6P}$ and $\gamma_{Al}$ are the surface free energies of the sexiphenyl 
substrate and of the Al overlayer, respectively. The $\gamma_{Al / 6P}$ is the 
interface free energy, and $\theta$ is the contact angle of the Al island. 
With $\gamma_{Al} = 1.1$ Jm$^{-2}$, $\gamma_{6P} \approx 0.03$ Jm$^{ - 2}$ [194], 
and $\gamma_{6P / Al}\sim 0.17$ Jm$^{-2}$ [180], Eq. (\ref{eq12}) leads to the 
contact angle $\theta$ of 97 degree. 

\begin{figure}[tb]
\begin{center}
\includegraphics[width=5.0cm,clip]{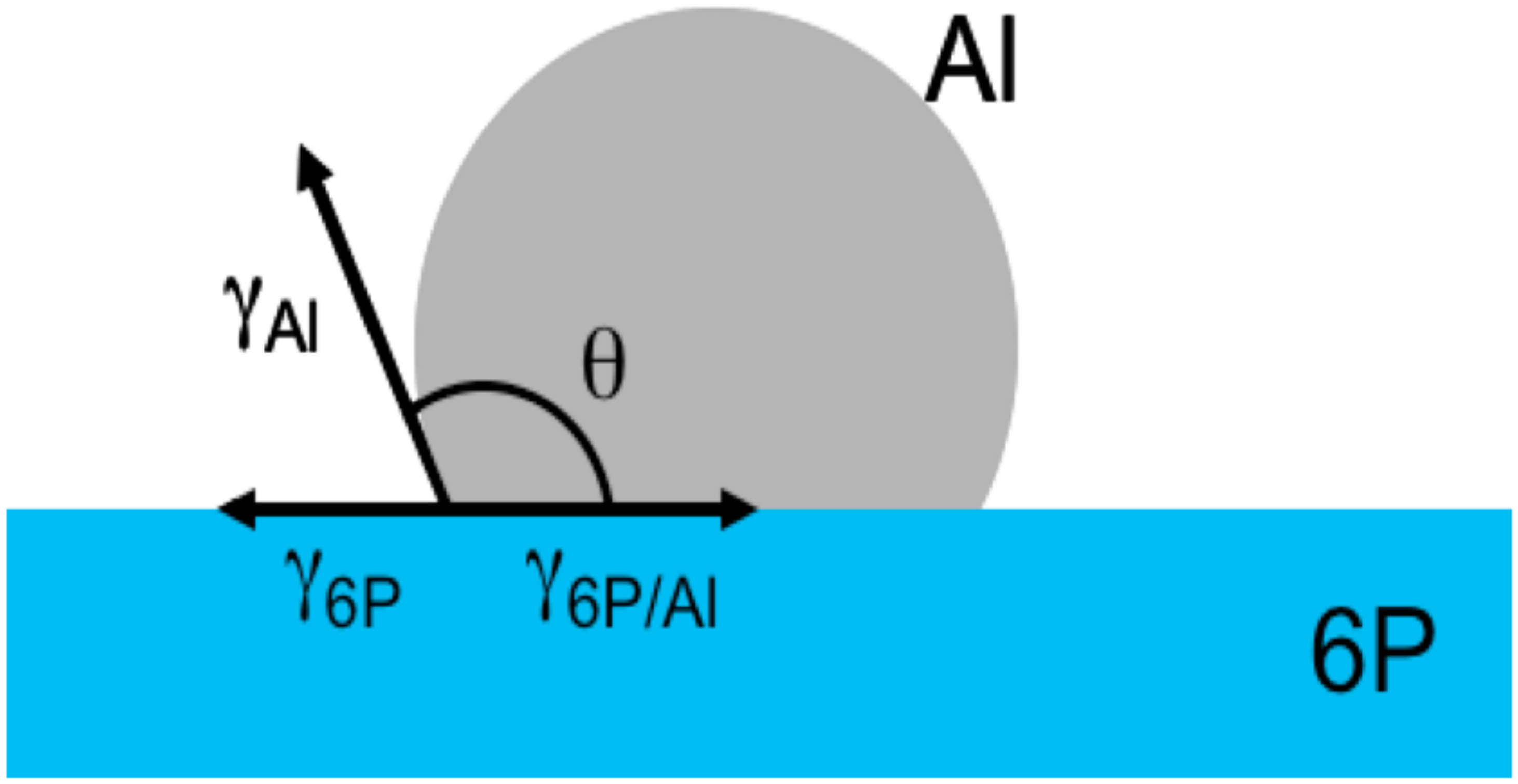}
\caption{Aluminium island on a surface of sexiphenyl.}
\vspace*{-0.5cm}
\end{center}
\end{figure}
\begin{figure}[!tbp]
\begin{center}
\includegraphics[width=6.0cm,clip]{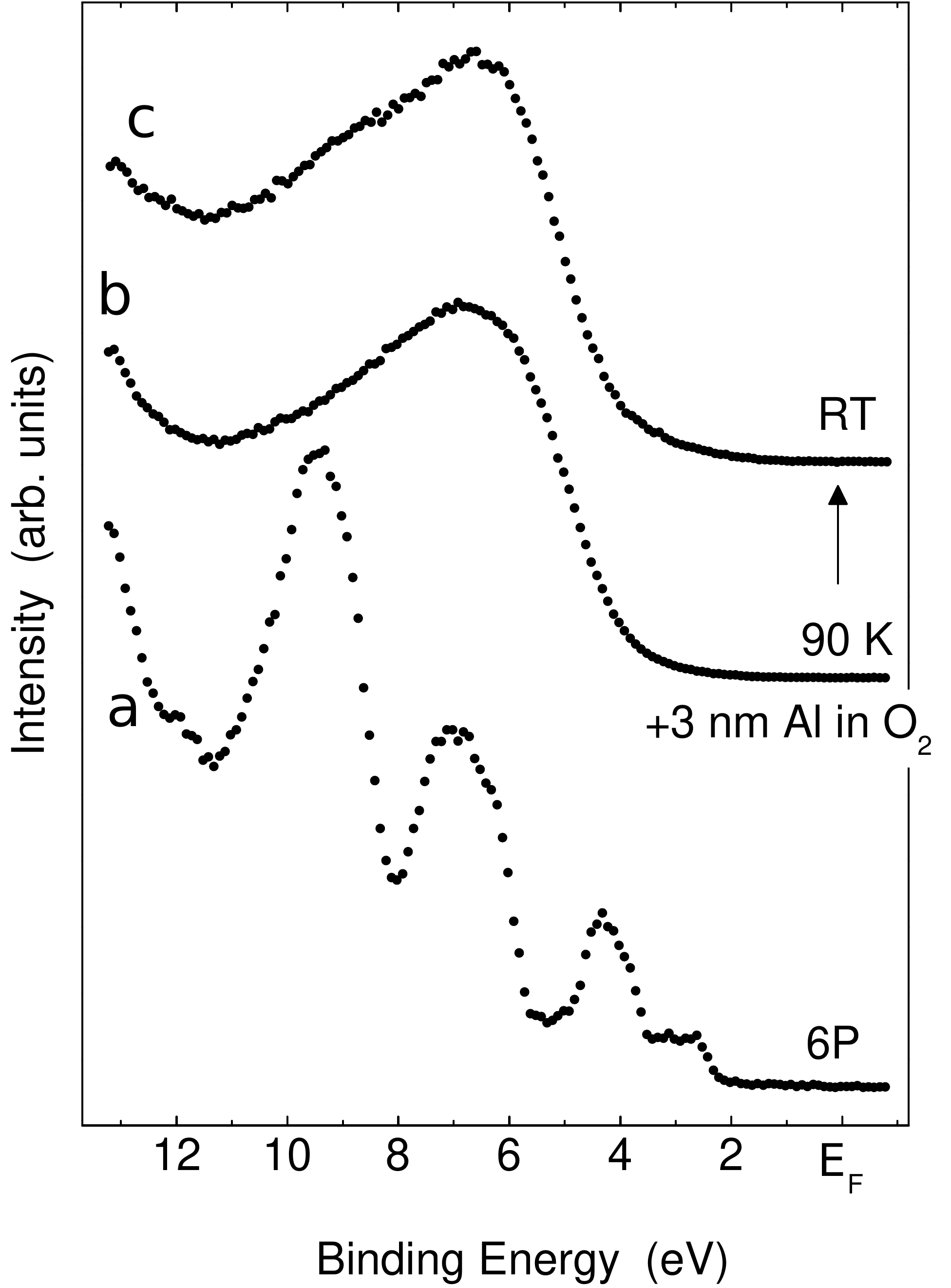}
\caption{Photoemission spectrum (a) of $15$ nm of sexiphenyl, (b) after adding
$3$ nm of Al evaporated at oxygen pressure of $10^{-7}$ mbar at $90$ K, and (c) upon
warming to RT. Reprinted with permission from J. Ivanco {\it et al.}, Appl.
Phys. Lett. {\bf 85} (2004) 585. Copyright 2004, {\it AIP Publishing LLC}.}
\vspace*{-0.5cm}
\end{center}
\end{figure}

Formation of 3D aluminium clusters naturally imposes a heavy departure from 
the laminar growth manifested by the markedly delayed onset of Al-related 
photoemission features; it accordingly leads to the underestimate of the 
amount of metal located on the surface (bottom curve in Fig.~4.5),
and---more importantly---it avoids the formation of continuous 6P/Al interface. 
The high $\gamma$ of aluminium, which precludes the wetting of the 
low-$\gamma$ 6P surface, can be diminished by oxidation of Al, since oxidation 
lowers $\gamma$. This is shown in Fig.~4.7, where only 3 nm of Al, however 
evaporated in high vacuum (with the oxygen partial pressure $10^{-7}$ 
mbar) ensures a continuous wetting layer eliminating the signal of 
underlying 6P film also at RT [180].

\begin{figure}[tb]
\begin{center}
\includegraphics[width=13.4cm,clip]{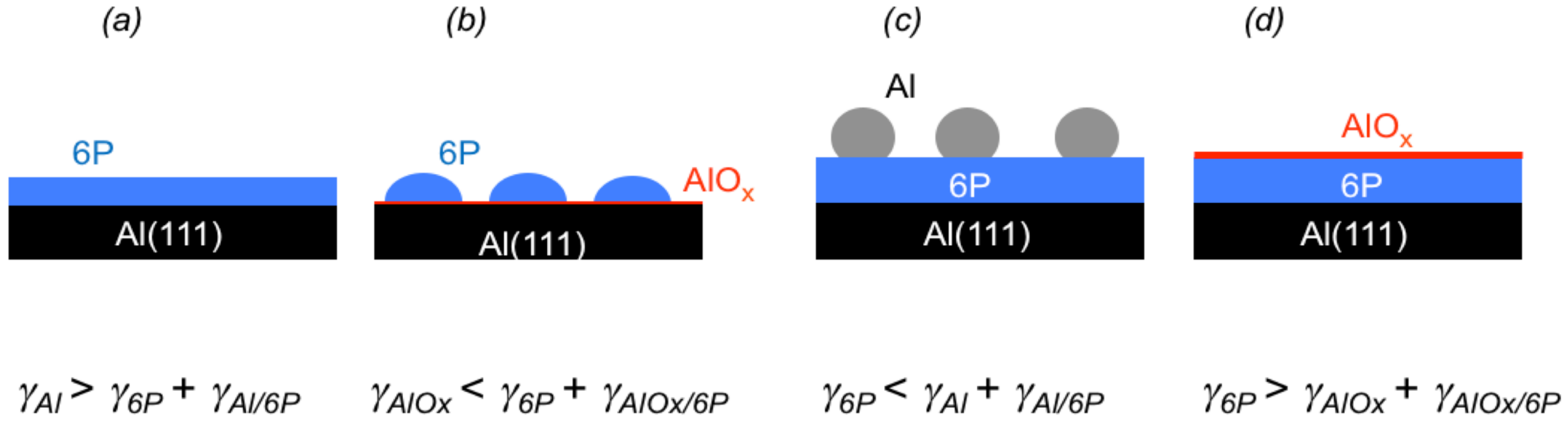}
\caption{Various process sequences for the preparation of an interface between
materials with distinct free surface energies $\gamma$, such as Al, AlO$_x$,
and 6P, whereas $\gamma_{Al} > \gamma_{AlO_x} \approx \gamma_{6P}$.}
\vspace*{-0.5cm}
\end{center}
\end{figure}

Various process sequences examined in Figs.~4.5 and 4.7 are summarized in 
Fig.~4.8. Due to the inequality $\gamma_{Al}\gg\gamma_{6P}$, 3D Al islands are 
favoured to grow on sexiphenyl surface [Fig.~4.8(c)], and, conversely, 
sexiphenyl wets the Al surface (panel $a$). If Al is oxidized ($e.g$., by 
evaporation in the ambient with higher oxygen partial pressure, such as in 
high vacuum), the decrease in aluminium surface free energy can reverse the 
inequality and result in wetting of the sexiphenyl surface (panel $d$). For 
sexiphenyl tends to grow on AlO$_x$ in Stransky-Krastanov or Volmer-Weber 
growth fashion, apparently $\gamma_{6P}$ approximately equals or is higher 
than $\gamma_{AlO_{x}}$ (panel $b$).

\newpage
\section{Summary}

The review addresses electronic, chemical, and geometric properties of 
electronically active molecular film and related interfaces probed by 
photoemission, which are pursued with the consideration to the molecular 
orientation. It accents several issues necessary to consider while employing 
photoemission characterization for examination of molecular films and 
related interfaces. In essence: 
\begin{itemize}
\item[$\bullet$] Dependence of electronic properties on the molecular orientation 
necessitates the extension of the standard discernment of growth fashions 
(laminar, islanding). Neglecting to consider the molecular orientation can 
lead to erroneous conclusions on the presence of band bending as the effect 
of varying work function implied by the orientational transition of 
molecules masquerades as the band bending. 
\item[$\bullet$] Molecular films can be characterized by their intrinsic work function 
analogously to inorganic elemental electronic materials ($e.g.$ metals). The 
intrinsic work function of molecular films is an experimentally accessible 
and it is an essential parameter in reference to the proposed model on 
energy level alignment (ELA); the model accounts the equalization of 
electrochemical potentials of materials forming a contact to be the 
controlling mechanism for the ELA at the film/substrate interface. 
\item[$\bullet$] The vacuum-level offset $\Delta E_{vac}$ measured by photoemission upon the 
film growth converges to the difference between the initial work functions 
of the substrate and the intrinsic work function of the film with the 
increasing film thickness.
\item[$\bullet$] Interaction-strength model employed to rationalize the lying and upright 
molecular orientation seems invalid and the molecular orientation of 
rod-like molecules is determined by the morphology of the substrate surface.
\item[$\bullet$] Due to markedly higher surface free energy of metal compared to organic, 
metal does not wet the organics` surface in the absence of chemical reaction 
and the formation of 3D metal nanoclusters are favoured at onset of the 
interface formation. Neglecting the metal morphology on photoemission 
characterization in this stage can eventuate in erroneous conclusions on the 
interfacial chemical and electronic structure. 
\end{itemize}

\section*{Acknowledgement}
\addcontentsline{toc}{section}{Acknowledgement}

The supports of the Slovak Research and Development Agency under pro\-ject No. 
APVV-0096-11 and from Grant Agency VEGA Bratislava under project 
No.2/0162/12 are acknowledged.

\newpage
\fancyhead[LO]{References}

\end{document}